\documentclass[12pt,english,onecolumn]{IEEEtran}
\usepackage[T1]{fontenc}
\usepackage[latin1]{inputenc}
\usepackage{subfigure}
\usepackage{amsmath}
\usepackage{graphicx}
\usepackage{amssymb}

\makeatletter
 \usepackage{verbatim}
 \newtheorem{problem}{Problem}
 \newtheorem{thm}{Theorem}
 \newtheorem{cor}{Corollary}
 \newtheorem{prop}{Proposition}
 \newtheorem{lemma}{Lemma}
 \newtheorem{example}{Example}

\usepackage{cite}
\usepackage{url}

\usepackage{babel}
\makeatother
\begin{document}

\title{On the Information Rate of MIMO Systems with Finite Rate Channel
State Feedback Using Beamforming and Power On/Off Strategy }

\author{Wei Dai$^{1}$, Youjian Liu$^{1}$, Vincent K.N. Lau$^{2}$ and Brian
Rider$^{1}$\\
dai@colorado.edu, eugeneliu@ieee.org, knlau@ieee.org and brider@euclid.colorado.edu\\
$^{1}$University of Colorado at Boulder\\
$^{2}$University of Hong Kong}

\maketitle
\begin{abstract}
It is well known that Multiple-Input Multiple-Output (MIMO) systems
have high spectral efficiency, especially when channel state information
at the transmitter (CSIT) is available. When CSIT is obtained by feedback,
it is practical to assume that the channel state feedback rate is
finite and the CSIT is not perfect. For such a system, we consider
beamforming and power on/off strategy for its simplicity and near
optimality, where power on/off means that a beamforming vector (beam)
is either turned on with a constant power or turned off. The main
contribution of this paper is to accurately evaluate the information
rate as a function of the channel state feedback rate. Name a beam
turned on as an \emph{on-beam} and the minimum number of the transmit
and receive antennas as the \emph{dimension} of a MIMO system. We
prove that the ratio of the optimal number of on-beams and the system
dimension converges to a constant for a given signal-to-noise ratio
(SNR) when the numbers of transmit and receive antennas approach infinity
simultaneously and when beamforming is perfect. Asymptotic formulas
are derived to evaluate this ratio and the corresponding information
rate per dimension. The asymptotic results can be accurately applied
to finite dimensional systems and suggest a power on/off strategy
with a constant number of on-beams. For this suboptimal strategy,
we take a novel approach to introduce \emph{power efficiency factor},
which is a function of the feedback rate, to quantify the effect of
imperfect beamforming. By combining power efficiency factor and the
asymptotic formulas for perfect beamforming case, the information
rate of the power on/off strategy with a constant number of on-beams
is accurately characterized.
\end{abstract}

\section{Introduction\label{sec:Introduction}}


This paper considers multiple-input multiple-output (MIMO) systems
with finite rate channel state feedback. Multiple-antenna wireless
communication systems, also known as MIMO systems, have high spectral
efficiency. It is also well known that the capacity of MIMO systems
with channel state information (CSI) at the transmitter (CSIT) is
generally higher than the systems without it. When perfect CSI is
available at both transmitter and receiver (CSITR), the MIMO channel
can be viewed as a set of parallel sub-channels. The transmission
power on each sub-channel obeys water filling principle \cite{Telatar_EuroTele99_Capacity_MIMO}.
If CSIT is obtained from channel state feedback, however, perfect
CSIT requires infinite feedback rates, which is not practical. On
the other hand, in practical systems such as UMTS-HSDPA \cite{Standard_3Gpp},
there is a control field which can be used to carry a certain number
of channel state feedback bits on a per-fading block basis. It is
reasonable to consider MIMO systems with finite rate channel state
feedback. 

For a given feedback rate, this paper tries to answer two basic questions,
how much benefit the feedback can bring and how to exploit the feedback
to achieve that benefit. It is difficult to answer these two questions
in general. To achieve or calculate the information rate for a given
feedback rate, the optimal transmission strategy and the optimal feedback
strategy need to be found. It has been shown in \cite{Lau_IT04_Capacity_Memoryless_Block_Fading,Lau_COM04_Desing_MIMO_block_fading_feedback}
that the design of transmission and feedback strategies is an unconventional
optimization problem. For memoryless channels, it is proved that the
information theoretic limit can be achieved by memoryless transmission
and feedback strategies. However, the explicit forms of the optimal
strategies are still unknown. Lloyd algorithm is resorted to obtain
suboptimal numerical solution in \cite{Lau_IT04_Capacity_Memoryless_Block_Fading,Lau_COM04_Desing_MIMO_block_fading_feedback}.

On the other hand, the optimization problems can be simplified if
the transmission strategy is restricted to power on/off strategy (with
beamforming). In a general setting, the optimal transmission strategy
is to choose the covariance matrix of the transmitted Gaussian coded
symbols according to the current feedback \cite{Lau_IT04_Capacity_Memoryless_Block_Fading,Lau_COM04_Desing_MIMO_block_fading_feedback}.
By the singular value decomposition, the covariance matrix can be
decomposed to a unitary matrix and a non-negative diagonal matrix
which are called as beamforming matrix and power control matrix respectively.
We describe each column vector of the beamforming matrix as a \emph{beam}
and the diagonal element corresponding to a beam as the \emph{power}
on that beam. The power on/off strategy means that a beam is either
turned on, i.e., its power is a positive constant $P_{\mathrm{on}}$,
or turned off, i.e., its power is zero. As we will show later, the
power on/off and beamforming assumption simplifies the analysis. Although
power on/off is suboptimal, it has been shown in \cite{Goldsmith_Comm01_Degrees_freedom_adaptive_modulation}
and \cite{Chow_GlobeCom92_Bandwidth_optimization} that power on/off
can achieve performance close to water filling power control for single
antenna systems and parallel Gaussian channels respectively. This
paper will show that power on/off is near optimal for MIMO channels
as well.

The main contribution of this paper is to accurately characterize
the information rate of the power on/off strategy with finite rate
channel state feedback. Name a beam turned on as an \emph{on-beam}.
The optimization problem corresponding to power on/off strategy is
to find the optimal number of on-beams, which is related to power
control, and the directions of the on-beams, which is called as beamforming,
according to the channel realization. Both power control and beamforming
have influence on the overall information rate. By analyzing these
two effects separately, this paper is able to characterize the overall
information rate accurately.

To isolate the effect of beamforming, we first discuss the perfect
beamforming case. Perfect beamforming means that the beamforming matrix
at the transmitter changes the MIMO channel to parallel channels without
interference. We analyze this case by asymptotics, where the numbers
of the transmit and receive antennas approach infinity simultaneously.
The derived asymptotic results are as follows.

\begin{itemize}
\item Define the minimum number of transmit and receive antennas as the
\emph{dimension} of a MIMO system. We prove that the ratio of the
optimal number of on-beams and the system dimension converges to a
constant for a given signal-to-noise ratio (SNR) and perfect beamforming.
This result suggest a \emph{power on/off strategy with a constant
number of on-beams}. The assumption of a constant number of on-beams
is crucial to analyze the effect of imperfect beamforming.
\item We also prove that the optimal number of on-beams is a non-decreasing
function of SNR.
\item We derive asymptotic formulas to simplify the calculations. By following
the method developed in \cite{Popescu_TComm00_Capacity_Random_Signature_MIMO,Hughes_SSC02_Gaussian_approximation_capacity_MIMO_Rayleigh_fading},
we derive asymptotic formulas to evaluate the optimal number of on-beams
and the corresponding information rate, which are obtained by simulation
traditionally. Furthermore, for the CSITR case, asymptotic formulas
are derived to calculate the Lagrange multiplier required for water
filling power control and the corresponding channel capacity for the
first time.
\end{itemize}
It is noteworthy that the asymptotic results are accurate enough for
MIMO systems with finite many antennas.

Then we quantify the effect of imperfect beamforming accurately by
assuming a constant number of on-beams. There are many works studying
similar problems. Some works add some structures to make the MIMO
system equivalent to a single-input single-output (SISO) system. The
structures could be single receive antenna \cite{Sabharwal_IT03_Beamforming_MIMO,Zhou_TC05_Multiantenna_adaptive_modulation_beamforming,Zhou_WCom05_Quantifying_power_loss_transmit_beamforming,Rao_spcom04_VQ_MIMO_CSI_feedback,Rao_asil04_MIMO_VQ_feedback}
and single beam for a single data stream \cite{Love_IT03_Grassman_Beamforming_MIMO,Heath_asil04_Lower_bound_Outage_limited_feedback_MIMO,Heath_ITW04_upper_bound_SNR_FeedbackMIMO,Heathl_allerton04_Performance_bounds_limited_feedback_MIMO}.
For MIMO systems with multiple beams, transmit antenna subset selection
is viewed as a special case. Different antenna selection criteria
are proposed in \cite{Love_asilomar03_dual_mode_antenna_selection_linear_receivers,Love_allerton03_multi_mode_antenna_selection_spatial_multiplexing}
and the effect on information rate is analyzed for extreme SNR regimes
in \cite{Blum_CommLett02_MIMO_Antenna_Selection,Sanayei_2003_asymptotic_capacity_gain_transmit_antenna_selection},
whose analysis is hard to be generalized to other SNR regimes. For
general multiple beams, assuming that the transmitter knows some singular
vectors of the channel matrix perfectly, power allocation to maximize
information rate is discussed in \cite{Rao_TWC04_mutiple_antenna_channels_partial_CSIT}
and beamforming matrix selection to minimize Bit Error Rate (BER)
is proposed in \cite{Zhou_SP05_BER_Criterion_Codebook_Finite_Rate_Multiplexing_linear_receiver}.
More practically, if the information about the channel state is obtained
through a finite rate feedback, it is reasonable to assume that the
transmitter only knows quantized information about the channel state.
The popular strategy is to construct a finite size beamforming codebook
and select a beamforming matrix for transmission according to the
channel state feedback. The algorithms to construct a beamforming
codebook are proposed in \cite{Hochwald_IT00_design_unitary_spacetime_codes,Rao_pimrc04ltr_Channel_Quantization_MIMO_MISO,Rao_wcnc04_efficient_feedback_MIMO_slowly_fading}.
The beamforming codebook design criteria and the beamforming matrix
feedback criteria, which are often coupled, are discussed in \cite{Love_SP05_Limited_feedback_unitary_precoding,Love_SP05_multi_mode_precoding_MIMO,Love_IT05sub_limited_feedback_unitary_precoding,Zhou_SP05_BER_Criterion_Codebook_Finite_Rate_Multiplexing_linear_receiver,Rao_icc05_MIMO_spatial_multiplexing_limit_feedback}.
Based on Grassmann manifolds, the effect of finite beamforming on
performance is analyzed in \cite{Love_IT05sub_limited_feedback_unitary_precoding,Love_SP05_Limited_feedback_unitary_precoding}
and refined later in \cite{Heath_ICASSP05_Quantization_Grassmann_Manifold},
all of which are based on Barg's formula \cite{Barg_IT02_Bounds_Grassmann_Manifold}
which is only valid for MIMO systems with asymptotically large number
of transmit antennas but fixed finite receive antennas. Applied for
all MIMO systems, the performance of finite beamforming is analyzed
for high SNR region in \cite{Rao_icc05_MIMO_spatial_multiplexing_limit_feedback},
which is difficult to be generalized to other SNR regimes. Valid for
all SNR regimes, the information rate is quantified in \cite{Honig_Allerton03_Benefits_Limited_Feedback_Wireless_Channels,Honig_MiliComm03_Asymptotic_MIMO_Limited_Feedback}
by letting the numbers of transmit and receive antennas approach infinity
simultaneously and applying extreme order statistics. The proposed
formula over-estimates the performance. A correction of the result
is in \cite{Honig_discussion}. In the presenting paper, we take a
novel approach by introducing the \emph{power efficiency factor} to
quantify the effect of imperfect beamforming. The power efficiency
factor can be calculated using a closed form formula derived in \cite{Dai_05_Quantization_Grassmannian_manifold},
which is valid for MIMO systems with arbitrary number of antennas.
As a result, the information rate is accurately analyzed as a function
of feedback rate. The analysis matches the simulations almost perfectly
for all SNR regimes.

Finally, we show the near optimality of the power on/off strategy
with a constant number of on-beams by comparing it with a general
power on/off strategy. For a general power on/off strategy, we derive
the optimal feedback strategy for a given arbitrary beamforming codebook.
Then we are able to compare the two different power on/off strategies
numerically. Simulations show that a constant number of on-beams is
near optimal for all SNR regimes. Therefore, power on/off strategy
with a constant number of on-beams provides a simple but near optimal
solution.


This paper is organized as follows. The system model and the related
design problem are outlined in Section \ref{sec:Preliminaries}, where
preliminary knowledge about random matrices and Stiefel and Grassmann
manifolds are also presented. In Section \ref{sec:Power-On/Off-Perfect-Beamforming},
the power on/off strategy with a constant number of on-beams is derived
as the optimal solution for perfect beamforming. Section \ref{sec:StrategyDesign_with_beamforming_codebook}
considers the effect of imperfect beamforming due to finite rate channel
state feedback. Section \ref{sec:Performance-Comparision} shows that
a constant number of on-beams is also near optimal for imperfect beamforming.
Conclusions are given in Section \ref{sec:Conclusions}.

\section{Preliminaries\label{sec:Preliminaries}}


In this section, we first describe the system model. Then we present
some preliminary knowledge about random matrices and Stiefel and Grassmann
manifolds.


In this paper, we use $\mathbb{Z}^{+}$ to denote the set of positive
integers, $\mathbb{R}^{k}$ and $\mathbb{C}^{k}$ to denote the $k$-dimensional
real and complex vector spaces respectively, $\mathbb{C}^{k\times l}$
to denote the vector space of $k\times l$ complex matrices, $\mathbf{I}_{k}$
to denote the $k\times k$ identity matrix, $\mathbf{A}^{\dagger}$
to denote the conjugate transpose of a matrix $\mathbf{A}$, $\mathrm{tr}\left(\cdot\right)$
to denote the trace of a matrix, $\mathrm{rank}\left(\cdot\right)$
to denote the rank of a matrix, $\left\Vert \cdot\right\Vert _{F}$
to denote the matrix Frobenius norm, $\left|\cdot\right|$ to denote
the determinant of a matrix or the cardinality of a set according
to its context, $\mathrm{E}_{X}\left[\cdot\right]$ to denote the
expectation with respect to the random variable $X$, $\arg\;\max$
and $\arg\;\min$ to denote the functions that return the global maximizer
and minimizer respectively.

\subsection{System Model and the Corresponding Design Problem\label{sub:System-Model}}

\begin{figure}
\includegraphics[%
  clip]{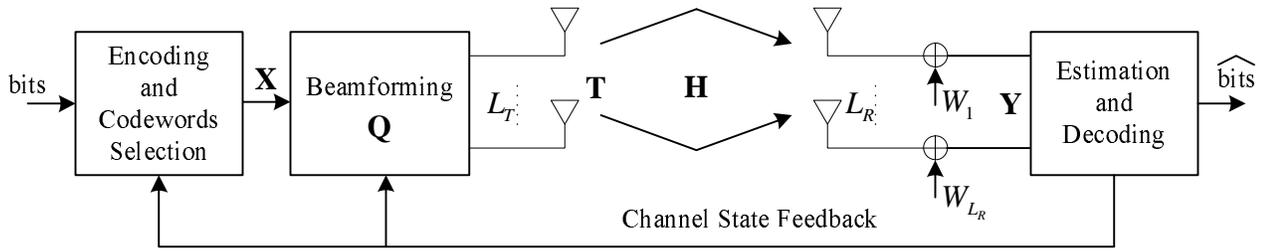}

\caption{\label{cap:Fig-system-model}System model}
\end{figure}

A communication system with $L_{T}$-transmit antennas and $L_{R}$-receive
antennas is shown in Fig. \ref{cap:Fig-system-model}. Let $\mathbf{T}\in\mathbb{C}^{L_{T}\times1}$
be the transmitted signal, $\mathbf{Y}\in\mathbb{C}^{L_{R}\times1}$
be the received signal, $\mathbf{H}\in\mathbb{C}^{L_{R}\times L_{T}}$
be the channel state matrix and $\mathbf{Z}\in\mathbb{C}^{L_{R}\times1}$
be the Gaussian noise with zero mean. The system model can be expressed
as \[
\mathbf{Y}=\mathbf{HT}+\mathbf{Z}\]
 where $\mathrm{E}\left[\mathbf{Z}\mathbf{Z}^{\dagger}\right]=\mathbf{I}_{L_{R}}$.
In this paper, the Rayleigh flat fading channel is considered: the
entries of $\mathbf{H}$ are independent and identically distributed
(i.i.d.) circularly symmetric complex Gaussian variables with zero
mean and unit variance ($\mathcal{CN}\left(0,1\right)$) and $\mathbf{H}$
is i.i.d. for each channel use%
\footnote{This is a suitable model for the block fading channel when the channel
state can be estimated and fed back at the beginning of each fading
block.%
}. At the beginning of each channel use, the channel state $\mathbf{H}$
is assumed to be perfectly estimated at the receiver, then quantized
to finite bits and fed back to the transmitter through a feedback
channel. The feedback channel is assumed to be error-free and zero-delay.
The rate of the feedback is up to $R_{\mathrm{fb}}$ bits/channel
use. After receiving the channel state feedback, the transmitter transmits
the encoded signal according to the current feedback%
\footnote{For i.i.d. channel states, the memoryless transmission and feedback
strategy can achieve the information theoretic limit provided by the
finite rate channel state feedback \cite{Lau_IT04_Capacity_Memoryless_Block_Fading}.%
}.

The general design problem for finite rate channel state feedback
is difficult to solve. It is well known that the optimal transmitted
signal should be circular symmetric Gaussian signal with zero mean
and covariance matrix adapted according to the feedback \cite{Lau_IT04_Capacity_Memoryless_Block_Fading}.
Define the covariance matrix of the transmitted signal as $\mathbf{\Sigma}\triangleq E\left[\mathbf{T}\mathbf{T}^{\dagger}\right]$,
the codebook of the covariance matrices as \[
\mathcal{B}_{\mathbf{\Sigma}}=\left\{ \mathbf{\Sigma}_{i}\in\mathbb{C}^{L_{T}\times L_{T}}:\;1\leq i\leq2^{R_{\mathrm{fb}}}\right\} \]
and the feedback function $\varphi\left(\cdot\right)$ as a mapping
from the space of $\mathbf{H}$ to a index set $\left\{ i:\;1\leq i\leq2^{R_{\mathrm{fb}}}\right\} $.
The corresponding optimization problem is to find the optimal codebook
$\mathcal{B}_{\mathbf{\Sigma}}$ and the optimal feedback function
$\varphi\left(\cdot\right)$ to maximize the information rate\[
\underset{\mathcal{B}_{\mathbf{\Sigma}}}{\max}\;\underset{\varphi\left(\cdot\right)}{\max}\;\mathrm{E}_{\mathbf{H}}\left[\mathrm{ln}\left|\mathbf{I}+\mathbf{H\Sigma}_{\varphi\left(\mathbf{H}\right)}\mathbf{H}^{\dagger}\right|\right],\]
with the average power constraint%
\footnote{The average power constraint $\rho$ is also the average received
SNR because the variance of Gaussian noise is normalized to 1.%
} $\rho$\[
\mathrm{E}_{\mathbf{H}}\left[\mathrm{tr}\left(\mathbf{\Sigma}_{\varphi\left(\mathbf{H}\right)}\right)\right]\leq\rho.\]
It has been shown in \cite{Lau_IT04_Capacity_Memoryless_Block_Fading}
that the design of covariance codebook and the design of feedback
function are two coupled optimization problems and difficult to solve. 

To obtain analytic solution that reflects the influence of feedback
rate on the information rate, we simplify the general problem to suboptimal
power on/off strategy (with beamforming). In the later parts of this
paper, we'll show that power on/off strategy is near optimal. Denote
the singular value decomposition of the covariance matrix as $\mathbf{\Sigma}=\mathbf{QPQ}^{\dagger}$
where the matrices $\mathbf{Q}$ and $\mathbf{P}$ are called as beamforming
matrix and power control matrix respectively. Describe the column
vectors of $\mathbf{Q}$ as \emph{beams}. Name the beam corresponding
to a positive power as an \emph{on-beam}. The statistics of the transmitted
signal is uniquely determined by the on-beams and the power on them.
In our power on/off model, every on-beam corresponds to a constant
power $P_{\mathrm{on}}$. The transmitted Gaussian signal $\mathbf{T}$
can be expressed as\begin{eqnarray*}
\mathbf{T} & = & \mathbf{QX}\end{eqnarray*}
where $\mathbf{X}$ is random Gaussian vector with zero mean and covariance
matrix $P_{\mathrm{on}}\mathbf{I}_{s}$, $s$ is the number of on-beams
and the beamforming matrix $\mathbf{Q}\in\mathbb{C}^{L_{T}\times s}$
is composed of the $s$ on-beams and satisfies $\mathbf{Q}^{\dagger}\mathbf{Q}=\mathbf{I}_{s}$.
The system model for power on/off strategy is given by \begin{eqnarray*}
\mathbf{Y} & = & \mathbf{HQX}+\mathbf{Z}.\end{eqnarray*}

The optimization problem for power on/off strategy is stated in Problem
\ref{pro:(Power-On/off-Strategy-Design)}. Since the number of on-beams
$s$ is the rank of the beamforming matrix $\mathbf{Q}$, the feedback
only needs to specify $\mathbf{Q}$. Denote the codebook of beamforming
matrices as $\mathcal{B}=\left\{ \mathbf{Q}_{i}\in\mathbb{C}^{L_{T}\times s}:\;\mathbf{Q}_{i}^{\dagger}\mathbf{Q}_{i}=\mathbf{I}_{s},\;0\leq s\leq L_{T},\;1\leq i\leq2^{R_{\mathrm{fb}}}\right\} $.
The feedback function is a mapping from the space of $\mathbf{H}$
to the index set $\left\{ i:\;1\leq i\leq2^{R_{\mathrm{fb}}}\right\} $.

\begin{problem}
\label{pro:(Power-On/off-Strategy-Design)}(Power On/Off Strategy
Design Problem) Find the optimal beamforming codebook $\mathcal{B}$,
feedback function $\varphi\left(\cdot\right)$ and $P_{\mathrm{on}}$
to maximize the information rate,\[
\underset{P_{\mathrm{on}}}{\max}\;\underset{\mathcal{B}}{\max}\;\underset{\varphi\left(\cdot\right)}{\max\;}\mathrm{E}_{\mathbf{H}}\left[\mathrm{ln}\left|\mathbf{I}+P_{\mathrm{on}}\mathbf{HQ}_{\varphi\left(\mathbf{H}\right)}\mathbf{Q}_{\varphi\left(\mathbf{H}\right)}^{\dagger}\mathbf{H}^{\dagger}\right|\right],\]
with the average power constraint\[
\mathrm{E}_{\mathbf{H}}\left[P_{\mathrm{on}}\mathrm{tr}\left(\mathbf{Q}_{\varphi\left(\mathbf{H}\right)}\mathbf{Q}_{\varphi\left(\mathbf{H}\right)}^{\dagger}\right)\right]=P_{\mathrm{on}}\mathrm{E}_{\mathbf{H}}\left[s\right]\leq\rho,\]
where $s=s\left(\mathbf{H}\right)=\mathrm{rank}\left(\mathbf{Q}_{\varphi\left(\mathbf{H}\right)}\right)$
is the number of on-beams for a channel realization $\mathbf{H}$.
\end{problem}

As we will show later, the power on/off assumption is the key to decouple
the beamforming codebook design and feedback function design.

\subsection{Random Matrix Theory\label{sub:Random-Matrix-Theory}}


In this subsection, we review relevant results on the spectra of large
random matrices. Recall that $\mathbf{H}$ is an $L_{R}\times L_{T}$
random matrix with i.i.d. complex Gaussian entries with zero mean
and unit variance. Define $m\triangleq\min\left\{ L_{T},L_{R}\right\} $
and $n\triangleq\max\left\{ L_{T},L_{R}\right\} $. Define \[
\mathbf{W}\triangleq\left\{ \begin{array}{cc}
\frac{1}{m}\mathbf{H}\mathbf{H}^{\dagger} & \mathrm{if}\; L_{R}<L_{T}\\
\frac{1}{m}\mathbf{H}^{\dagger}\mathbf{H} & \mathrm{if}\; L_{R}\geq L_{T}\end{array}\right..\]
 Let $\left\{ \lambda_{i}\right\} $ be the set of the eigenvalues
of $\mathbf{W}$. Define the empirical eigenvalue distribution of
$\mathbf{W}$ as\[
F\left(\lambda\right)\triangleq\frac{1}{m}\left|\left\{ j:\;\lambda_{j}<\lambda\right\} \right|.\]
 Then as $m$ and $n$ approach infinity simultaneously with $\tau\triangleq\frac{n}{m}$
fixed, \begin{eqnarray}
\underset{\left(n,m\right)\rightarrow\infty}{\lim}\frac{dF\left(\lambda\right)}{d\lambda}=\left\{ \begin{array}{ll}
\frac{1}{2\pi\lambda}\sqrt{\left(\lambda^{+}-\lambda\right)\left(\lambda-\lambda^{-}\right)} & \mathrm{for}\;\lambda\in\left[\lambda^{-},\lambda^{+}\right]\\
0 & \mathrm{otherwise}\end{array}\right.\label{eq:dF-dlambda-asymptotic}\end{eqnarray}
almost surely where $\lambda^{\pm}=\left(\sqrt{\tau}\pm1\right)^{2}$
\cite{Hughes_IT02_Asymptotic_Rayleigh_MIMO}. Furthermore, consider
a spectral statistical function with the form \[
g\left(\mathbf{W}\right)=\frac{1}{m}\sum_{i=1}^{m}g\left(\lambda_{i}\right).\]
If $g$ is continuous and bounded on $\left[\lambda^{-},\lambda^{+}\right]$,
then \begin{eqnarray}
\underset{\left(n,m\right)\rightarrow\infty}{\lim}g\left(\mathbf{W}\right) & = & \int g\left(\lambda\right)dF\left(\lambda\right)\label{eq:g-lambda-asymptotic}\end{eqnarray}
almost surely \cite{Popescu_TComm00_Capacity_Random_Signature_MIMO,Hughes_SSC02_Gaussian_approximation_capacity_MIMO_Rayleigh_fading,Hughes_IT02_Asymptotic_Rayleigh_MIMO}.

\subsection{Stiefel and Grassmann Manifolds\label{sub:Grassmannian-Manifold}}


Stiefel manifold and Grassmann manifold are the geometric objects
relevant to the beamforming codebook design. The \emph{Stiefel manifold}
$\mathcal{S}_{L_{T},s}\left(\mathbb{C}\right)$ (where $L_{T}\geq s$)
is the set of all complex unitary $L_{T}\times s$ matrices $\mathcal{S}_{L_{T},s}\left(\mathbb{C}\right)=\left\{ \mathbf{Q}\in\mathbb{C}^{L_{T}\times s}:\;\mathbf{Q}^{\dagger}\mathbf{Q}=\mathbf{I}_{s}\right\} $.
Define an equivalence relation on the Stiefel manifold, i.e., two
matrices $\mathbf{P},\mathbf{Q}\in\mathcal{S}_{L_{T},s}\left(\mathbb{C}\right)$
are equivalent if their column vectors span the same subspace. The
\emph{Grassmann manifold} $\mathcal{G}_{L_{T},s}\left(\mathbb{C}\right)$
is defined as the quotient space of $\mathcal{S}_{L_{T},s}\left(\mathbb{C}\right)$
with respect to this equivalent relation. It can also be viewed as
the set of all the $s$-dimensional planes through the origin in the
$L_{T}$-dimensional Euclidean space \cite{Conway_96_PackingLinesPlanes,Tse_IT02_Communication_on_Grassmann_Manifold}.
A generator matrix $\mathbf{Q}\in\mathcal{S}_{L_{T},s}\left(\mathbb{C}\right)$
for an $s$-plane $\mathcal{Q}\in\mathcal{G}_{L_{T},s}\left(\mathbb{C}\right)$
is defined as the matrix whose columns span $\mathcal{Q}$. The generator
matrix is not unique. If $\mathbf{Q}$ is a generator matrix for an
$s$-dimensional plane $\mathcal{Q}\in\mathcal{G}^{L_{T},s}\left(\mathbb{C}\right)$,
then $\mathbf{QU}$ with $\mathbf{U}\in\mathcal{S}_{s,s}$ is also
a generator matrix of the same plane $\mathcal{Q}$ \cite{Conway_96_PackingLinesPlanes}. 

This paper considers the projection Frobenius metric (chordal distance)
on the Grassmann manifold because it is relevant to the the performance
analysis of power on/off strategy. The chordal distance between two
$s$-planes $\mathcal{Q}_{1},\mathcal{Q}_{2}\in\mathcal{G}_{L_{T},s}\left(\mathbb{C}\right)$
can be defined by their generator matrices,\begin{eqnarray}
d_{c} & \triangleq & \frac{1}{\sqrt{2}}\left\Vert \mathbf{Q}_{1}\mathbf{Q}_{1}^{\dagger}-\mathbf{Q}_{2}\mathbf{Q}_{2}^{\dagger}\right\Vert _{F}\nonumber \\
 & = & =s-\mathrm{trace}\left(\left(\mathbf{Q}_{1}^{\dagger}\mathbf{Q}_{2}\right)\left(\mathbf{Q}_{1}^{\dagger}\mathbf{Q}_{2}\right)^{\dagger}\right),\label{eq:Dc_def_equivalent}\end{eqnarray}
where $\mathbf{Q}_{1}$ and $\mathbf{Q}_{2}$ are the generator matrices
of $\mathcal{Q}_{1}$ and $\mathcal{Q}_{2}$ respectively \cite{Conway_96_PackingLinesPlanes}.
Since the chordal distance is independent with the choice of the generator
matrices, it is well defined \cite{Conway_96_PackingLinesPlanes}. 

The invariant measure and the uniform distribution play a crucial
role in the statistics on $\mathcal{S}_{L_{T},s}\left(\mathbb{C}\right)$
and $\mathcal{G}_{L_{T},s}\left(\mathbb{C}\right)$. Let $\mathcal{M}$
be a measurable set in $\mathcal{S}_{L_{T},s}\left(\mathbb{C}\right)$
or $\mathcal{G}_{L_{T},s}\left(\mathbb{C}\right)$, a measure $\zeta$
is called invariant if \[
\zeta\left(\mathbf{A}\mathcal{M}\right)=\zeta\left(\mathcal{M}\right)=\zeta\left(\mathcal{M}\mathbf{B}\right)\]
 for arbitrary $L_{T}\times L_{T}$ unitary matrix $\mathbf{A}$ and
$s\times s$ unitary matrix $\mathbf{B}$. The invariant probability
measure defines the uniform distribution on $\mathcal{S}_{L_{T},s}\left(\mathbb{C}\right)$
or $\mathcal{G}_{L_{T},s}\left(\mathbb{C}\right)$ \cite{Muirhead_book82_multivariate_statistics,Barg_IT02_Bounds_Grassmann_Manifold}.

\section{Power On/Off Strategy with Perfect Beamforming\label{sec:Power-On/Off-Perfect-Beamforming}}


To isolate the effect of power on/off from the effect of imperfect
beamforming, this section discusses the perfect beamforming case.
The effect of imperfect beamforming will be treated in Section \ref{sec:StrategyDesign_with_beamforming_codebook}.

In this section and throughout, the following notations are used.
Define $m=\min\left(L_{T},L_{R}\right)$ and $n=\max\left(L_{T},L_{R}\right)$.
Define the normalized number of on-beams as $\bar{s}\triangleq\frac{1}{m}s$
and the normalized on-power as $\bar{P}_{\mathrm{on}}=mP_{\mathrm{on}}$.
Define $\mathbf{W}\triangleq\frac{1}{m}\mathbf{HH}^{\dagger}$ if
$L_{R}<L_{T}$ or $\mathbf{W}\triangleq\frac{1}{m}\mathbf{H}^{\dagger}\mathbf{H}$
if $L_{R}\geq L_{T}$. Denote the $i^{\mathrm{th}}$ largest eigenvalue
of $\mathbf{W}$ by $\lambda_{i}$.

To analyze the perfect beamforming case, Section \ref{sub:Design-with-Perfect-Beamforming}
describes the corresponding optimization problem, Section \ref{sub:Asymptotic-analysis}
solves the optimization problem by letting $L_{T}$ and $L_{R}$ approach
infinity simultaneously, and Section \ref{sub:MIMO-finite-many-antennas}
shows that the asymptotic solution is near optimal for MIMO systems
with finite many antennas.

\subsection{The Design Problem with Perfect Beamforming\label{sub:Design-with-Perfect-Beamforming}}


The definition for perfect beamforming is given as follows. Consider
the singular value decomposition of the channel state matrix $\mathbf{H}=\mathbf{U\Lambda V}^{\dagger}$.
Perfect beamforming means that for $\forall\mathbf{H}\in\mathbb{C}^{L_{R}\times L_{T}}$
and $1\leq s\leq L_{T}$, there exists $\mathbf{Q}\in\mathcal{B}$
such that the $s$ columns of the beamforming matrix $\mathbf{Q}\in\mathbb{C}^{L_{T}\times s}$
are some columns of the right singular-vector matrix $\mathbf{V}$,
i.e., $\mathbf{V}^{\dagger}\mathbf{Q}\in\mathbb{C}^{L_{T}\times s}$
is with elements either 1 or 0.

With perfect beamforming, the optimization problem can be simplified.
Suppose that $P_{\mathrm{on}}$ and $s=s\left(\mathbf{H}\right)$
are given. For a channel realization $\mathbf{H}$, the optimal feedback
beamforming matrix is $\mathbf{Q}_{\varphi\left(\mathbf{H}\right)}=\mathbf{V}_{s}$%
\footnote{Rigorously speaking, the beamforming matrix $\mathbf{Q}=\mathbf{V}_{s}\mathbf{U}$
for any $s\times s$ unitary matrix $\mathbf{U}$ is optimal. %
} where $\mathbf{V}_{s}$ is composed of the right singular vectors
corresponding to the largest $s$ singular values of $\mathbf{H}$.
Then, the mutual information between the transmitted signal and the
received signal is \begin{eqnarray}
\mathcal{I}\left(\mathbf{H}\right) & = & \ln\left|\mathbf{I}_{L_{R}}+P_{\mathrm{on}}\mathbf{HQ}_{\varphi\left(\mathbf{H}\right)}\mathbf{Q}_{\varphi\left(\mathbf{H}\right)}^{\dagger}\mathbf{H}^{\dagger}\right|\nonumber \\
 & = & \sum_{i=1}^{s}\ln\left(1+\bar{P}_{\mathrm{on}}\lambda_{i}\right).\label{eq:I_perfect_beamforming}\end{eqnarray}
 The corresponding optimization problem is stated as follows.

\begin{problem}
\label{pro:(Power-On/Off-Design-Perfect-Beamforming)}(Power On/Off
Design with Perfect Beamforming) Find the optimal $s=s\left(\mathbf{H}\right)$
(or $\bar{s}=\bar{s}\left(\mathbf{H}\right)$) function and $P_{\mathrm{on}}$
(or $\bar{P}_{\mathrm{on}}$) to maximize the information rate, \begin{eqnarray*}
\underset{P_{\mathrm{on}}}{\max}\;\underset{s\left(\cdot\right)}{\max}\;\mathrm{E}_{\mathbf{H}}\left[\sum_{i=1}^{s}\mathrm{ln}\left(1+\bar{P}_{\mathrm{on}}\lambda_{i}\right)\right],\end{eqnarray*}
with the power constraint\begin{eqnarray*}
\mathrm{E}_{\mathbf{H}}\left[sP_{\mathrm{on}}\right]=\bar{P}_{\mathrm{on}}\mathrm{E}_{\mathbf{H}}\left[\bar{s}\right]\leq\rho.\end{eqnarray*}

\end{problem}

The following theorem gives the form of the optimal $\bar{s}$ function
to solve Problem \ref{pro:(Power-On/Off-Design-Perfect-Beamforming)}.

\begin{thm}
\label{thm:optimal_s_perfect_beamforming}The optimal $\bar{s}$ function
to solve Problem \ref{pro:(Power-On/Off-Design-Perfect-Beamforming)}
is of the form \begin{equation}
\bar{s}=\frac{1}{m}\left|\left\{ k:\;\lambda_{k}\geq\kappa\right\} \right|\label{eq:s_optimal_form}\end{equation}
where $\kappa$ is the appropriate threshold chosen to satisfy the
average power constraint\[
\bar{P}_{\mathrm{on}}\mathrm{E}_{\mathbf{H}}\left[\bar{s}\right]=\rho.\]

\end{thm}
\begin{proof}
See Appendix \ref{sub:pf_optimal_s_perfect_beamforming}.
\end{proof}

The intuition behind the proof is that all the {}``good'' beams
(corresponding to $\lambda\geq\kappa$) and only the {}``good''
beams should be turned on. This intuition will be used in the proof
of Theorem \ref{thm:optimal-feedback-multirank-B} later.

Although the form of the optimal $\bar{s}$ function is given in (\ref{eq:s_optimal_form}),
it is difficult to find the key parameters (the optimal $\bar{P}_{\mathrm{on}}$
and $\kappa$) and the corresponding information rate $\mathcal{I}$.
Different from the water filling solution for CSITR case where the
Lagrange multiplier is uniquely determined by $\rho$ \cite{Telatar_EuroTele99_Capacity_MIMO},
power on/off strategy has uncountable many pairs of $\bar{P}_{\mathrm{on}}$
and $\kappa$ corresponding to the same $\rho$. Numerical search
may be employed to find the optimal $\bar{P}_{\mathrm{on}}$, $\kappa$
and the corresponding $\mathcal{I}$. However, if the numbers of transmit
and receive antennas approach infinity simultaneously, as we will
show in Section \ref{sub:Asymptotic-analysis}, the corresponding
key parameters and information rate can be explicitly computed.

\subsection{\label{sub:Asymptotic-analysis}MIMO Systems with Infinitely Many
Antennas}


This section provides explicit formulas to solve Problem \ref{pro:(Power-On/Off-Design-Perfect-Beamforming)}
by letting the numbers of transmit and receive antennas approach infinity
simultaneously. As a byproduct of the employed method, this section
also presents asymptotic formulas for the capacity of CSITR case.
According to the authors knowledge, the derived asymptotic formulas
are presented for the first time.

\subsubsection{\label{subsub:Asymptotic-Analysis-Power-on/off}Asymptotic Analysis
for Power On/off Strategy}

The main result of the asymptotic analysis is the following theorem,
which gives the optimal $\bar{s}$ function when the numbers of transmit
and receive antennas approach infinity simultaneously.

\begin{thm}
\label{thm:Optimal-s-function}Define $\tau\triangleq\frac{n}{m}$.
For a given SNR $\rho$, if $m$ and $n$ approach infinity simultaneously
with $\tau$ fixed, the optimal $\bar{s}$ function converges to a
constant, \[
\bar{s}_{\infty}\triangleq\underset{\left(n,m\right)\rightarrow\infty}{\lim}\bar{s}=\int_{\kappa}^{\lambda^{+}}f\left(\lambda\right)d\lambda,\]
almost surely, and the corresponding normalized information rate $\bar{\mathcal{I}}\triangleq\frac{1}{m}\mathcal{I}$
also converges to a constant,\begin{equation}
\bar{\mathcal{I}}_{\infty}\triangleq\underset{\left(n,m\right)\rightarrow\infty}{\lim}\bar{\mathcal{I}}=\int_{\kappa}^{\lambda^{+}}\mathrm{ln}\left(1+\frac{\rho}{\bar{s}_{\infty}}\lambda\right)f\left(\lambda\right)d\lambda,\label{eq:I_asymptotic_lambda}\end{equation}
 where \begin{eqnarray*}
f\left(\lambda\right) & = & \frac{1}{2\pi\lambda}\sqrt{\left(\lambda^{+}-\lambda\right)\left(\lambda-\lambda^{-}\right)},\end{eqnarray*}
 $\lambda^{\pm}\triangleq\left(\sqrt{\tau}\pm1\right)^{2}$ and $\lambda^{+}>\kappa\geq\lambda^{-}$
is the appropriate constant chosen to maximize the normalized information
rate (\ref{eq:I_asymptotic_lambda}).
\end{thm}
\begin{proof}
Recall the optimal $\bar{s}$ function in (\ref{eq:s_optimal_form}).
According to (\ref{eq:dF-dlambda-asymptotic}) in Section \ref{sub:Random-Matrix-Theory},
\begin{eqnarray*}
\underset{\left(n,m\right)\rightarrow\infty}{\lim}\bar{s} & = & \underset{\left(n,m\right)\rightarrow\infty}{\lim}\frac{1}{m}\left|\left\{ k:\;\lambda_{k}\geq\kappa\right\} \right|\\
 & = & \underset{\left(n,m\right)\rightarrow\infty}{\lim}1-F\left(\lambda\right)\\
 & = & \int_{\kappa}^{\lambda^{+}}f\left(\lambda\right)d\lambda\end{eqnarray*}
almost surely. 

For any positive constant $\bar{P}_{\mathrm{on}}$ and a channel realization
$\mathbf{H}$, according to the random matrix theory in (\ref{eq:g-lambda-asymptotic}),
the normalized mutual information between the transmitted signal and
the received signal converges to a constant, \begin{eqnarray*}
\underset{\left(n,m\right)\rightarrow\infty}{\lim}\bar{\mathcal{I}}\left(\mathbf{H}\right) & = & \underset{\left(n,m\right)\rightarrow\infty}{\lim}\frac{1}{m}\sum_{i=1}^{s}\mathrm{ln}\left(1+\bar{P}_{\mathrm{on}}\lambda_{i}\right)\\
 & = & \int_{\kappa}^{\lambda^{+}}\mathrm{ln}\left(1+\bar{P}_{\mathrm{on}}\lambda\right)f\left(\lambda\right)d\lambda,\end{eqnarray*}
almost surely. Thus the normalized information rate converges to a
constant \begin{eqnarray*}
\underset{\left(n,m\right)\rightarrow\infty}{\lim}\bar{\mathcal{I}} & = & \underset{\left(n,m\right)\rightarrow\infty}{\lim}\mathrm{E}_{\mathbf{H}}\left[\bar{\mathcal{I}}\left(\mathbf{H}\right)\right]\\
 & = & \int_{\kappa}^{\lambda^{+}}\mathrm{ln}\left(1+\bar{P}_{\mathrm{on}}\lambda\right)f\left(\lambda\right)d\lambda.\end{eqnarray*}
Furthermore, an elementary calculation shows that the choice of $\bar{P}_{\mathrm{on}}=\rho/\bar{s}_{\infty}$
satisfies the average power constraint. Therefore, we have \[
\bar{\mathcal{I}}_{\infty}=\underset{\left(n,m\right)\rightarrow\infty}{\lim}\bar{\mathcal{I}}=\int_{\kappa}^{\lambda^{+}}\mathrm{ln}\left(1+\frac{\rho}{\bar{s}_{\infty}}\lambda\right)f\left(\lambda\right)d\lambda.\]

Finally, $\bar{s}_{\infty}$, $\bar{P}_{\mathrm{on}}$ and $\bar{\mathcal{I}}_{\infty}$
are all functions of $\kappa$, the optimization problem is to choose
appropriate $\kappa$ to maximize $\bar{\mathcal{I}}_{\infty}$. 
\end{proof}

This theorem proves that the optimal normalized number of on-beams
$\bar{s}$ converges to a constant independent of the specific channel
realization for a given SNR requirement. The principle behind this
theorem is same as that of channel hardening \cite{Hochwald_IT04_Channel_Hardening}:
the characteristic of a MIMO channel turns to be deterministic as
the numbers of transmit and receive antennas approach infinity.

To find explicit formulas to calculate the key parameters and the
corresponding performance, we need the following variable change\begin{equation}
\lambda\left(t\right)=\frac{1}{y}\left(1+y-2\sqrt{y}\mathrm{cos}\left(t\right)\right),\label{eq:variable_change}\end{equation}
where $y\triangleq\frac{m}{n}=\frac{1}{\tau}$ and $t\in\left[0,\pi\right]$.
After the variable change, the asymptotic empirical density function
of $t$ can be written as \begin{eqnarray}
f_{T}\left(t\right)=\left\{ \begin{array}{ll}
\frac{1}{\pi}\cdot\frac{1-\cos\left(2t\right)}{1+y-2\sqrt{y}\cos\left(t\right)} & \mathrm{if}\; y<1\\
\frac{1+\cos\left(t\right)}{\pi} & \mathrm{if}\; y=1\end{array}\right.\label{eq:f_T}\end{eqnarray}
Define $a$ such that $\lambda\left(a\right)=\kappa$ where $\kappa$
is the optimal threshold in Theorem \ref{thm:Optimal-s-function}.
Then we have the following corollary according to Theorem \ref{thm:Optimal-s-function}.

\begin{cor}
\label{cor:Optimal-s-function}If $m$ and $n$ approach infinity
simultaneously with $y\triangleq\frac{m}{n}$ fixed, the optimal $\bar{s}$
function converges to a constant, \begin{equation}
\bar{s}_{\infty}=\underset{\left(n,m\right)\rightarrow\infty}{\lim}\bar{s}=\int_{a}^{\pi}f_{T}\left(t\right)dt,\label{eq:S_asymptotic_t}\end{equation}
almost surely and the corresponding $\bar{\mathcal{I}}$ converges
to a constant, \begin{equation}
\bar{\mathcal{I}}_{\infty}=\underset{\left(n,m\right)\rightarrow\infty}{\lim}\bar{\mathcal{I}}=\int_{a}^{\pi}\ln\left(1+\frac{\rho}{y\bar{s}}\left(1+y-2\sqrt{y}\cos\left(t\right)\right)\right)f_{T}\left(t\right)dt,\label{eq:I_asymptotic_t}\end{equation}
where $a\in\left[0,\pi\right]$ is chosen to maximize the normalized
information rate (\ref{eq:I_asymptotic_t}).
\end{cor}

Since the variable change (\ref{eq:variable_change}) is invertible,
to find the optimal $\kappa$ in Theorem \ref{thm:Optimal-s-function}
is equivalent to find the optimal $a$ in Corollary \ref{cor:Optimal-s-function}.
The following theorem gives a method to find the optimal $a$.

\begin{thm}
\label{thm:condition-for-optimal-a}If $m$ and $n$ approach infinity
simultaneously with $y\triangleq\frac{m}{n}$ fixed, then $\frac{d\bar{\mathcal{I}}_{\infty}}{da}=0$
has at most one solution in the domain of $\left(0,\pi\right)$. The
optimal $a$ to maximize $\bar{\mathcal{I}}_{\infty}$ is either the
unique solution of $\frac{d\bar{\mathcal{I}}_{\infty}}{da}=0$ in
$\left(0,\pi\right)$ if it exists, or 0 if $\frac{d\bar{\mathcal{I}}_{\infty}}{da}\neq0$
for all $a\in\left(0,\pi\right)$.
\end{thm}
\begin{proof}
See Appendix \ref{sub:Proof-of-Theorem-condition-for-optimal-a}.
\end{proof}

The following corollaries show how the optimal $a$ and the optimal
$\bar{s}_{\infty}$ change when the average power constraint $\rho$
increases. The results will be applied to MIMO systems with finite
many antennas in Section \ref{sub:MIMO-finite-many-antennas}.

\begin{cor}
\label{cor:relationship-optimal-a-and-rho}If $m$ and $n$ approach
infinity simultaneously with $y\triangleq\frac{m}{n}$ fixed, the
optimal $a$ to maximize $\bar{\mathcal{I}}_{\infty}$ is a non-increasing
function of $\rho$. 
\end{cor}
\begin{proof}
See Appendix \ref{sub:Pf-of-Cor-relationship-optimal-a-and-rho}.
\end{proof}
\begin{cor}
\label{cor:relationship-optimal-s-and-rho}If $m$ and $n$ approach
infinity simultaneously with $y\triangleq\frac{m}{n}$ fixed, the
optimal number of on-beams $\bar{s}_{\infty}$ to maximize $\bar{\mathcal{I}}_{\infty}$
is a nondecreasing function of $\rho$. 
\end{cor}
\begin{proof}
Note that $\bar{s}_{\infty}=\int_{a}^{\pi}f_{T}\left(t\right)dt$
which is a monotone decreasing function of $a$. This corollary follows
Corollary \ref{cor:relationship-optimal-a-and-rho}.
\end{proof}

Based on the above asymptotic results, the design problem for perfect
beamforming (Problem \ref{pro:(Power-On/Off-Design-Perfect-Beamforming)})
can be solved. According to Theorem \ref{thm:condition-for-optimal-a},
the asymptotic optimal threshold, say $a_{\infty}$, can be found
by checking $\frac{d\bar{\mathcal{I}}_{\infty}}{da}$. The corresponding
optimal normalized number of on-beams $\bar{s}_{\infty}$ and the
normalized information rate $\bar{\mathcal{I}}_{\infty}$ can be computed
by substituting $a_{\infty}$ into (\ref{eq:S_asymptotic_t}) and
(\ref{eq:I_asymptotic_t}) respectively.

However, the calculations involve integrals, which may be computational
complex. To simplify the computation, Propositions \ref{pro:calculation-s}-\ref{pro:calculation-dCda}
express the integrals as some special functions which are defined
by infinite series. Generally, the calculation of the series is much
easier than numerical integrals. To make the expressions clear, the
following notations are used.

\begin{eqnarray}
r\triangleq\sqrt{y},\label{eq:def_r}\end{eqnarray}
 \begin{eqnarray}
\alpha\triangleq\frac{\bar{s}_{\infty}y}{\rho},\label{eq:def_alpha}\end{eqnarray}
\begin{eqnarray}
w\triangleq\frac{1}{2}\left(1+y+\alpha+\sqrt{\left(1+y+\alpha\right)^{2}-4y}\right),\label{eq:def_w}\end{eqnarray}
\begin{eqnarray}
u\triangleq\frac{1}{2r}\left(1+y+\alpha-\sqrt{\left(1+y+\alpha\right)^{2}-4y}\right),\label{eq:def_u}\end{eqnarray}
\begin{eqnarray}
\theta_{r}\triangleq\tan^{-1}\left(\frac{r\sin\left(a\right)}{1-r\cos\left(a\right)}\right),\label{eq:def_theta_r}\end{eqnarray}
for $r\cos\left(a\right)\neq1$ and\begin{eqnarray}
\theta_{u}\triangleq\tan^{-1}\left(\frac{u\sin\left(a\right)}{1-u\cos\left(a\right)}\right),\label{eq:def_theta_u}\end{eqnarray}
for $u\cos\left(a\right)\neq1$. There are also three special functions
defined by series. The first one is called Dilogarithm in literature
\cite{Andrews_1999_Special_Functions} and defined as \begin{eqnarray}
\mathrm{Li}_{2}\left(x\right)\triangleq\sum_{n=1}^{\infty}\frac{x^{n}}{n^{2}},\label{eq:def_Li_2}\end{eqnarray}
for $\left|x\right|\leq1$. We define the other two as \begin{eqnarray}
\mathrm{Sr}_{1}\left(u,r,t\right)\triangleq\sum_{l=1}^{\infty}\frac{r^{l}e^{ilt}}{l}\left(\sum_{k=1}^{l-1}\frac{\left(\frac{u}{r}\right)^{k}}{k}+\frac{1}{r^{2l}}\sum_{k=l}^{\infty}\frac{r^{2k}\left(\frac{u}{r}\right)^{k}}{k}\right)\label{eq:Sr_1}\end{eqnarray}
and \begin{eqnarray}
\mathrm{Sr}_{2}\left(r,t\right)\triangleq\sum_{l=1}^{\infty}\frac{r^{l}e^{ilt}}{l}\left(\frac{1}{r^{2l}}\sum_{k=l}^{\infty}\frac{r^{2k}}{k}\right)\label{eq:Sr_2}\end{eqnarray}
for $\left|u\right|<1$, $\left|r\right|<1$ and $\left|\frac{u}{r}\right|<1$.

\begin{prop}
\label{pro:calculation-s}If $m$ and $n$ approach infinity simultaneously
with $y\triangleq\frac{m}{n}$ fixed, the normalized number of on-beams
$\bar{s}_{\infty}$ (as a function of $a$) is given by \begin{eqnarray*}
\bar{s}_{\infty}=\left\{ \begin{array}{ll}
\frac{1}{\pi}\left\{ \pi-a-\frac{1}{r}\sin\left(a\right)+\frac{1-r^{2}}{r^{2}}\theta_{r}\right\}  & \mathrm{if}\: y<1\\
\frac{1}{\pi}\left\{ \pi-a-\sin\left(a\right)\right\}  & \mathrm{if}\: y=1\end{array}\right..\end{eqnarray*}

\end{prop}
\begin{proof}
See Appendix \ref{sub:Calculation-of-s}.
\end{proof}
\begin{prop}
\label{pro:calculation-C}If $m$ and $n$ approach infinity simultaneously
with $y\triangleq\frac{m}{n}$ fixed, the normalized information rate
$\bar{\mathcal{I}}_{\infty}$ (as a function of the threshold $a$)
is given by \begin{eqnarray*}
\bar{\mathcal{I}}_{\infty}=\left\{ \begin{array}{ll}
\left[\ln\left(w\right)-\ln\left(\alpha\right)\right]\bar{s}_{\infty}+J_{0}+J_{1}+J_{2} & \mathrm{if}\: y<1\\
\left[\ln\left(w\right)-\ln\left(\alpha\right)\right]\bar{s}_{\infty}+J_{0}+J_{1} & \mathrm{if}\: y=1\end{array}\right.\end{eqnarray*}
where \begin{eqnarray*}
J_{0}=\frac{1}{\pi r}\left\{ \sin\left(a\right)\left[1-\ln\left(1+u^{2}-2u\cos\left(a\right)\right)\right]-u\left(\pi-a\right)-\left(\frac{1}{u}-u\right)\theta_{u}\right\} \end{eqnarray*}
\begin{eqnarray*}
J_{1}=\frac{1+r^{2}}{2\pi r^{2}}i\left[\mathrm{Li}_{2}\left(ue^{-ia}\right)-\mathrm{Li}_{2}\left(ue^{ia}\right)\right]\end{eqnarray*}
and\begin{eqnarray*}
J_{2}=\frac{1-r^{2}}{2\pi r^{2}}\left[-2\ln\left(1-ur\right)\left(\pi-a-\theta_{r}\right)+i\mathrm{Sr}_{1}\left(u,r,a\right)-i\mathrm{Sr}_{1}\left(u,r,-a\right)\right].\end{eqnarray*}

\end{prop}
\begin{proof}
See Appendix \ref{sub:Calculation-of-C}.
\end{proof}
\begin{prop}
\label{pro:calculation-dCda}If $m$ and $n$ approach infinity simultaneously
with $y\triangleq\frac{m}{n}$ fixed, $\frac{d\bar{\mathcal{I}}_{\infty}}{da}$
is given by \begin{eqnarray*}
\frac{d\bar{\mathcal{I}}_{\infty}}{da}=\frac{J_{3}}{\pi}\cdot\left[1-\ln\left(1+\frac{\rho}{\bar{s}_{\infty}y}\left(1+r^{2}-2r\cos\left(a\right)\right)\right)-\frac{y}{\rho}I^{d}\right],\end{eqnarray*}
where \begin{eqnarray*}
J_{3}=\left\{ \begin{array}{ll}
\frac{1-\cos\left(2a\right)}{1+r^{2}-2r\cos\left(a\right)} & \mathrm{if}\: y<1\\
1+\cos\left(a\right) & \mathrm{if}\: y=1\end{array}\right.\end{eqnarray*}
and \begin{eqnarray*}
I^{d}=\left\{ \begin{array}{ll}
\frac{1}{\pi w\left(1-ur\right)}\left[\pi-a-\frac{1-u^{2}}{u\left(r-u\right)}\theta_{u}+\frac{1-r^{2}}{r\left(r-u\right)}\theta_{r}\right] & \mathrm{if}\: y<1\\
\frac{\pi-a}{\pi w\left(1-u\right)}-\frac{\left(1+u\right)\theta_{u}}{\pi wu\left(1-u\right)} & \mathrm{if}\: y=1.\end{array}\right.\end{eqnarray*}

\end{prop}
\begin{proof}
See Appendix \ref{sub:Calculation-of-dCda}.
\end{proof}

Following the method in \cite{Popescu_TComm00_Capacity_Random_Signature_MIMO,Hughes_SSC02_Gaussian_approximation_capacity_MIMO_Rayleigh_fading},
Proposition \ref{pro:calculation-s}-\ref{pro:calculation-dCda} provide
close form formulas to evaluate $\bar{s}_{\infty}$, $\bar{\mathcal{I}}_{\infty}$
and $\frac{d\bar{\mathcal{I}}_{\infty}}{da}$. In \cite{Popescu_TComm00_Capacity_Random_Signature_MIMO,Hughes_SSC02_Gaussian_approximation_capacity_MIMO_Rayleigh_fading},
the closed form of the capacity is derived for CSIR only case, where
all $L_{T}$ available beams are turned on. The results in \cite{Popescu_TComm00_Capacity_Random_Signature_MIMO,Hughes_SSC02_Gaussian_approximation_capacity_MIMO_Rayleigh_fading}
can be viewed as a special case of Proposition \ref{pro:calculation-C}
where $a=0$.

\subsubsection{\label{subsub:Asymptotic-Analysis-CSITR}Asymptotic Analysis for
CSITR Case}

To compare power on/off strategy with water filling power control
(corresponds to CSITR case), we present asymptotic formulas to evaluate
the CSITR capacity. As far as the authors know, these asymptotic results
are presented for the first time.

It is well known that water filling power control can achieve the
capacity assuming perfect CSIT \cite{Telatar_EuroTele99_Capacity_MIMO}.
Let $m=\mathrm{min}\left(L_{R},L_{T}\right)$, $n=\mathrm{max}\left(L_{R},L_{T}\right)$,
$\tau=\frac{n}{m}$. When $m$ and $n$ approach infinity simultaneously
with the ratio $\tau$ fixed, according to (\ref{eq:g-lambda-asymptotic}),
the normalized capacity is given by\[
\bar{C}_{\infty}\triangleq\underset{\left(m,n\right)\rightarrow\infty}{\lim}\;\frac{1}{m}C=\int_{\max\left(\lambda^{-},\frac{1}{\nu}\right)}^{\lambda^{+}}\mathrm{ln}\left(\lambda\nu\right)f\left(\lambda\right)d\lambda,\]
 where $f\left(\lambda\right)=\frac{1}{2\pi\lambda}\sqrt{\left(\lambda^{+}-\lambda\right)\left(\lambda-\lambda^{-}\right)}$,
$\lambda^{\pm}=\left(\sqrt{\tau}\pm1\right)^{2}$ and $\nu$ is the
Lagrange multiplier chosen to satisfy the average power constraint,
\[
\rho=\int_{\max\left(\lambda^{-},\frac{1}{\nu}\right)}^{\lambda^{+}}\left(\nu-\frac{1}{\lambda}\right)f\left(\lambda\right)d\lambda.\]

To derive closed forms for the integrals, consider the same variable
change as in (\ref{eq:variable_change}). Then the asymptotic normalized
capacity $\bar{C}_{\infty}$ is given by\begin{eqnarray*}
\bar{C}_{\infty}=\int_{a}^{\pi}\mathrm{ln}\left(\frac{\nu}{y}\left(1+y-2\sqrt{y}\cos\left(t\right)\right)\right)f_{T}\left(t\right)dt,\end{eqnarray*}
where\[
a=\left\{ \begin{array}{ll}
\cos^{-1}\left(\frac{1+y-\frac{y}{\nu}}{2\sqrt{y}}\right) & \mathrm{if}\;\lambda^{-}\leq\frac{1}{\nu}\leq\lambda^{+}\\
0 & \mathrm{if}\;\frac{1}{\nu}<\lambda^{-}\end{array}\right.,\]
 $\nu$ is the Lagrange multiplier chosen to satisfy the average power
constraint\begin{eqnarray*}
\rho=\int_{a}^{\pi}\left(\nu-\frac{y}{1+y-2\sqrt{y}\cos\left(t\right)}\right)f_{T}\left(t\right)dt,\end{eqnarray*}
 and $f_{T}\left(t\right)$ is given in (\ref{eq:f_T}).


The following propositions give the closed forms for the average power
and the normalized capacity as a function of $\nu$. To make the presentation
clearer, the notations in (\ref{eq:def_r}-\ref{eq:Sr_2}) are used. 

\begin{prop}
\label{pro:rho_CSITR}If $m$ and $n$ approach infinity simultaneously
with $y=\frac{m}{n}$ fixed, the relationship between the power constraint
$\rho$ and the Lagrange multiplier $\nu$ is given by\begin{eqnarray*}
\rho=\nu\bar{s}_{\infty}-J_{4},\end{eqnarray*}
where \[
\bar{s}_{\infty}=\int_{a}^{\pi}f_{T}\left(t\right)dt,\]
 \begin{eqnarray*}
J_{4}=\left\{ \begin{array}{ll}
\frac{1}{\pi}\left[\frac{r^{2}}{1-r^{2}}\left(\pi-a\right)-\frac{1+r^{2}}{1-r^{2}}\theta_{r}+\frac{i}{2}\left(\frac{1}{1-re^{-ia}}-\frac{1}{1-re^{ia}}\right)\right] & \mathrm{if}\; y<1\\
\frac{1}{2\pi}\left[\pi-a-\frac{2}{\tan\left(\frac{a}{2}\right)}\right] & \mathrm{if}\; y=1\end{array}\right.\end{eqnarray*}
and\begin{eqnarray*}
a=\left\{ \begin{array}{ll}
\cos^{-1}\left(\frac{1+y-\frac{y}{\nu}}{2\sqrt{y}}\right) & \mathrm{if}\;\lambda^{-}\leq\frac{1}{\nu}\leq\lambda^{+}\\
0 & \mathrm{if}\;\frac{1}{\nu}<\lambda^{-}\end{array}\right..\end{eqnarray*}

\end{prop}
\begin{proof}
See Appendix \ref{sub:Calculation-of-rho-CSITR}.
\end{proof}

\begin{prop}
\label{pro:C_CSITR}If $m$ and $n$ approach infinity simultaneously
with $y=\frac{m}{n}$ fixed, the normalized capacity $\bar{C}_{\infty}$
is given by \begin{eqnarray*}
\bar{C}_{\infty}=\left\{ \begin{array}{ll}
\ln\left(\frac{\nu}{y}\right)\bar{s}_{\infty}+J_{5}+J_{6}+J_{7} & \mathrm{if}\; y<1\\
\ln\left(\frac{\nu}{y}\right)\bar{s}_{\infty}+J_{5}+J_{6} & \mathrm{if}\; y=1\end{array}\right.,\end{eqnarray*}
where \[
\bar{s}_{\infty}=\int_{a}^{\pi}f_{T}\left(t\right)dt,\]
 \[
J_{5}=\frac{1}{\pi r}\left\{ \sin\left(a\right)\left[1-\ln\left(1+r^{2}-2r\cos\left(a\right)\right)\right]-r\left(\pi-a\right)-\left(\frac{1}{r}-r\right)\theta_{r}\right\} ,\]
\[
J_{6}=\frac{1+r^{2}}{2\pi r^{2}}i\left[\mathrm{Li}_{2}\left(re^{-ia}\right)-\mathrm{Li}_{2}\left(re^{ia}\right)\right],\]
\begin{eqnarray*}
J_{7} & = & -\frac{1-r^{2}}{2\pi r^{2}}\left\{ \frac{i}{2}\left[\ln^{2}\left(1-re^{-ia}\right)-\ln^{2}\left(1-re^{ia}\right)\right]\right.\\
 &  & \left.+2\ln\left(1-r^{2}\right)\left(\pi-a-\theta_{r}\right)+i\left[\mathrm{Sr}_{2}\left(r,-a\right)-\mathrm{Sr}_{2}\left(r,a\right)\right]\right\} ,\end{eqnarray*}
and \begin{eqnarray*}
a=\left\{ \begin{array}{ll}
\cos^{-1}\left(\frac{1+y-\frac{y}{\nu}}{2\sqrt{y}}\right) & \mathrm{if}\;\lambda^{-}\leq\frac{1}{\nu}\leq\lambda^{+}\\
0 & \mathrm{if}\;\frac{1}{\nu}<\lambda^{-}.\end{array}\right..\end{eqnarray*}

\end{prop}
\begin{proof}
See Appendix \ref{sub:Calculation-of-C-CSITR}.
\end{proof}

Based on the above propositions, the Lagrange multiplier $\nu$ and
the corresponding normalized capacity $\bar{C}_{\infty}$ can be easily
computed for a given SNR requirement $\rho$.

\subsection{\label{sub:MIMO-finite-many-antennas}MIMO Systems with Finite Many
Antennas}


The asymptotic results in Section \ref{sub:Asymptotic-analysis} can
be applied to MIMO systems with finite many antennas. It is often
the case that the asymptotic results are accurate enough for MIMO
systems with finite many antennas \cite{Popescu_TComm00_Capacity_Random_Signature_MIMO,Hughes_IT02_Asymptotic_Rayleigh_MIMO,Hughes_SSC02_Gaussian_approximation_capacity_MIMO_Rayleigh_fading,Wang_IT04_Outage_Mutual_Information_space_time_MIMO,Hochwald_IT04_Channel_Hardening}.
So are the asymptotic results in Section \ref{sub:Asymptotic-analysis}.
Theorem \ref{thm:Optimal-s-function} proves that the optimal normalized
number of on-beams $\bar{s}$ converges to a constant asymptotically.
We will show that a constant $\bar{s}$ is near optimal for MIMO systems
with finite many antennas. Moreover, according to the asymptotic result
in Corollary \ref{cor:relationship-optimal-s-and-rho}, the optimal
$\bar{s}$ is a nondecreasing function as the average $\rho$ increases.
It is consistent with the results in \cite{Blum_CommLett02_MIMO_Antenna_Selection,Sanayei_2003_asymptotic_capacity_gain_transmit_antenna_selection},
which consider the special case of transmit antenna selection and
show that at most one beam should be turned on when $\rho$ is small
enough and $m$ beams should be on when $\rho$ is sufficiently large.
Importantly though, the results in this paper is more general.

Before applying the asymptotic results, however, it is worthy to note
note the difference between the asymptotic case and the case of finite
many antennas. In asymptotic case, $\bar{s}$ can be any rational
number in $\left[0,1\right]$. On the other hand, in the case of finite
many antennas, $\bar{s}$ can only take finite many discrete values,
$\bar{s}\in\left\{ \frac{1}{m},\frac{2}{m},\cdots,1\right\} $ where
$m=\min\left(L_{R},L_{T}\right)$ is the dimension of the MIMO system.

To apply the asymptotic results to the finite case, we use the following
procedure.

\begin{enumerate}
\item For a given MIMO system with $L_{T}$-transmit antennas and $L_{R}$-receive
antennas, define $m=\min\left(L_{R},L_{T}\right)$, $n=\max\left(L_{R},L_{T}\right)$
and $y=\frac{m}{n}$. According to the asymptotic analysis and formulas
in Section \ref{sub:Asymptotic-analysis}, evaluate the asymptotic
optimal threshold $a_{\infty}$ and the asymptotic optimal normalized
number of on-beams $\bar{s}_{\infty}$ for a given average SNR requirement
$\rho$.
\item If $\bar{s}_{\infty}<\frac{1}{m}$, then go to 3). Otherwise, we choose
the optimal $\bar{s}$ as the one corresponding to the larger $\bar{\mathcal{I}}$
from the adjacent discrete values to $\bar{s}_{\infty}$. Specifically,
let $\bar{s}_{1}=\frac{1}{m}\left\lceil m\bar{s}_{\infty}\right\rceil $
and $\bar{s}_{2}=\frac{1}{m}\left\lfloor m\bar{s}_{\infty}\right\rfloor $
where $\left\lceil \cdot\right\rceil $ denotes the ceil function
and $\left\lfloor \cdot\right\rfloor $ represents the floor function.
Compare the corresponding performance $\bar{\mathcal{I}}$ (evaluated
by substituting the corresponding $a$ into the asymptotic formula
for $\bar{\mathcal{I}}_{\infty}$ in Proposition \ref{pro:calculation-C})
and choose the better one as the optimal $\bar{s}$. According the
Theorem \ref{thm:Optimal-s-function}, the $m\bar{s}$ beams corresponding
to the largest $m\bar{s}$ eigenvalues of $\mathbf{W}$ are always
turned on independent of the specific channel state realization $\mathbf{H}$.
The power on each on-beam is $P_{\mathrm{on}}=\frac{\rho}{m\bar{s}}$
and the corresponding $\bar{\mathcal{I}}$ can be evaluated by asymptotic
formula for $\bar{\mathcal{I}}_{\infty}$.
\item If $\bar{s}_{\infty}<\frac{1}{m}$, then at most one beam should be
turned on. Put $\bar{s}=\frac{1}{m}$ and $P_{\mathrm{on}}=\frac{\rho}{m\bar{s}_{\infty}}$.
We turn on/off the strongest beam, which corresponds to the largest
eigenvalue of $\mathbf{W}$, according to the following threshold
test,\[
\lambda_{1}\overset{\mathrm{on}}{\underset{\mathrm{off}}{\gtrless}}\kappa\]
where $\kappa=\frac{1}{y}\left(1+y-2\sqrt{y}\cos\left(a_{\infty}\right)\right)$.
\end{enumerate}


The power on/off strategy designed according to the above procedure
is called \emph{power on/off strategy with a constant number of on-beams}.
When the given average SNR $\rho$ is large enough so that $\bar{s}_{\infty}\geq\frac{1}{m}$,
a constant number of on-beams are turned on independent of the specific
channel realization $\mathbf{H}$. The only exception happens when
$\rho$ is so low that $\bar{s}_{\infty}\leq\frac{1}{m}$, where either
the strongest beam is turned on, when $\lambda_{1}\geq\kappa$, or
no beams is on, when $\lambda_{1}<\kappa$. Although this strategy
is designed according to the asymptotic results, it is near-optimal
for MIMO systems with finite many antennas according to the simulation
results.

\begin{figure}
\subfigure[Information rate]{\includegraphics[%
  clip,
  scale=0.8]{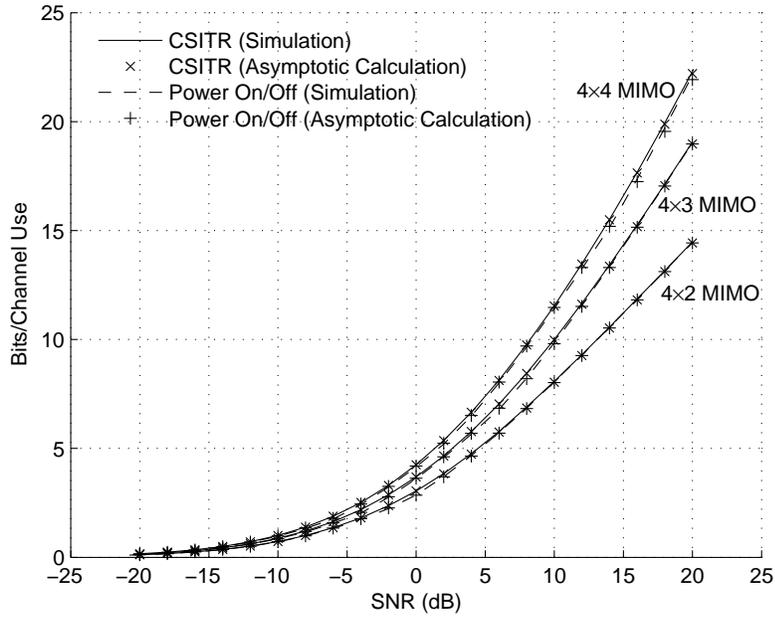}}

\subfigure[Relative performance]{\includegraphics[%
  clip,
  scale=0.8]{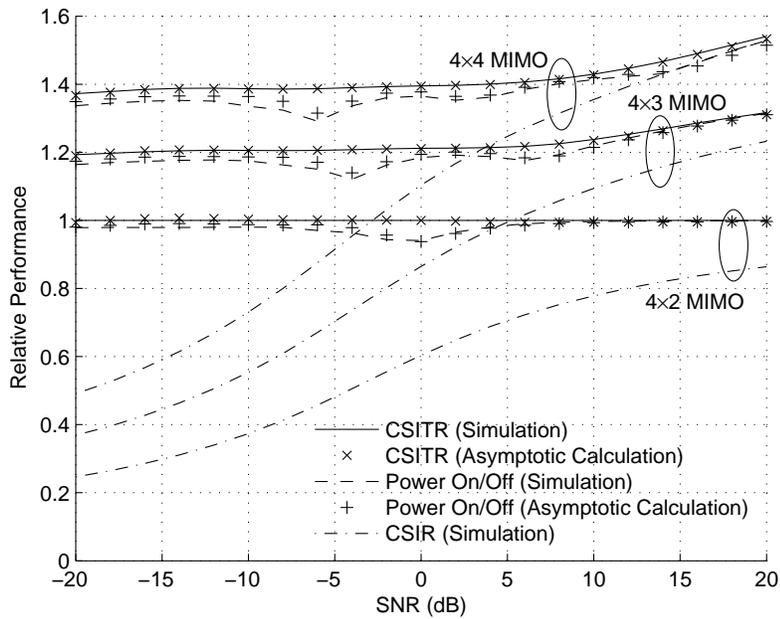}}

\caption{\label{cap:Perf_SNR_perfect_beam}Information rate v.s. SNR for perfect
beamforming}
\end{figure}

\begin{figure}
\subfigure[Information rate]{\includegraphics[%
  clip,
  scale=0.8]{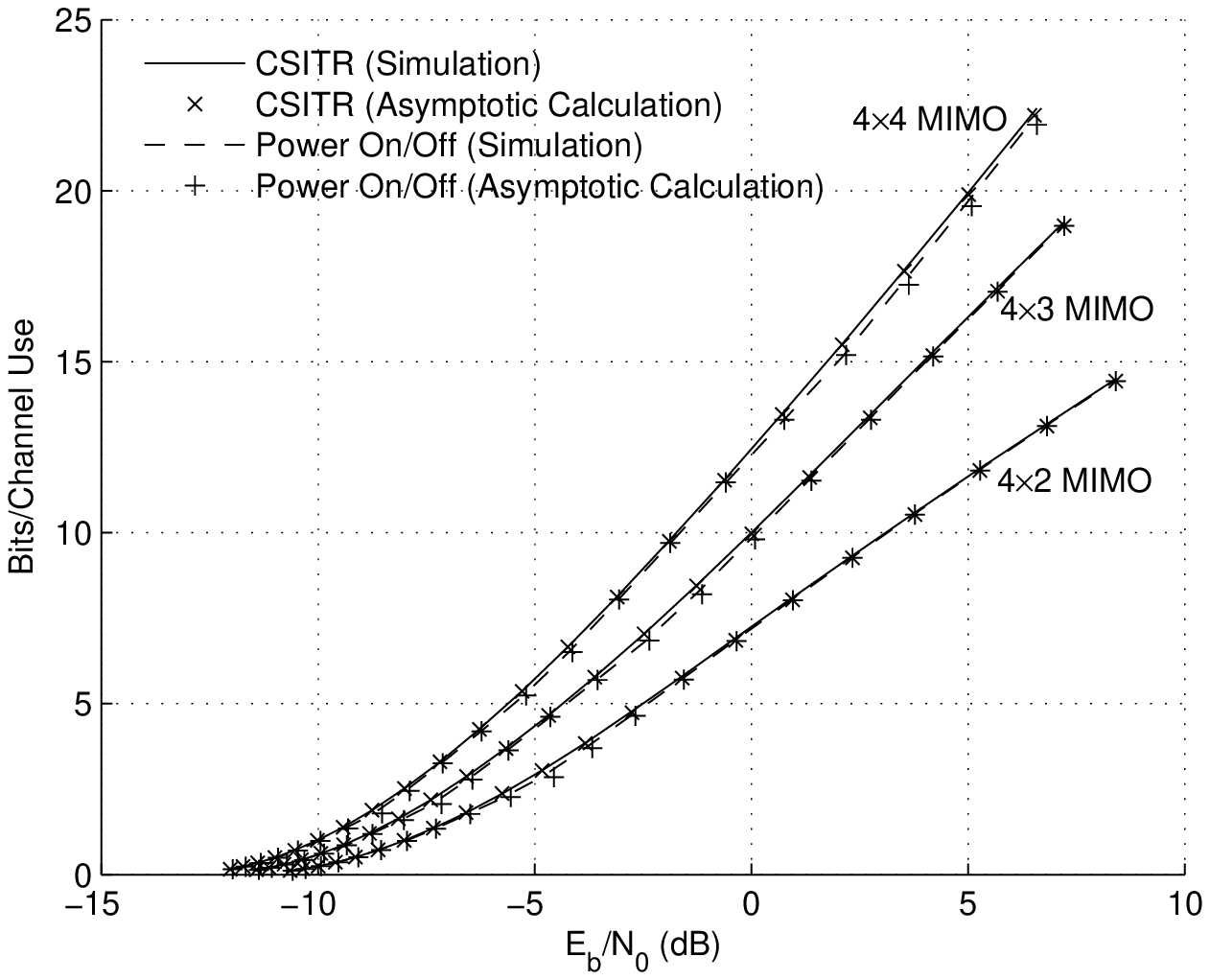}}

\subfigure[Relative performance]{\includegraphics[%
  clip,
  scale=0.8]{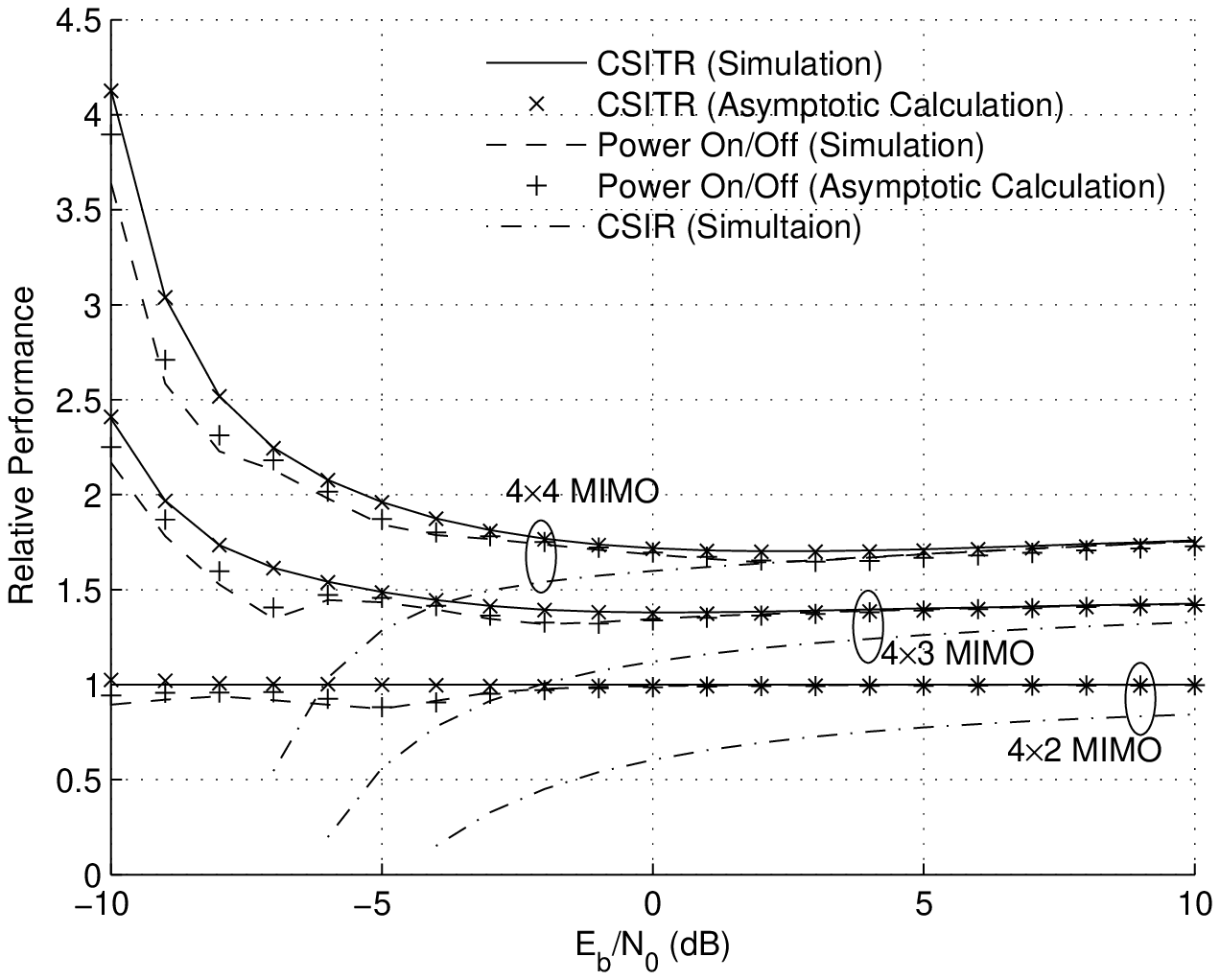}}

\caption{\label{cap:Perf_EbN0_perfect_beam}Information rate v.s. $\mathrm{E_{b}/N_{0}}$
for perfect beamforming}
\end{figure}

Simulation results are given in Fig. \ref{cap:Perf_SNR_perfect_beam}
and Fig. \ref{cap:Perf_EbN0_perfect_beam}. The information rate v.s.
SNR is presented in Fig. \ref{cap:Perf_SNR_perfect_beam}(a) while
Fig. \ref{cap:Perf_EbN0_perfect_beam}(a) shows the information rate
v.s. $\mathrm{E_{b}/N_{0}}$. Different MIMO systems with $4\times2$,
$4\times3$ and $4\times4$ antennas are considered. The solid line
and the dashed line are the simulated information rate for CSITR case
and power on/off strategy respectively. The {}``x'' marker and the
{}``+'' marker are the information rate calculated according to
asymptotic analysis for CSITR case and power on/off strategy respectively.
The difference among them is almost unnoticeable. To make the performance
difference clearer, we also define the relative performance as the
ratio of the considered information rate and the capacity of a $4\times2$
MIMO achieved by water filling power control with perfect CSITR. The
relative performance for different MIMO systems is given in Fig. \ref{cap:Perf_SNR_perfect_beam}(b)
and \ref{cap:Perf_EbN0_perfect_beam}(b). The simulation results show
that power on/off strategy (dashed lines) can achieve more than 90\%
of the capacity provided by water filling power control (solid lines)
and has significant gain comparing to CSIR case (dash-dot lines) at
low SNR. Note that there are several vales in the relative performance
curves. This is due to the fact that $\bar{s}$ can only take discrete
values. Furthermore, the performance evaluated by asymptotic analysis
({}``x'' markers for CSITR case and {}``+'' markers for power
on/off strategy) is very close to the simulated performance. In conclusion,
the power on/off strategy is near optimal for all SNR regimes and
the corresponding performance can be well characterized by asymptotic
analysis.

Since the asymptotic results are accurate for the finite many antennas
case, we can also conclude that the information rate achieved by power
on/off strategy or water filling power allocation grows linearly with
the system dimension $m=\min\left(L_{R},L_{T}\right)$ for a given
SNR. That is, for a given $L_{T}\times L_{R}$ MIMO system, the normalized
information rate $\bar{\mathcal{I}}$ and the normalized capacity
$\bar{C}$ are constants determined by the SNR $\rho$. The total
information rate is that constant multiplied by the dimension $m$.

\section{\label{sec:StrategyDesign_with_beamforming_codebook}Power On/Off
Strategy with a Finite Size Beamforming Codebook}


This section is devoted to quantify the effect of imperfect beamforming
due to finite rate feedback. Comparing to the capacity for perfect
CSITR case, the performance loss of power on/off strategy with finite
rate feedback comes from power on/off and imperfect beamforming. While
Section \ref{sec:Power-On/Off-Perfect-Beamforming} characterizes
the information rate of power on/off strategy for perfect beamforming,
this section will characterize the overall information rate by quantifying
the effect of imperfect beamforming.

Recall the power on/off strategy optimization problem in Problem \ref{pro:(Power-On/off-Strategy-Design)}.
Since the power on/off strategy with a constant number of on-beams
is simple and near optimal, we focus on the effect of imperfect beamforming
when the number of on-beams is a constant. For a constant number of
on-beams, the beamforming codebook contains beamforming matrices of
the same rank. Specifically, let the optimal number of on-beams be
$s$ and the asymptotic optimal normalized number of on-beams be $\bar{s}_{\infty}$.
When $\bar{s}_{\infty}\geq\frac{1}{m}$ (true for most SNR regimes),\[
\mathcal{B}=\left\{ \mathbf{Q}_{i}\in\mathbb{C}^{L_{T}\times s}:\;\mathbf{Q}_{i}^{\dagger}\mathbf{Q}_{i}=\mathbf{I}_{s},\;1\leq i\leq2^{R_{\mathrm{fb}}}\right\} .\]
The only exception happens when the required SNR is so low that $\bar{s}_{\infty}<\frac{1}{m}$
(see Section \ref{sub:MIMO-finite-many-antennas} for details), where
the codebook contains beamforming vectors and one extra index for
the case that the transmitter is turned off. In this case,\[
\mathcal{B}=\left\{ \mathbf{Q}_{i}\in\mathbb{C}^{L_{T}\times1}:\;\mathbf{Q}_{i}^{\dagger}\mathbf{Q}_{i}=1,\;1\leq i\leq2^{R_{\mathrm{fb}}}-1\right\} \cup\left\{ \mathbf{Q}_{\phi}\right\} \]
where $\mathbf{Q}_{\phi}$ is the artificial notation for the the
case that the transmitter is turned off (no beam is on). Since there
is no beamforming when no beam is on, the matrix $\mathbf{Q}_{\phi}$
has no effect in the analysis of imperfect beamforming. Thus the effect
of imperfect beamforming can be analyzed for \[
\mathcal{B}=\left\{ \mathbf{Q}_{i}\in\mathbb{C}^{L_{T}\times1}:\;\mathbf{Q}_{i}^{\dagger}\mathbf{Q}_{i}=1,\;1\leq i\leq2^{R_{\mathrm{fb}}}-1\right\} ,\]
 which can be viewed as a codebook containing $2^{R_{\mathrm{fb}}}-1$
beamforming matrices of rank $1$. Call a beamforming codebook containing
beamforming matrices of the same rank as a \emph{single rank beamforming
codebook}. The power on/off strategy optimization problem (Problem
\ref{pro:(Power-On/off-Strategy-Design)}) is simplified to design
a single rank beamforming codebook $\mathcal{B}$ with size $2^{R_{\mathrm{fb}}}$
or $2^{R_{\mathrm{fb}}}-1$ and the corresponding feedback function
$\varphi\left(\cdot\right)$ to maximize the corresponding information
rate.

To solve the optimization problem and make the performance analysis
tractable, an asymptotic optimal feedback function is introduced and
discussed in Section \ref{sub:Feedback-strategy}. The effect of a
single rank beamforming codebook with this asymptotic optimal feedback
function is well characterized in Section \ref{sub:Effect_Beamforming_Codebook}.

\subsection{\label{sub:Feedback-strategy}Feedback Function}


This subsection considers the feedback function for a given single
rank beamforming codebook. 

The optimal feedback function is given as follows. When the number
of on-beams is a constant $s$, the transmitter transmits a constant
power $sP_{\mathrm{on}}$. For a given single rank beamforming codebook,
it is easy to verify that the optimal feedback function $\varphi^{*}\left(\cdot\right)$
is given by\[
\varphi^{*}\left(\mathbf{H}\right)=\arg\;\underset{1\leq i\leq\left|\mathcal{B}\right|}{\max}\;\ln\left(I_{L_{R}}+P_{\mathrm{on}}\mathbf{HQ}_{i}\mathbf{Q}_{i}^{\dagger}\mathbf{H}^{\dagger}\right).\]

However, this paper considers a suboptimal but asymptotic optimal
feedback function because the corresponding performance can be well
analyzed. Consider the singular value decomposition that $\mathbf{H}=\mathbf{U\Lambda V}^{\dagger}$.
Define $\mathbf{V}_{s}$ as the $L_{T}\times s$ matrix composed by
the singular vectors in $\mathbf{V}$ corresponding to the largest
$s$ singular values. Then both $\mathbf{V}_{s}$ and a beamforming
matrix $\mathbf{Q}\in\mathcal{B}$ can be viewed as generator matrices
of $s$-planes in Grassmann manifold $\mathcal{G}_{L_{T},s}\left(\mathbb{C}\right)$
(see Section \ref{sub:Grassmannian-Manifold} for relative definitions).
Denote the planes generated by $\mathbf{V}_{s}$ and $\mathbf{Q}$
as $\mathcal{P}\left(\mathbf{V}_{s}\right)$ and $\mathcal{P}\left(\mathbf{Q}\right)$
respectively. The feedback function $\hat{\varphi}\left(\cdot\right)$
is defined as\begin{eqnarray}
\hat{\varphi}\left(\mathbf{H}\right) & \triangleq & \arg\;\underset{1\leq i\leq\left|\mathcal{B}\right|}{\min}\; d_{c}\left(\mathcal{P}\left(\mathbf{Q}_{i}\right),\mathcal{P}\left(\mathbf{V}_{s}\right)\right)\nonumber \\
 & = & \arg\;\underset{1\leq i\leq\left|\mathcal{B}\right|}{\max}\;\mathrm{tr}\left(\left(\mathbf{V}_{s}^{\dagger}\mathbf{Q}_{i}\right)\left(\mathbf{V}_{s}^{\dagger}\mathbf{Q}_{i}\right)^{\dagger}\right),\label{eq:suboptimal_feedback}\end{eqnarray}
 where $d_{c}$ is the chordal distance between two elements in $\mathcal{G}_{L_{T},s}\left(\mathbb{C}\right)$%
\footnote{Ties, the case that $\exists\mathbf{Q}_{1},\mathbf{Q}_{2}\in\mathcal{B}$
such that $\mathbf{Q}_{1}\neq\mathbf{Q}_{2}$ but $d_{c}\left(\mathcal{P}\left(\mathbf{Q}_{1}\right),\mathcal{P}\left(\mathbf{V}_{s}\right)\right)=\underset{\mathbf{Q}\in\mathcal{B}}{\min}\; d_{c}\left(\mathcal{P}\left(\mathbf{Q}\right),\mathcal{P}\left(\mathbf{V}_{s}\right)\right)=d_{c}\left(\mathcal{P}\left(\mathbf{Q}_{2}\right),\mathcal{P}\left(\mathbf{V}_{s}\right)\right)$,
can be broken arbitrarily because the probability of ties is zero.%
}.

The feedback function (\ref{eq:suboptimal_feedback}) is asymptotic
optimal. When the size of $\mathcal{B}$ approaches infinity, the
beamforming codebook $\mathcal{B}$ can be constructed so that the
chordal distance between $\mathcal{P}\left(\mathbf{Q}_{\hat{\varphi}\left(\mathbf{H}\right)}\right)$
and $\mathcal{P}\left(\mathbf{V}_{s}\right)$ approaches zero for
any given $\mathbf{V}_{s}$. The information rate achieved by the
suboptimal feedback function approaches that of perfect beamforming,
which is also the limit that the optimal feedback strategy can achieve.

Theorem \ref{thm:u-identity} shows a nice property of the asymptotic
feedback strategy, which will be used to quantify the effect of a
given single rank beamforming codebook in Section \ref{sub:Effect_Beamforming_Codebook}.
The following lemma is used in the proof of Theorem \ref{thm:u-identity}.

\begin{lemma}
\label{lem:u-identisy}Let $\mathbf{A}\in\mathbb{C}^{k\times k}$
be a Hermitian matrix. If $\mathbf{A}=\mathbf{UA}\mathbf{U}^{\dagger}$
for any $k\times k$ unitary matrix $\mathbf{U}$, then $\mathbf{A}=\mu\mathbf{I}$
for some constant $\mu\in\mathbb{R}$.
\end{lemma}
\begin{proof}
For any Hermitian $\mathbf{A}$, there exists a $k\times k$ unitary
$\mathbf{U}$ such that $\mathbf{UA}\mathbf{U}^{\dagger}=\mathbf{\Lambda}$
where $\mathbf{\Lambda}$ is diagonal and with real diagonal elements.
But $\mathbf{UA}\mathbf{U}^{\dagger}=\mathbf{A}$, then $\mathbf{A}$
is diagonal and real. Furthermore, put $\mathbf{U}$ as a permutation
matrix, it is easy to verify that the diagonal elements are identical. 
\end{proof}

\begin{thm}
\label{thm:u-identity}Let $\mathcal{B}$ be a single rank beamforming
codebook with rank $s$ where $1\leq s\leq L_{T}$. Let $\mathbf{V}_{s}$
be a random matrix uniformly distributed on the Stiefel manifold $\mathcal{S}_{L_{T},s}\left(\mathbb{C}\right)$.
Let \[
\hat{\varphi}\left(\mathbf{V}_{s}\right)=\arg\;\underset{1\leq i\leq\left|\mathcal{B}\right|}{\min}\; d_{c}\left(\mathcal{P}\left(\mathbf{Q}_{i}\right),\mathcal{P}\left(\mathbf{V}_{s}\right)\right).\]
Then \[
E_{\mathbf{V}_{s}}\left[\mathbf{V}_{s}^{\dagger}\mathbf{Q}_{\hat{\varphi}\left(\mathbf{V}_{s}\right)}\mathbf{Q}_{\hat{\varphi}\left(\mathbf{V}_{s}\right)}^{\dagger}\mathbf{V}_{s}\right]=\mu\mathbf{I}\]
 where \[
\mu\triangleq1-\frac{1}{s}E_{\mathbf{V}_{s}}\left[d_{c}^{2}\left(\mathcal{P}\left(\mathbf{Q}_{\hat{\varphi}\left(\mathbf{V}_{s}\right)}\right),\mathcal{P}\left(\mathbf{V}_{s}\right)\right)\right]\]
 is a non-negative constant.
\end{thm}
\begin{proof}
For any $s\times s$ unitary matrix $\mathbf{U}$,\begin{eqnarray*}
\mathbf{Q}_{\hat{\varphi}\left(\mathbf{V}_{s}\mathbf{U}\right)} & = & \arg\;\underset{1\leq i\leq\left|\mathcal{B}\right|}{\max}\;\mathrm{tr}\left(\left(\mathbf{Q}_{i}^{\dagger}\mathbf{V}_{s}\mathbf{U}\right)^{\dagger}\left(\mathbf{Q}_{i}^{\dagger}\mathbf{V}_{s}\mathbf{U}\right)\right)\\
 & = & \arg\;\underset{1\leq i\leq\left|\mathcal{B}\right|}{\max}\;\mathrm{tr}\left(\left(\mathbf{Q}_{i}^{\dagger}\mathbf{V}_{s}\right)^{\dagger}\left(\mathbf{Q}_{i}^{\dagger}\mathbf{V}_{s}\right)\right)\\
 & = & \mathbf{Q}_{\hat{\varphi}\left(\mathbf{V}_{s}\right)}.\end{eqnarray*}
Since $\mathbf{V}_{s}$ is uniformly distributed on $\mathcal{S}_{L_{T},s}\left(\mathbb{C}\right)$,
$\mathbf{V}_{s}\mathbf{U}$ is uniformly distributed on $\mathcal{S}_{L_{T},s}\left(\mathbb{C}\right)$
as well \cite{Muirhead_book82_multivariate_statistics}. Then, \begin{eqnarray*}
 &  & \mathbf{U}^{\dagger}E_{\mathbf{V}_{s}}\left[\mathbf{V}_{s}^{\dagger}\mathbf{Q}_{\hat{\varphi}\left(\mathbf{V}_{s}\right)}\mathbf{Q}_{\hat{\varphi}\left(\mathbf{V}_{s}\right)}^{\dagger}\mathbf{V}_{s}\right]\mathbf{U}\\
 &  & =E_{\mathbf{V}_{s}}\left[\mathbf{U}^{\dagger}\mathbf{V}_{s}^{\dagger}\mathbf{Q}_{\hat{\varphi}\left(\mathbf{V}_{s}\right)}\mathbf{Q}_{\hat{\varphi}\left(\mathbf{V}_{s}\right)}^{\dagger}\mathbf{V}_{s}\mathbf{U}\right]\\
 &  & \overset{\left(a\right)}{=}E_{\mathbf{V}_{s}}\left[\left(\mathbf{V}_{s}\mathbf{U}\right)^{\dagger}\mathbf{Q}_{\hat{\varphi}\left(\mathbf{V}_{s}\mathbf{U}\right)}\mathbf{Q}_{\hat{\varphi}\left(\mathbf{V}_{s}\mathbf{U}\right)}^{\dagger}\left(\mathbf{V}_{s}\mathbf{U}\right)\right]\\
 &  & \overset{\left(b\right)}{=}E_{\mathbf{V}_{s}\mathbf{U}}\left[\left(\mathbf{V}_{s}\mathbf{U}\right)^{\dagger}\mathbf{Q}_{\hat{\varphi}\left(\mathbf{V}_{s}\mathbf{U}\right)}\mathbf{Q}_{\hat{\varphi}\left(\mathbf{V}_{s}\mathbf{U}\right)}^{\dagger}\left(\mathbf{V}_{s}\mathbf{U}\right)\right]\\
 &  & \overset{\left(c\right)}{=}E_{\mathbf{V}_{s}}\left[\mathbf{V}_{s}^{\dagger}\mathbf{Q}_{\hat{\varphi}\left(\mathbf{V}_{s}\right)}\mathbf{Q}_{\hat{\varphi}\left(\mathbf{V}_{s}\right)}^{\dagger}\mathbf{V}_{s}\right],\end{eqnarray*}
where
\begin{quote}
(a) follows from the fact that $\mathbf{Q}_{\hat{\varphi}\left(\mathbf{V}_{s}\mathbf{U}\right)}=\mathbf{Q}_{\hat{\varphi}\left(\mathbf{V}_{s}\right)}$,\\
(b) follows from the fact that $\mathbf{V}_{s}\mathbf{U}$ and $\mathbf{V}_{s}$
have the same distribution, and \\
(c) follows from the variable change from $\mathbf{V}_{s}\mathbf{U}$
to $\mathbf{V}_{s}$.
\end{quote}
Therefore, $E_{\mathbf{V}_{s}}\left[\mathbf{V}_{s}^{\dagger}\mathbf{Q}_{\hat{\varphi}\left(\mathbf{V}_{s}\right)}\mathbf{Q}_{\hat{\varphi}\left(\mathbf{V}_{s}\right)}^{\dagger}\mathbf{V}_{s}\right]=\mu\mathbf{I}$
for some constant $\mu$ according to Lemma \ref{lem:u-identisy}.
The constant $\mu$ is non-negative because $\mathbf{V}_{s}^{\dagger}\mathbf{Q}_{\hat{\varphi}\left(\mathbf{V}_{s}\right)}\mathbf{Q}_{\hat{\varphi}\left(\mathbf{V}_{s}\right)}^{\dagger}\mathbf{V}_{s}$
is non-negative definite.

Furthermore, an elementary calculation shows that \[
\mu=1-\frac{1}{s}E_{\mathbf{V}_{s}}\left[d_{c}^{2}\left(\mathcal{P}\left(\mathbf{Q}_{\hat{\varphi}\left(\mathbf{V}_{s}\right)}\right),\mathcal{P}\left(\mathbf{V}_{s}\right)\right)\right].\]

\end{proof}

The constant $\mu$ in the above theorem is related to the average
distortion (defined by squared chordal distance) of a quantization
on the Grassmann manifold. Particularly, we are interested in the
maximum $\mu$ achievable given a codebook size. This problem is solved
in \cite{Dai_05_Quantization_Grassmannian_manifold} and we state
the result as the following.

\begin{thm}
\label{thm:minDc_achievable}Let $\mathcal{B}$ be a single rank beamforming
codebook with rank $s$ where $1\leq s\leq L_{T}$. Denote the size
of $\mathcal{B}$ by $K$. Let $\mathcal{V}$ be a random plane uniformly
distributed on the Grassmann manifold $\mathcal{G}_{L_{T},s}\left(\mathbb{C}\right)$.
Define the average squared chordal distance as \[
\overline{d_{c}^{2}}\left(\mathcal{B}\right)\triangleq E_{\mathcal{V}}\left[\underset{\mathbf{Q}\in\mathcal{B}}{\min}\; d_{c}^{2}\left(\mathcal{P}\left(\mathbf{Q}\right),\mathcal{V}\right)\right].\]
 The minimum average squared chordal distance achievable for a given
$K$, say $\overline{d_{c}^{2}}_{\inf}$, is defined as\[
\overline{d_{c}^{2}}_{\inf}\triangleq\underset{\mathcal{B}:\;\left|\mathcal{B}\right|=K}{\inf}\;\overline{d_{c}^{2}}\left(\mathcal{B}\right).\]
Assume that $K$ is large. $\overline{d_{c}^{2}}_{\inf}$ can be bounded
by \begin{equation}
\frac{t}{t+1}\eta^{-\frac{1}{t}}2^{-\frac{\log_{2}K}{t}}\lesssim\overline{d_{c}^{2}}_{\inf}\lesssim\frac{\Gamma\left(\frac{1}{t}\right)}{t}\eta^{-\frac{1}{t}}2^{-\frac{\log_{2}K}{t}},\label{eq:dc_bounds}\end{equation}
where $t=s\left(L_{T}-s\right)$ is the number of the real dimensions
of $\mathcal{G}_{L_{T},s}\left(\mathbb{C}\right)$, \begin{equation}
\eta=\left\{ \begin{array}{ll}
\frac{1}{t!}\prod_{i=1}^{s}\frac{\left(L_{T}-i\right)!}{\left(s-i\right)!} & \mathrm{if}\;1\leq s\leq\frac{L_{T}}{2}\\
\frac{1}{t!}\prod_{i=1}^{L_{T}-s}\frac{\left(L_{T}-i\right)!}{\left(L_{T}-s-i\right)!} & \mathrm{if}\;\frac{L_{T}}{2}\leq s\leq L_{T}\end{array}\right.,\label{eq:dc_constant}\end{equation}
and the symbol $\lesssim$ denotes the \emph{main order inequality}.
\end{thm}

Although this theorem is for asymptotically large $K$, the bounds
(\ref{eq:dc_bounds}) are accurate enough for relatively small $K$.
For example, it is shown in \cite{Dai_05_Quantization_Grassmannian_manifold}
that the bounds are tight for $K\geq10$ when $L_{T}=4$, $s=2$ .
Furthermore, as the number of real dimensions of the Grassmann manifold
($2t$) approaches infinity, the lower bound and the upper bound are
asymptotically equal.

It is noteworthy that Theorem \ref{thm:minDc_achievable} holds for
Grassmann manifolds with arbitrary dimensions. In \cite{Heath_ICASSP05_Quantization_Grassmann_Manifold},
approximations to $\bar{d}_{c,\inf}^{2}$ are developed for $s=1$
case and the case that $s\geq1$ is fixed and $L_{T}$ is asymptotically
large. Indeed, the approximation in \cite{Heath_ICASSP05_Quantization_Grassmann_Manifold}
for $s=1$ is a lower bound on $\bar{d}_{c,\inf}^{2}$. The approximation
in \cite{Heath_ICASSP05_Quantization_Grassmann_Manifold} for fixed
$s$ and asymptotically large $L_{T}$ is neither a lower bound nor
an upper bound. A detailed comparison of Theorem \ref{thm:minDc_achievable}
and the results of \cite{Heath_ICASSP05_Quantization_Grassmann_Manifold}
can be found in \cite{Dai_05_Quantization_Grassmannian_manifold}.

Apply Theorem \ref{thm:minDc_achievable} to Theorem \ref{thm:u-identity},
the maximum $\mu$ achievable, say $\mu_{\sup}$, can be upper and
lower bounded. This result about the suboptimal feedback function
will be employed to analyze the effect of a single rank beamforming
codebook on information rate in Section \ref{sub:Effect_Beamforming_Codebook}.

\subsection{\label{sub:Effect_Beamforming_Codebook}Effect of a Beamforming Codebook}

In this section, the effect of a single rank beamforming codebook
is accurately quantified. Thus the overall performance of power on/off
strategy with finite rate feedback can be well characterized by combining
the asymptotic results in Section \ref{sec:Power-On/Off-Perfect-Beamforming}
and the effect of a single rank beamforming codebook. 

A lower bound to the information rate is derived first. For a channel
state realization $\mathbf{H}$, let $\lambda_{i}$ be the $i^{\mathrm{th}}$
largest eigenvalue of $\mathbf{H}^{\dagger}\mathbf{H}$ and $\mathbf{v}_{i}$
be the eigenvector corresponding to $\lambda_{i}$. Then $\mathbf{H}^{\dagger}\mathbf{H}=\mathbf{V\Lambda V}^{\dagger}$
where $\mathbf{V}=\left[\mathbf{v}_{1},\mathbf{v}_{2},\cdots,\mathbf{v}_{L_{T}}\right]$
and $\mathbf{\Lambda}=\mathrm{diag}\left[\lambda_{1},\lambda_{2},\cdots,\lambda_{L_{T}}\right]$.
For a given optimal number of on-beams $s$ such that $1\leq s\leq L_{T}$,
define $\mathbf{V}_{s}\triangleq\left[\mathbf{v}_{1},\mathbf{v}_{2},\cdots,\mathbf{v}_{s}\right]$
and $\mathbf{\Lambda}_{s}\triangleq\mathrm{diag}\left[\lambda_{1},\lambda_{2},\cdots,\lambda_{s}\right]$.
Then,

\begin{eqnarray*}
\mathbf{V\Lambda V}^{\dagger} & \geq & \mathbf{V}\left[\begin{array}{cc}
\mathbf{\Lambda}_{s}\\
 & \mathbf{0}\end{array}\right]\mathbf{V}^{\dagger}\\
 & = & \mathbf{V}_{s}\mathbf{\Lambda}_{s}\mathbf{V}_{s}^{\dagger},\end{eqnarray*}
where two matrices $\mathbf{A}\;\mathrm{and}\;\mathbf{B}$ have the
relationship $\mathbf{A}\geq\mathbf{B}$ if $\mathbf{A}-\mathbf{B}$
is non-negative definite. Let $\mathbf{Q}_{\hat{\varphi}\left(\mathbf{H}\right)}$
be the feedback beamforming matrix given by the feedback function
(\ref{eq:suboptimal_feedback}). We have\[
\mathbf{I}_{s}+P_{\mathrm{on}}\mathbf{Q}_{\hat{\varphi}\left(\mathbf{H}\right)}^{\dagger}\mathbf{V\Lambda V}^{\dagger}\mathbf{Q}_{\hat{\varphi}\left(\mathbf{H}\right)}\geq\mathbf{I}_{s}+P_{\mathrm{on}}\mathbf{Q}_{\hat{\varphi}\left(\mathbf{H}\right)}^{\dagger}\mathbf{V}_{s}\mathbf{\Lambda}_{s}\mathbf{V}_{s}^{\dagger}\mathbf{Q}_{\hat{\varphi}\left(\mathbf{H}\right)}.\]
Moreover, the matrices on both sides of the above inequality are positive
definite. Because $\mathbf{A}\geq\mathbf{B}$ implies $\left|\mathbf{A}\right|\geq\left|\mathbf{B}\right|$
for any two positive definite matrices $\mathbf{A}$ and $\mathbf{B}$
\cite{Muirhead_book82_multivariate_statistics}, we have \begin{eqnarray*}
\mathrm{ln}\left|\mathbf{I}+P_{\mathrm{on}}\mathbf{Q}_{\hat{\varphi}\left(\mathbf{H}\right)}^{\dagger}\mathbf{V\Lambda V}^{\dagger}\mathbf{Q}_{\hat{\varphi}\left(\mathbf{H}\right)}\right| & \geq & \mathrm{ln}\left|\mathbf{I}+P_{\mathrm{on}}\mathbf{Q}_{\hat{\varphi}\left(\mathbf{H}\right)}^{\dagger}\mathbf{V}_{s}\mathbf{\Lambda}_{s}\mathbf{V}_{s}^{\dagger}\mathbf{Q}_{\hat{\varphi}\left(\mathbf{H}\right)}\right|.\end{eqnarray*}
Therefore, the information rate is lower bounded by\begin{eqnarray}
\bar{\mathcal{I}} & = & \frac{1}{m}\mathrm{E}_{\mathbf{H}}\left[\mathrm{ln}\left|\mathbf{I}_{L_{R}}+P_{\mathrm{on}}\mathbf{H}\mathbf{Q}_{\hat{\varphi}\left(\mathbf{H}\right)}\mathbf{Q}_{\hat{\varphi}\left(\mathbf{H}\right)}^{\dagger}\mathbf{H}^{\dagger}\right|\right]\nonumber \\
 & = & \frac{1}{m}\mathrm{E}_{\mathbf{H}}\left[\mathrm{ln}\left|\mathbf{I}_{s}+P_{\mathrm{on}}\mathbf{Q}_{\hat{\varphi}\left(\mathbf{H}\right)}^{\dagger}\mathbf{H}^{\dagger}\mathbf{H}\mathbf{Q}_{\hat{\varphi}\left(\mathbf{H}\right)}\right|\right]\nonumber \\
 & = & \frac{1}{m}\mathrm{E}_{\mathbf{H}}\left[\mathrm{ln}\left|\mathbf{I}_{s}+P_{\mathrm{on}}\mathbf{Q}_{\hat{\varphi}\left(\mathbf{H}\right)}^{\dagger}\mathbf{V\Lambda V}^{\dagger}\mathbf{Q}_{\hat{\varphi}\left(\mathbf{H}\right)}\right|\right]\nonumber \\
 & \geq & \frac{1}{m}\mathrm{E}_{\mathbf{H}}\left[\mathrm{ln}\left|\mathbf{I}_{s}+P_{\mathrm{on}}\mathbf{Q}_{\hat{\varphi}\left(\mathbf{H}\right)}^{\dagger}\mathbf{V}_{s}\mathbf{\Lambda}_{s}\mathbf{V}_{s}^{\dagger}\mathbf{Q}_{\hat{\varphi}\left(\mathbf{H}\right)}\right|\right].\label{eq:capacity-lbd}\end{eqnarray}
 The lower bound is tight under high feedback rate assumption. For
the perfect beamforming case, the lower bound is indeed the information
rate itself.

Based on this lower bound, an approximation to the information rate
can be obtained. Since entries of $\mathbf{H}$ are i.i.d. $\mathcal{CN}\left(0,1\right)$,
$\mathbf{V}_{s}$ is uniformly (isotropically) distributed on $\mathcal{S}_{L_{T},s}\left(\mathbb{C}\right)$
and independent with $\mathbf{\Lambda}_{s}$ \cite{Muirhead_book82_multivariate_statistics,Tse_IT02_Communication_on_Grassmann_Manifold}.
By the lower bound in (\ref{eq:capacity-lbd}), we have\begin{eqnarray}
 &  & \frac{1}{m}\mathrm{E}_{\mathbf{H}}\left[\mathrm{ln}\left|\mathbf{I}_{s}+P_{\mathrm{on}}\mathbf{Q}_{\hat{\varphi}\left(\mathbf{H}\right)}^{\dagger}\mathbf{V}_{s}\mathbf{\Lambda}_{s}\mathbf{V}_{s}^{\dagger}\mathbf{Q}_{\hat{\varphi}\left(\mathbf{H}\right)}\right|\right]\nonumber \\
 &  & \overset{\left(a\right)}{=}\frac{1}{m}\mathrm{E}_{\mathbf{\Lambda}_{s}}\left[\mathrm{E}_{\mathbf{V}_{s}}\left[\mathrm{ln}\left|\mathbf{I}_{s}+P_{\mathrm{on}}\mathbf{V}_{s}^{\dagger}\mathbf{Q}_{\hat{\varphi}\left(\mathbf{H}\right)}\mathbf{Q}_{\hat{\varphi}\left(\mathbf{H}\right)}^{\dagger}\mathbf{V}_{s}\mathbf{\Lambda}_{s}\right|\right]\right]\nonumber \\
 &  & \overset{\left(b\right)}{\leq}\frac{1}{m}\mathrm{E}_{\mathbf{\Lambda}_{s}}\left[\mathrm{ln}\left|\mathbf{I}_{s}+P_{\mathrm{on}}\mathrm{E}_{\mathbf{V}_{s}}\left[\mathbf{V}_{s}^{\dagger}\mathbf{Q}_{\hat{\varphi}\left(\mathbf{H}\right)}\mathbf{Q}_{\hat{\varphi}\left(\mathbf{H}\right)}^{\dagger}\mathbf{V}_{s}\right]\mathbf{\Lambda}_{s}\right|\right]\nonumber \\
 &  & \overset{\left(c\right)}{=}\frac{1}{m}\mathrm{E}_{\mathbf{H}}\left[\mathrm{ln}\left|\mathbf{I}+\mu P_{\mathrm{on}}\mathbf{\Lambda}_{s}\right|\right]\label{eq:capacity-approx}\end{eqnarray}
where 

\begin{figure}
\includegraphics[%
  clip,
  scale=0.8]{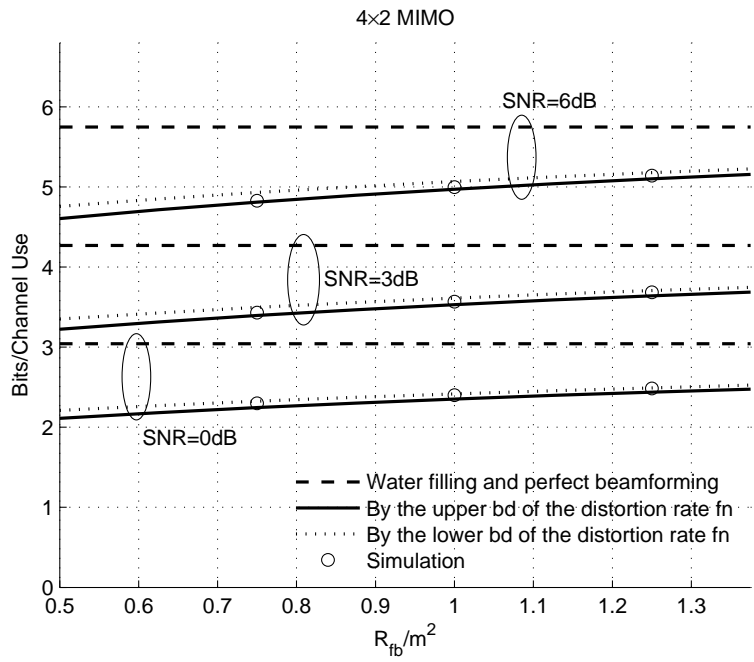}

\caption{\label{cap:Fig-capacity-approx}Information rate of finite size single
rank beamforming codebooks}
\end{figure}

\begin{quote}
(a) holds because $\left|\mathbf{I}+\mathbf{AB}\right|=\left|\mathbf{I}+\mathbf{BA}\right|$
and $\mathbf{V}_{s}$ is independent with $\mathbf{\Lambda}_{s}$, 

(b) follows from the concavity of $\ln\left|\cdot\right|$ function
\cite[prob. 2 on pg. 237]{Cover_book91_information_theory}, and 

(c) follows from Theorem \ref{thm:u-identity} where $\mu$ is defined.
\end{quote}
Although the approximation (\ref{eq:capacity-approx}) is neither
an upper bound nor a lower bound to the normalized information rate,
it gives a good characterization under high feedback rate assumption.
In fact, for a $4\times2$ MIMO system with feedback rate $R_{\mathrm{fb}}=4$bits/channel
use, the information rate calculated by (\ref{eq:capacity-approx})
is very close to that evaluated by Monte Carlo simulation (see Fig.
\ref{cap:Fig-capacity-approx}).

The constant $\mu$ is called as \emph{power efficiency factor.} The
effect of a finite size beamforming codebook can be viewed as decreasing
the $P_{\mathrm{on}}$ in (\ref{eq:I_perfect_beamforming}) to $\mu P_{\mathrm{on}}$
in (\ref{eq:capacity-approx}). Thus for a given codebook size, the
beamforming codebook should be designed to maximize the corresponding
power efficiency factor $\mu$, or equivalently, to minimize the average
squared chordal distance $\overline{d_{c}^{2}}$. However, it may
be computational complex to design a codebook to minimize $\overline{d_{c}^{2}}$
directly. In \cite{Love_SP05_Limited_feedback_unitary_precoding},
a criterion of maximizing the minimum chordal distance between any
pair of beamforming matrices (max-min criterion) is proposed to achieve
small $\overline{d_{c}^{2}}$. In this paper, we adopt the max-min
criterion to design beamforming codebook for simplicity. Assuming
that a beamforming codebook is well designed, the maximum $\mu$ achievable
can be tightly upper and lower bounded as functions of the codebook
size according to Theorem \ref{thm:minDc_achievable}. Note that $K=2^{R_{\mathrm{fb}}}$
when $\bar{s}_{\infty}\geq\frac{1}{m}$ and $K=2^{R_{\mathrm{fb}}}-1\approx2^{R_{\mathrm{fb}}}$
when $\bar{s}_{\infty}<\frac{1}{m}$. We have\begin{equation}
1-\frac{\Gamma\left(\frac{1}{t}\right)}{st}\eta^{-\frac{1}{t}}2^{-\frac{R_{\mathrm{fb}}}{t}}\lesssim\mu_{\sup}\lesssim1-\frac{t}{s\left(t+1\right)}\eta^{-\frac{1}{t}}2^{-\frac{R_{\mathrm{fb}}}{t}},\label{eq:u-bounds}\end{equation}
where $s$ is the rank of the single rank beamforming codebook, $t=s\left(L_{T}-s\right)$
and $\eta$ is given in (\ref{eq:dc_constant}). Comparing the imperfect
beamforming case to the perfect beamforming case, the effective power
loss $1-\mu_{\sup}$ decays exponentially with a rate proportional
to $R_{\mathrm{fb}}/m^{2}$ (specifically, the exact rate of exponential
decay is $\frac{m^{2}}{s\left(L_{T}-s\right)}\cdot\frac{R_{\mathrm{fb}}}{m^{2}}$).
Thus for practical MIMO systems where $m$ is not large, a few bits
may be enough to achieve a performance close to CSITR.

According to the above results, the information rate of a power on/off
strategy with a well-designed single rank beamforming codebook can
be well characterized. For a given $L_{T}\times L_{R}$ MIMO system
with finite rate channel state feedback up to $R_{\mathrm{fb}}$ bits/channel
use, $\mu_{\sup}$ can be estimated according to (\ref{eq:u-bounds})
for all $s$'s such that $1\leq s\leq L_{T}$. Substitute the bounds
on $\mu_{\sup}$ into the information rate approximation (\ref{eq:capacity-approx})
and then use the the asymptotic formulas in Section \ref{subsub:Asymptotic-Analysis-Power-on/off}
for perfect beamforming case. The optimal number of on-beams $s$
and the corresponding information rate $\mathcal{I}$ can be calculated. 

Fig. \ref{cap:Fig-capacity-approx} gives the simulation results for
a $4\times2$ MIMO system. The performance curves are plotted as functions
of $R_{\mathrm{fb}}/m^{2}$. The simulated information rate (circles)
is compared to the information rate characterized by the lower bound
(solid lines) and the upper bound (dotted lines) of $\overline{d_{c}^{2}}_{\inf}$.
The simulation results show that the information rate characterized
by the bounds (\ref{eq:dc_bounds}) matches the actual performance
almost perfectly. Note that the previous approximation proposed in
\cite{Honig_MiliComm03_Asymptotic_MIMO_Limited_Feedback,Honig_Allerton03_Benefits_Limited_Feedback_Wireless_Channels},
which is based on asymptotic analysis and Gaussian approximation,
overestimates the information rate (a correction of the result is
in \cite{Honig_discussion}). Our characterization is more accurate.

\section{\label{sec:Performance-Comparision}Performance Comparison}


While we have shown that power on/off strategy with a constant number
of on-beams is near optimal for perfect beamforming in Section \ref{sec:Power-On/Off-Perfect-Beamforming},
this section will show that a constant number of on-beams are near
optimal when beamforming is imperfect as well.

To show the near optimality of a constant number of on-beams, the
single rank beamforming codebooks are compared to multi-rank beamforming
codebooks, which may contain beamforming matrices of different ranks.
For a multi-rank beamforming codebook \[
\mathcal{B}=\left\{ \mathbf{Q}_{i}\in\mathbb{C}^{L_{T}\times s}:\;\mathbf{Q}_{i}^{\dagger}\mathbf{Q}_{i}=\mathbf{I}_{s},\;0\leq s\leq L_{T},\;1\leq i\leq2^{R_{\mathrm{fb}}}\right\} ,\]
 define a single rank sub-code with rank $s$ as \[
\mathcal{B}_{s}\triangleq\left\{ \mathbf{Q}_{i}:\;\mathbf{Q}_{i}\in\mathcal{B},\;\mathrm{rank}\left(\mathbf{Q}_{i}\right)=s\right\} \]
 where $0\leq s\leq L_{T}$. The multi-rank beamforming codebook $\mathcal{B}$
can be viewed as a union of the single rank sub-codes $\mathcal{B}=\bigcup_{s=0}^{L_{T}}\mathcal{B}_{s}$.
The corresponding power on/off strategy design problem is to find
the optimal multi-rank beamforming codebook $\mathcal{B}$, feedback
function $\varphi\left(\cdot\right)$ and constant $P_{\mathrm{on}}$
to maximize the information rate with a power constraint $\rho$,
as stated in Problem \ref{pro:(Power-On/off-Strategy-Design)}.

It is difficult to solve the optimization problem for multi-rank beamforming
codebooks. However, for a given multi-rank beamforming codebook $\mathcal{B}$
and a given $P_{\mathrm{on}}$, the following theorem gives the explicit
form of the optimal feedback function, say $\tilde{\varphi}\left(\cdot\right)$,
to avoid exhaustive search in all possible feedback functions. The
intuition behind is same as the intuition which we learnt from Theorem
\ref{thm:optimal_s_perfect_beamforming}: all the {}``good'' beams
and only the {}``good'' beams should be turned on. The particular
aspect of the following theorem is that {}``good'' beams need to
be reasonably defined.

\begin{thm}
\label{thm:optimal-feedback-multirank-B}Consider the power on/off
strategy with a given multi-rank beamforming codebook $\mathcal{B}=\bigcup_{s=0}^{L_{T}}\mathcal{B}_{s}$
and a given $P_{\mathrm{on}}$. For a given channel realization $\mathbf{H}$,
define $\mathcal{I}_{s}\left(\mathbf{H}\right)$ as the largest mutual
information achievable for a non-empty sub-code $\mathcal{B}_{s}$\[
\mathcal{I}_{s}\left(\mathbf{H}\right)=\underset{\mathbf{Q}_{i}\in\mathcal{B}_{s}}{\max}\;\ln\left|\mathbf{I}_{L_{R}}+P_{\mathrm{on}}\mathbf{HQ}_{i}\mathbf{Q}_{i}^{\dagger}\mathbf{\mathbf{H}}^{\dagger}\right|,\]
where $\mathcal{B}_{s}\neq\phi$, $0\leq s\leq L_{T}$ ($\mathcal{I}_{0}\left(\mathbf{H}\right)=0$).
Denote the optimal feedback function as $\tilde{\varphi}\left(\cdot\right)$.
Then \[
\tilde{\varphi}\left(\mathbf{H}\right)=\arg\;\underset{i:\;\mathbf{Q}_{i}\in\mathcal{B}_{\tilde{s}\left(\mathbf{H}\right)}}{\max}\;\ln\left|\mathbf{I}_{L_{R}}+P_{\mathrm{on}}\mathbf{H}\mathbf{Q}_{i}\mathbf{Q}_{i}^{\dagger}\mathbf{H}^{\dagger}\right|,\]
where \[
\tilde{s}\left(\mathbf{H}\right)\triangleq\max\left\{ s:\;\mathcal{B}_{s}\neq\phi,\;\mathcal{I}_{s}\left(\mathbf{H}\right)-\mathcal{I}_{t}\left(\mathbf{H}\right)\geq(s-t)\kappa\;\textrm{for }\mathbf{all}\; t\;\textrm{s.t.}\;0\leq t<s,\;\mathcal{B}_{t}\neq\phi\right\} \]
and $\kappa$ is the appropriate threshold to satisfy the average
power constraint\begin{eqnarray*}
E_{\mathbf{H}}\left[\tilde{s}\left(\mathbf{H}\right)P_{\mathrm{on}}\right]=\rho.\end{eqnarray*}

\end{thm}
\begin{proof}
See Appendix \ref{sub:Proof-of-multirank-B}.
\end{proof}

The following examples are direct applications of Theorem \ref{thm:optimal-feedback-multirank-B}.

\begin{example}
Let $\mathcal{B}=\left\{ \mathbf{I}_{L_{T}},\mathbf{Q}_{\phi}\right\} $
where $\mathbf{Q}_{\phi}$ is the artificial notion for the case that
the transmitter is turned off. Then the optimal power on/off function
is to turn on all transmit antennas if \[
\ln\left(\mathbf{I}_{L_{R}}+P_{\mathrm{on}}\mathbf{H}\mathbf{H}^{\dagger}\right)\geq\kappa L_{T}\]
and turn off the transmitter if \[
\ln\left(\mathbf{I}_{L_{R}}+P_{\mathrm{on}}\mathbf{H}\mathbf{H}^{\dagger}\right)<\kappa L_{T}\]
where $\kappa$ is an appropriate chosen threshold to satisfy\[
L_{T}P_{\mathrm{on}}\Pr\left\{ \ln\left(\mathbf{I}+P_{\mathrm{on}}\mathbf{H}\mathbf{H}^{\dagger}\right)\geq\kappa L_{T}\right\} =\rho.\]

\end{example}

\begin{figure}
\subfigure[Information rate]{\includegraphics[%
  clip,
  scale=0.8]{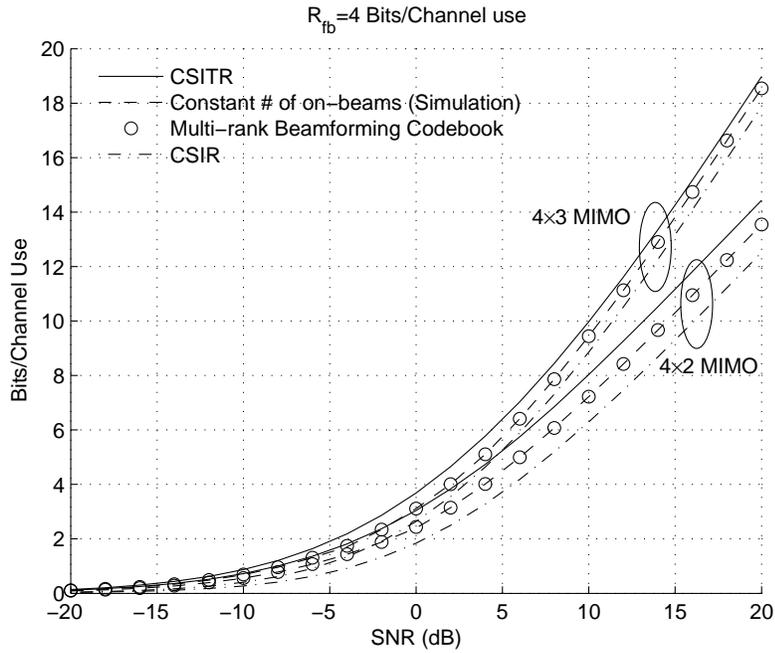}}

\subfigure[Relative performance]{\includegraphics[%
  clip,
  scale=0.8]{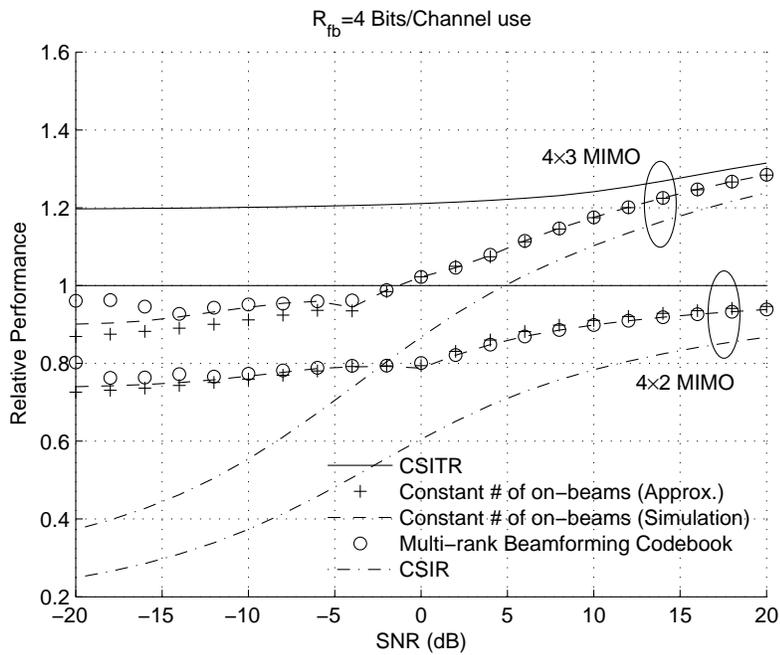}}

\caption{\label{cap:Fig-Perf-Compare}Comparison of single rank beamforming
codebooks and multi-rank beamforming codebooks}
\end{figure}

\begin{example}
Let $\left|\mathcal{B}\right|\rightarrow\infty$ and $\mathcal{B}$
is constructed so that the beamforming is asymptotically perfect.
It is easy to verify that the optimal feedback function given by Theorem
\ref{thm:optimal-feedback-multirank-B} is same as the one given in
Theorem \ref{thm:optimal_s_perfect_beamforming} for perfect beamforming
case.


Although the optimal feedback function for a multi-rank beamforming
codebook is given in Theorem \ref{thm:optimal-feedback-multirank-B},
it is difficult to find the optimal multi-rank beamforming codebook
$\mathcal{B}$, the optimal $P_{\mathrm{on}}$ and the corresponding
information rate. In our simulation, we try different multi-rank codebooks
and different $P_{\mathrm{on}}$'s and then choose the best one. Specifically,
denote $K_{s}$ as the size of the sub-code $\mathcal{B}_{s}$, $K_{s}\triangleq\left|\mathcal{B}_{s}\right|$.
We try all possible combinations of $\left[K_{0},K_{1},\cdots,K_{L_{T}}\right]$'s
such that $K_{s}\in\mathbb{Z}^{+}\cup\left\{ 0\right\} $ and $\sum_{s=0}^{L_{T}}K_{s}\leq2^{R_{\mathrm{fb}}}$.
For each $\left[K_{0},K_{1},\cdots,K_{L_{T}}\right]$, we construct
the sub-codes $\mathcal{B}_{s}$'s such that $\left|\mathcal{B}_{s}\right|=K_{s}$
for $s=0,1,\cdots,L_{T}$ according to the max-min criterion in \cite{Love_SP05_Limited_feedback_unitary_precoding}.
The ultimate multi-rank beamforming codebook is given by $\mathcal{B}=\bigcup_{s=0}^{L_{T}}\mathcal{B}_{s}$.
For every multi-rank codebook $\mathcal{B}$, we try different $P_{\mathrm{on}}$'s
and search for the optimal one. The optimal multi-rank codebook $\mathcal{B}$
is chosen from the codebooks corresponding to all possible $\left[K_{0},K_{1},\cdots,K_{L_{T}}\right]$'s.

Fig. \ref{cap:Fig-Perf-Compare} shows the simulation results. Fig.
\ref{cap:Fig-Perf-Compare}(a) compares the information rates of single
rank beamforming codebooks and multi-rank beamforming codebooks. Fig.
\ref{cap:Fig-Perf-Compare}(b) presents the relative performance,
which is defined as the ratio of the considered information rate and
the capacity of a $4\times2$ MIMO system with perfect CSITR. We also
present the information rate characterization by the upper bound of
$\overline{d_{c}^{2}}_{\inf}$ (Section \ref{sub:Effect_Beamforming_Codebook}).
Simulations show that single rank beamforming codebooks (dashed lines)
achieve almost the same information rate of multi-rank beamforming
codebooks (circles). The performance difference is noticeable in very
low SNR regime. This is because the power on/off strategy with a single
rank beamforming codebook is designed according to the asymptotic
distribution of eigenvalues of $\frac{1}{m}\mathbf{H}\mathbf{H}^{\dagger}$
while the key parameters ($P_{\mathrm{on}}$ and $\kappa$) of power
on/off strategy with multi-rank beamforming codebooks are numerically
optimized according to the actual distribution of $\frac{1}{m}\mathbf{H}\mathbf{H}^{\dagger}$.
According to the simulation results, power on/off strategy with constant
number of on-beams provides a simple but near-optimal solution for
finite rate channel state feedback.
\end{example}

\section{\label{sec:Conclusions}Conclusions}


This paper accurately characterizes the information rate of the power
on/off strategy with finite rate channel state feedback. According
to asymptotic analysis, the power on/off strategy with a constant
number of on-beams is employed and studied. Simulations show that
this strategy is near optimal for all SNR regimes. We derive asymptotic
formulas for perfect beamforming case and introduce the power efficiency
factor to quantify the effect of imperfect beamforming. By combining
a formula for power efficiency factor and the asymptotic formulas
for perfect beamforming, we characterize  the corresponding information
rate accurately for all SNR regimes. 

An important point that is not mentioned in this paper is the complexity
of selecting the feedback beamforming matrix in a codebook, which
may involve exhaustive search. To avoid exhaustive search, beamforming
codebooks with certain structure may be considered in future so that
the matrix selection can be more efficient by employing the structure
of the codebook.

\appendix

\subsection{\label{sub:pf_optimal_s_perfect_beamforming}Proof of Theorem \ref{thm:optimal_s_perfect_beamforming}}

Let's start with the single input single output (SISO) case. In SISO
case, $\mathbf{H}$ is a scalar and $\frac{1}{m}\mathbf{H}^{\dagger}\mathbf{H}$
has only one eigenvalue, i.e., $\lambda=\left|\mathbf{H}\right|^{2}$.
Denote the corresponding cumulative distribution function (CDF) as
$F_{\Lambda}\left(\lambda\right)$. Define $\Omega$ as the set of
$\lambda$ corresponding to the case that the transmitter is turned
on. Then any deterministic power on/off strategy can be uniquely defined
by $\Omega$. Thus the optimization problem is to choose an appropriate
Lebesgue measurable set $\Omega\subset\mathbb{R}^{+}\cup\left\{ 0\right\} $
to maximize \[
\int_{\Omega}\log\left(1+P_{\mathrm{on}}\lambda\right)dF_{\Lambda}\left(\lambda\right)\]
with the power constraint \[
\int_{\Omega}dF_{\Lambda}\left(\lambda\right)=\rho/P_{\mathrm{on}}.\]
Since $F_{\Lambda}\left(\lambda\right)$ is continuous, there exists
an $\Omega$ to satisfy the power constraint. The optimization problem
is well defined. 

Define $\Omega^{*}=\left\{ \lambda:\;\lambda\geq\kappa\right\} $
such that $\int_{\Omega^{*}}dF_{\Lambda}\left(\lambda\right)=\rho/P_{on}$.
For any Lebesgue measurable set $\Omega\subset\mathbb{R}^{+}\cup\left\{ 0\right\} $
such that $\int_{\Omega}dF_{\Lambda}\left(\lambda\right)=\rho/P_{\mathrm{on}}$,
\begin{eqnarray*}
 &  & \int_{\Omega^{*}}\log\left(1+P_{\mathrm{on}}\lambda\right)dF_{\Lambda}\left(\lambda\right)-\int_{\Omega}\log\left(1+P_{\mathrm{on}}\lambda\right)dF_{\Lambda}\left(\lambda\right)\\
 &  & \;=\int_{\Omega^{*}-\Omega}\log\left(1+P_{\mathrm{on}}\lambda\right)dF_{\Lambda}\left(\lambda\right)-\int_{\Omega-\Omega^{*}}\log\left(1+P_{\mathrm{on}}\lambda\right)dF_{\Lambda}\left(\lambda\right)\\
 &  & \;\overset{\left(a\right)}{\geq}\int_{\Omega^{*}-\Omega}\log\left(1+P_{\mathrm{on}}\kappa\right)dF_{\Lambda}\left(\lambda\right)-\int_{\Omega-\Omega^{*}}\log\left(1+P_{\mathrm{on}}\kappa\right)dF_{\Lambda}\left(\lambda\right)\\
 &  & \;=\log\left(1+P_{\mathrm{on}}\kappa\right)\left[\int_{\Omega^{*}-\Omega}dF_{\Lambda}\left(\lambda\right)-\int_{\Omega-\Omega^{*}}dF_{\Lambda}\left(\lambda\right)\right]\\
 &  & \;\overset{\left(b\right)}{=}0,\end{eqnarray*}
where 

\begin{quote}
(a) follows from the facts that $\lambda\geq\kappa$ when $\lambda\in\Omega^{*}-\Omega$
and $\lambda<\kappa$ when $\lambda\in\Omega-\Omega^{*}$, and

(b) holds because $\int_{\Omega^{*}}dF_{\Lambda}\left(\lambda\right)=\int_{\Omega}dF_{\Lambda}\left(\lambda\right)$
implies $\int_{\Omega^{*}-\Omega}dF_{\Lambda}\left(\lambda\right)=\int_{\Omega-\Omega^{*}}dF_{\Lambda}\left(\lambda\right)$.
\end{quote}
Therefore, $\Omega^{*}$ is the optimal set and the power on/off strategy
defined by $\Omega^{*}$ is optimal.

The proof for MIMO case follows the same idea. For an $L_{T}\times L_{R}$
MIMO system, denote the vector of the ordered $L_{T}$ eigenvalues
of $\mathbf{H}^{\dagger}\mathbf{H}$ as $\mathbf{\lambda}=\left[\lambda_{1},\cdots,\lambda_{L_{T}}\right]$
where $\lambda_{1}\geq\lambda_{2}\geq\cdots\geq\lambda_{L_{T}}\geq0$
and the corresponding multivariate CDF as $F_{\mathbf{\Lambda}}\left(\mathbf{\lambda}\right)$.
Define\[
\Omega_{k}=\left\{ \mathbf{\lambda}:\;\mathrm{the\; eigen\; channel\; corresponding\; to\;}\lambda_{k}\;\mathrm{is\; on}\right\} \]
where $1\leq k\leq L_{T}$. Then any deterministic power on/off strategy
can be uniquely defined by $\Omega_{k}$'s where $1\leq k\leq L_{T}$.
The optimization problem is to choose Lebesgue measurable sets $\Omega_{k}\subset\left(\mathbb{R}^{+}\cup\left\{ 0\right\} \right)^{m}$,
$k=1,2,\cdots,L_{T}$, to maximize \[
\sum_{k=1}^{L_{T}}\int_{\Omega_{k}}\ln\left(1+P_{\mathrm{on}}\lambda_{k}\right)dF_{\mathbf{\Lambda}}\left(\mathbf{\lambda}\right)\]
with the power constraint \[
\sum_{k=1}^{L_{T}}\int_{\Omega_{k}}dF_{\mathbf{\Lambda}}\left(\mathbf{\lambda}\right)=\rho/P_{\mathrm{on}}.\]
 Since $F_{\mathbf{\Lambda}}\left(\mathbf{\lambda}\right)$ is continuous,
there exist $\Omega_{k}$'s to satisfy the power constraint. The optimization
problem is well defined. 

Define $\Omega_{k}^{*}=\left\{ \mathbf{\lambda}:\;\lambda_{1}\geq\cdots\geq\lambda_{k}\geq\kappa\right\} $'s
where $\kappa$ is chosen to satisfy the power constraint. For any
Lebesgue measurable sets $\Omega_{k}\subset\left(\mathbb{R}^{+}\cup\left\{ 0\right\} \right)^{m}$'s
satisfying the power constraint,\begin{eqnarray*}
 &  & \sum_{k=1}^{L_{T}}\int_{\Omega_{k}^{*}}\ln\left(1+P_{\mathrm{on}}\lambda_{k}\right)dF_{\mathbf{\Lambda}}\left(\mathbf{\lambda}\right)\\
 &  & \quad\quad-\sum_{k=1}^{L_{T}}\int_{\Omega_{k}}\ln\left(1+P_{\mathrm{on}}\lambda_{k}\right)dF_{\mathbf{\Lambda}}\left(\mathbf{\lambda}\right)\\
 &  & \geq\sum_{k=1}^{L_{T}}\int_{\Omega_{k}^{*}-\Omega_{k}}\ln\left(1+P_{\mathrm{on}}\kappa\right)dF_{\mathbf{\Lambda}}\left(\mathbf{\lambda}\right)\\
 &  & \quad\quad-\sum_{k=1}^{L_{T}}\int_{\Omega_{k}-\Omega_{k}^{*}}\ln\left(1+P_{\mathrm{on}}\kappa\right)dF_{\mathbf{\Lambda}}\left(\mathbf{\lambda}\right)\\
 &  & =0,\end{eqnarray*}
where the inequality follows the facts that $\lambda_{k}\geq\kappa$
when $\lambda_{k}\in\Omega_{k}^{*}-\Omega_{k}$ and $\lambda_{k}<\kappa$
when $\lambda_{k}\in\Omega_{k}-\Omega_{k}^{*}$, and the last line
holds because of the power constraint. Therefore, the power on/off
strategy defined by $\Omega_{k}^{*}$'s is optimal.

\subsection{\label{sub:Proof-of-Theorem-condition-for-optimal-a}Proof of Theorem
\ref{thm:condition-for-optimal-a}}

The following lemma is needed to prove Theorem \ref{thm:condition-for-optimal-a}.

\begin{lemma}
\label{lem:unique-maximum}For a continuous and differentiable function
$h\left(x\right)$ defined on $\left(a,b\right)$, denote the first
derivative as $h^{'}\left(x\right)$. If $h\left(x\right)=0$ implies
$h^{'}\left(x\right)<0$, then $h\left(x\right)$ has at most one
zero in its domain. Furthermore, denote $x_{0}$ as the unique zero
if it exists, then $h\left(x\right)>0$ for all $x\in\left(a,x_{0}\right)$
and $h\left(x\right)<0$ for all $x\in\left(x_{0},b\right)$.
\end{lemma}
\begin{proof}
$h\left(x\right)$ has at most one zero. Let $x_{0}$ be a zero of
$h\left(x\right)$. Since $h^{'}\left(x_{0}\right)<0$ according to
the assumption, $\exists\epsilon>0$ such that $h\left(x_{0}+\epsilon\right)<0$,
$h\left(x_{0}-\epsilon\right)>0$ and $h\left(x\right)\neq0$ for
all $x\in\left(x_{0}-\epsilon,x_{0}+\epsilon\right)$ but $x_{0}$.
Now suppose that $x_{1}\in\left(a,b\right)$ be another zero of $h\left(x\right)$
adjacent to $x_{0}$. W.l.o.g, we assume that $x_{1}>x_{0}$. Then
$x_{0}<x_{0}+\epsilon/2<x_{1}$. Note that $h\left(x\right)$ is continuous.
$h\left(x\right)$ crosses the $x$ axis at $x_{1}$ from negative
to positive as $x$ increases. Thus $h^{'}\left(x_{1}\right)>0$.
It contradicts with the assumption.

Assume that $x_{0}$ is the unique zero if it exists. Because of the
continuity of $h\left(x\right)$, it is easy to verify that $h\left(x\right)>0$
for all $x\in\left(a,x_{0}\right)$ and $h\left(x\right)<0$ for all
$x\in\left(x_{0},b\right)$.
\end{proof}

To prove Theorem \ref{thm:condition-for-optimal-a}, we discuss two
cases. One case is that $\frac{d\bar{\mathcal{I}}_{\infty}}{da}$
has zeros in $\left(0,\pi\right)$ and the other case is that it has
no zero in $\left(0,\pi\right)$.

Evaluate $\frac{d\bar{\mathcal{I}}_{\infty}}{da}$ for an $a\in\left(0,\pi\right)$.
Denote \begin{eqnarray}
z\left(t\right) & = & \frac{\rho}{y\bar{s}_{\infty}}\left(1+y-2\sqrt{y}\cos\left(t\right)\right).\label{eq:zt}\end{eqnarray}
Then \begin{eqnarray*}
\frac{d\bar{\mathcal{I}}_{\infty}}{da} & = & \frac{d}{da}\left[\int_{a}^{\pi}\ln\left(1+\frac{\rho}{y\bar{s}_{\infty}}\left(1+y-2\sqrt{y}\cos\left(t\right)\right)\right)f_{T}\left(t\right)dt\right]\\
 & = & f_{T}\left(a\right)\left[1-\ln\left(1+z\left(a\right)\right)-\int_{a}^{\pi}\frac{1}{1+z\left(t\right)}\cdot\frac{f_{T}\left(t\right)}{\bar{s}_{\infty}}dt\right].\end{eqnarray*}
Define \begin{eqnarray}
J=1-\ln\left(1+z\left(a\right)\right)-\int_{a}^{\pi}\frac{1}{1+z\left(t\right)}\cdot\frac{f_{T}\left(t\right)}{\bar{s}_{\infty}}dt.\label{eq:J}\end{eqnarray}
Because $f_{T}\left(a\right)>0$ for all $a\in\left(0,\pi\right)$,
the sign of $\frac{d\bar{\mathcal{I}}_{\infty}}{da}$ is uniquely
determined by $J$ when $a\in\left(0,\pi\right)$. 

For the first case that $\frac{d\bar{\mathcal{I}}_{\infty}}{da}$
has zeros in $\left(0,\pi\right)$, we argue that $\frac{d\bar{\mathcal{I}}_{\infty}}{da}$
has a unique zero, say $a_{0}$, in $\left(0,\pi\right)$ and that
$\bar{\mathcal{I}}_{\infty}$ is maximized at $a_{0}$. This can be
accomplished by showing that $J=0$ implies $dJ/da<0$. Note that
\begin{eqnarray}
\frac{dJ}{da} & = & -\frac{\frac{1}{\bar{s}_{\infty}}z\left(a\right)+\frac{\rho}{y\bar{s}_{\infty}}2\sqrt{y}\sin\left(a\right)}{1+z\left(a\right)}+\frac{f_{T}\left(a\right)}{\bar{s}_{\infty}}\cdot\frac{1}{1+z\left(a\right)}\nonumber \\
 &  & -\frac{f_{T}\left(a\right)}{\bar{s}_{\infty}}\int_{a}^{\pi}\frac{1}{\left(1+z\left(t\right)\right)^{2}}\frac{f_{T}\left(t\right)}{\bar{s}_{\infty}}dt\nonumber \\
 & = & -\frac{f_{T}\left(a\right)}{\bar{s}_{\infty}}\left[\frac{z\left(a\right)-1}{z\left(a\right)+1}+\int_{a}^{\pi}\frac{1}{\left(1+z\left(t\right)\right)^{2}}\frac{f_{T}\left(t\right)}{\bar{s}_{\infty}}dt\right]\nonumber \\
 &  & -\frac{\frac{\rho}{y\bar{s}_{\infty}}2\sqrt{y}\sin\left(a\right)}{1+z\left(a\right)}.\label{eq:dJ-da}\end{eqnarray}
$J=0$ implies \begin{eqnarray*}
1-\ln\left(1+z\left(a\right)\right) & = & \int_{a}^{\pi}\frac{1}{1+z\left(t\right)}\frac{f_{T}\left(t\right)}{\bar{s}_{\infty}}dt.\end{eqnarray*}
Then \begin{eqnarray*}
\int_{a}^{\pi}\frac{1}{\left(1+z\left(t\right)\right)^{2}}\frac{f_{T}\left(t\right)}{\bar{s}_{\infty}}dt & \geq & \left(\int_{a}^{\pi}\frac{1}{1+z\left(t\right)}\frac{f_{T}\left(t\right)}{\bar{s}_{\infty}}dt\right)^{2}\\
 & = & \left(1-\ln\left(1+z\left(a\right)\right)\right)^{2},\end{eqnarray*}
where the inequality follows the fact that \begin{eqnarray*}
\int_{a}^{\pi}\left(\frac{1}{1+z\left(t\right)}-\int_{a}^{\pi}\frac{1}{1+z\left(t\right)}\frac{f_{T}\left(t\right)}{\bar{s}_{\infty}}dt\right)^{2}\frac{f_{T}\left(t\right)}{\bar{s}_{\infty}}dt & \geq & 0.\end{eqnarray*}
Thus, \begin{eqnarray*}
 &  & \frac{z\left(a\right)-1}{z\left(a\right)+1}+\int_{a}^{\pi}\frac{1}{\left(1+z\left(t\right)\right)^{2}}\frac{f_{T}\left(t\right)}{\bar{s}_{\infty}}dt\\
 &  & \geq\frac{z\left(a\right)-1}{z\left(a\right)+1}+\left(1-\ln\left(1+z\left(a\right)\right)\right)^{2}\\
 &  & >0,\end{eqnarray*}
where the last inequality follows the facts that $z\left(a\right)>0$
for $a\in\left(0,\pi\right)$ and that $\frac{x-1}{x+1}+\left(1-\ln\left(1+x\right)\right)^{2}>0$
for $x>0$, which can be verified by evaluating the first and second
derivatives. Therefore, $J=0$ implies that the first term of (\ref{eq:dJ-da})
is negative. It is also true that the last term of (\ref{eq:dJ-da})
is always negative for $a\in\left(0,\pi\right)$. We have shown that
$J=0$ implies $dJ/da<0$. According to Lemma \ref{lem:unique-maximum},
$J$ has a unique zero in $\left(0,\pi\right)$, say $a_{0}$, and
$J>0$ for $0<a<a_{0}$ and $J<0$ for $a_{0}<a<\pi$. Since the sign
of $\frac{d\bar{\mathcal{I}}_{\infty}}{da}$ is determined by $J$,
the same conclusion holds for $\frac{d\bar{\mathcal{I}}_{\infty}}{da}$.
Therefore, $\bar{\mathcal{I}}_{\infty}$ has the unique maximum point
$a_{0}$ in $\left(0,\pi\right)$. Furthermore, because of the continuity
of $\bar{\mathcal{I}}_{\infty}$, $a_{0}$ is also the unique maximum
point of $\bar{\mathcal{I}}_{\infty}$ in $\left[0,\pi\right]$.

For the second case that $\frac{d\bar{\mathcal{I}}_{\infty}}{da}$
has no zero in $\left(a,b\right)$, we show that $\bar{\mathcal{I}}_{\infty}$
is maximized at $a=0$. If $\frac{d\bar{\mathcal{I}}_{\infty}}{da}$
has no zero in $\left(a,b\right)$, $J$ has no zero in $\left(a,b\right)$.
But as $a\rightarrow\pi$, it can be verified that $z\left(a\right)\rightarrow+\infty$,
$\ln\left(1+z\left(a\right)\right)\rightarrow+\infty$ and $J\rightarrow-\infty$.
Then $J<0$ for $a\in\left(0,\pi\right)$ because of continuity. Therefore,
$\frac{d\bar{\mathcal{I}}_{\infty}}{da}<0$ for all $a\in\left(0,\pi\right)$
and $\bar{\mathcal{I}}_{\infty}$ is maximized at $a=0$.

\subsection{\label{sub:Pf-of-Cor-relationship-optimal-a-and-rho}Proof of Corollary
\ref{cor:relationship-optimal-a-and-rho}}

The proof follows the same idea in the proof of Theorem \ref{thm:condition-for-optimal-a}
(see Appendix \ref{sub:Proof-of-Theorem-condition-for-optimal-a}).
Let $J$ be defined in (\ref{eq:J}). Then the optimal $a$ to maximize
$\bar{\mathcal{I}}_{\infty}$, say $a_{0}$, should be either the
unique zero of $J$ if it exists, or $0$ if $J$ has no zero in $\left(0,\pi\right)$.
We first prove that $J=0$ implies $\frac{dJ}{d\rho}<0$ for a given
$a\in\left(0,\pi\right)$ and $\rho>0$. Then we show that $a_{0}$
is a non-decreasing function of $\rho$.

For a given $a\in\left(0,\pi\right)$ and $\rho>0$, we prove that
$J=0$ implies $\frac{dJ}{d\rho}<0$ as follows. Let $z\left(t\right)$
be defined in (\ref{eq:zt}). Note that $z\left(t\right)$ is a function
of $\rho$. Evaluation of $\frac{dJ}{d\rho}$ gives \begin{eqnarray}
\frac{dJ}{d\rho} & = & -\frac{1}{\rho}\left[\frac{z\left(a\right)}{1+z\left(a\right)}-\int_{a}^{\pi}\frac{z\left(t\right)}{\left(1+z\left(t\right)\right)^{2}}\frac{f_{T}\left(t\right)}{\bar{s}_{\infty}}dt\right]\nonumber \\
 & = & -\frac{1}{\rho}\left[\frac{z\left(a\right)}{1+z\left(a\right)}-\int_{a}^{\pi}\frac{z\left(t\right)}{1+z\left(t\right)}\frac{f_{T}\left(t\right)}{\bar{s}_{\infty}}dt\right.\nonumber \\
 &  & \left.+\int_{a}^{\pi}\frac{z^{2}\left(t\right)}{\left(1+z\left(t\right)\right)^{2}}\frac{f_{T}\left(t\right)}{\bar{s}_{\infty}}dt\right].\label{eq:noname01}\end{eqnarray}
$J=0$ implies \begin{equation}
\ln\left(1+z\left(a\right)\right)=\int_{a}^{\pi}\frac{z\left(t\right)}{1+z\left(t\right)}\frac{f_{T}\left(t\right)}{\bar{s}_{\infty}}dt<1,\label{eq:noname02}\end{equation}
where the inequality follows the fact that $\frac{z\left(t\right)}{1+z\left(t\right)}<1$.
Then we have $z\left(a\right)<e-1$. Furthermore, \begin{eqnarray}
\int_{a}^{\pi}\frac{z^{2}\left(t\right)}{\left(1+z\left(t\right)\right)^{2}}\frac{f_{T}\left(t\right)}{\bar{s}_{\infty}}dt & \geq & \frac{z^{2}\left(a\right)}{\left(1+z\left(a\right)\right)^{2}},\label{eq:noname03}\end{eqnarray}
where the inequality follows from $\frac{z\left(t\right)}{1+z\left(t\right)}\geq\frac{z\left(a\right)}{1+z\left(a\right)}$
for all $t\in\left(a,\pi\right)$. Note that the function $\frac{x}{1+x}-\ln\left(1+x\right)+\left(\frac{x}{1+x}\right)^{2}>0$
for $0<x<e-1$, which can be verified by checking its first and second
derivative. Substituting (\ref{eq:noname02}) and (\ref{eq:noname03})
into (\ref{eq:noname01}), we have shown that $J=0$ implies $\frac{dJ}{d\rho}<0$
for $a\in\left(0,\pi\right)$.

Now let $a_{0}$ maximize $\bar{\mathcal{I}}_{\infty}$ for an SNR
$\rho_{0}>0$, let $a_{1}$ maximize $\bar{\mathcal{I}}_{\infty}$
for an SNR $\rho_{1}>0$ and $\rho_{0}<\rho_{1}$. In the following,
we prove that $a_{1}\leq a_{0}$ by studying two cases: one is that
$a_{0}>0$ and the other is that $a_{0}=0$.

For the first case that $a_{0}>0$, we have $J\left|_{a_{0},\rho_{0}}\right.=0$
by Theorem \ref{thm:condition-for-optimal-a}. Since $J=0$ implies
$\frac{dJ}{d\rho}<0$ for $a=a_{0}$, $J\left|_{a_{0},\rho_{1}}\right.<0$
by Lemma \ref{lem:unique-maximum} in Appendix \ref{sub:Proof-of-Theorem-condition-for-optimal-a}.
But $a_{1}$ maximizes $\bar{\mathcal{I}}_{\infty}$ at $\rho_{1}$.
Then either $a_{1}=0$, or $a_{1}>0$ and $J\left|_{a_{1},\rho_{1}}\right.=0$
again by Theorem \ref{thm:condition-for-optimal-a}. If $a_{1}=0$,
$a_{1}<a_{0}$. If $a_{1}>0$, then $\frac{dJ}{da}\left|_{0<a<a_{1},\rho_{1}}\right.>0$
and $\frac{dJ}{da}\left|_{a_{1}<a<\pi,\rho_{1}}\right.<0$ according
to the proof in Appendix \ref{sub:Proof-of-Theorem-condition-for-optimal-a}.
Since we have shown $J\left|_{a_{0},\rho_{1}}\right.<0$, $a_{1}<a_{0}$.
Thus $a_{1}<a_{0}$ if $a_{0}>0$.

On the other hand, $a_{0}=0$ implies $a_{1}=a_{0}=0$. Suppose that
$a_{1}>a_{0}$, then $a_{1}\in\left(0,\pi\right)$, $J\left|_{a_{1},\rho_{1}}\right.=0$
and $J\left|_{a_{1},\rho_{0}}\right.>0$. Because $J\left|_{\pi,\rho_{0}}\right.\rightarrow-\infty$,
$\exists a'\in\left(a_{1},\pi\right)$ such that $J\left|_{a',\rho_{0}}\right.=0$.
According to Theorem \ref{thm:condition-for-optimal-a}, $a'$ maximizes
$\bar{\mathcal{I}}_{\infty}$ for $\rho_{0}$. It contradicts with
the assumption that $a_{0}=0$ maximizes $\bar{\mathcal{I}}_{\infty}$
for $\rho_{0}$. Therefore $0\leq a_{1}\leq a_{0}=0$ and thus $a_{1}=a_{0}=0$.

\subsection{\label{sub:Calculation-of-s}Calculation of $\bar{s}_{\infty}$}


Write the formula for $\bar{s}_{\infty}$ in (\ref{eq:S_asymptotic_t})
in another form. Recall the definition of $f_{T}\left(t\right)$ in
(\ref{eq:f_T}). It is easy to see that $f_{T}\left(t\right)=f_{T}\left(-t\right)$.
In order to use the symmetry, we define the integral range \begin{eqnarray}
I_{R} & = & \left[-\pi,-a\right]\cup\left[a,\pi\right].\label{eq:I_R}\end{eqnarray}
 Then the normalized number of on-beams $\bar{s}_{\infty}$ is given
by \begin{eqnarray*}
\bar{s}_{\infty}=\frac{1}{2}\int_{I_{R}}f_{T}\left(t\right)dt.\end{eqnarray*}

When $y<1$,

\begin{eqnarray*}
\bar{s}_{\infty} & = & \frac{1}{2}\int_{I_{R}}\frac{1}{2\pi}\cdot\frac{2-e^{2it}-e^{-2it}}{1+r^{2}-2r\cos\left(t\right)}dt\\
 & = & \frac{1}{4\pi\left(1-r^{2}\right)}\int_{I_{R}}\left(\frac{1}{1-re^{it}}+\frac{1}{1-re^{-it}}-1\right)\left(2-e^{2it}-e^{-2it}\right)dt,\end{eqnarray*}
where $r=\sqrt{y}$. Because of the symmetry of the integral range
and the integrand, we have \begin{eqnarray*}
\int_{I_{R}}\frac{1}{1-re^{it}}\left(2-e^{2it}-e^{-2it}\right)dt=\int_{I_{R}}\frac{1}{1-re^{-it}}\left(2-e^{2it}-e^{-2it}\right)dt.\end{eqnarray*}
Then\begin{eqnarray*}
\bar{s}_{\infty} & = & \frac{1}{4\pi\left(1-r^{2}\right)}\int_{I_{R}}\left(\frac{2}{1-re^{it}}-1\right)\left(2-e^{2it}-e^{-2it}\right)dt\\
 & = & \frac{1}{4\pi\left(1-r^{2}\right)}\int_{I_{R}}\left[-2\left(\frac{1}{r}-r\right)^{2}\frac{re^{it}}{1-re^{it}}\right.\\
 &  & \left.+e^{2it}+\frac{2}{r}e^{it}+2-2r^{2}-2re^{-it}-e^{-2it}\right]dt.\end{eqnarray*}
Note that \begin{eqnarray}
\int_{I_{R}}\frac{re^{it}}{1-re^{it}}dt=i\int_{I_{R}}d\ln\left(1-re^{it}\right)=i\ln\left(\frac{1-re^{-ia}}{1-re^{ia}}\right)=-2\theta_{r},\label{eq:int-theta-r}\end{eqnarray}
where \begin{eqnarray*}
\theta_{r}=\tan^{-1}\left(\frac{r\sin\left(a\right)}{1-r\cos\left(a\right)}\right).\end{eqnarray*}
Then\begin{eqnarray*}
\bar{s}_{\infty}=\frac{1}{\pi}\left(\left(\pi-a\right)-\frac{1}{r}\sin\left(a\right)+\frac{1-r^{2}}{r^{2}}\theta_{r}\right).\end{eqnarray*}

When $y=1$, it is easy to see that\begin{eqnarray*}
\bar{s}_{\infty}=\frac{1}{2\pi}\int_{I_{R}}\left(1+\cos\left(t\right)\right)dt=\frac{1}{\pi}\left(\pi-a-\sin\left(a\right)\right)\end{eqnarray*}

\subsection{\label{sub:Calculation-of-C}Calculation of $\bar{\mathcal{I}}_{\infty}$}

The normalized capacity is given by \begin{eqnarray*}
\bar{\mathcal{I}}_{\infty}=\frac{1}{2}\int_{I_{R}}\ln\left(1+\frac{\rho}{\bar{s}_{\infty}y}\left(1+y-2\sqrt{y}\cos\left(t\right)\right)\right)f_{T}\left(t\right)dt\end{eqnarray*}
where $f_{T}\left(t\right)$ is given in (\ref{eq:f_T}), the integral
range $I_{R}$ is defined in (\ref{eq:I_R}) and $\bar{s}_{\infty}$
can be calculated according to the Proposition \ref{pro:calculation-s}.

Define \begin{eqnarray}
\alpha & = & \frac{\bar{s}_{\infty}y}{\rho},\label{eq:alpha-def-appendix}\end{eqnarray}
 then \begin{eqnarray*}
 &  & \ln\left(1+\frac{\rho}{\bar{s}_{\infty}y}\left(1+y-2\sqrt{y}\cos\left(t\right)\right)\right)\\
 &  & =\ln\left(1+r^{2}+\alpha-2r\cos\left(t\right)\right)-\ln\left(\alpha\right),\end{eqnarray*}
where $r=\sqrt{y}$. Also define\begin{eqnarray}
w=\frac{1}{2}\left(1+y+\alpha+\sqrt{\left(1+y+\alpha\right)^{2}-4y}\right)\label{eq:w-def-appendix}\end{eqnarray}
and\begin{eqnarray}
u=\frac{1}{2\sqrt{y}}\left(1+y+\alpha-\sqrt{\left(1+y+\alpha\right)^{2}-4y}\right),\label{eq:u-def-appendix}\end{eqnarray}
then it is easy to verify that $u<1$ and \begin{eqnarray*}
 &  & \ln\left(1+r^{2}+\alpha-2r\cos\left(t\right)\right)\\
 &  & =\ln\left(w\right)+\ln\left(1+u^{2}-2u\cos\left(t\right)\right).\end{eqnarray*}
 Therefore, \begin{eqnarray*}
 &  & \ln\left(1+\frac{\rho}{\bar{s}_{\infty}y}\left(1+y-2\sqrt{y}\cos\left(t\right)\right)\right)\\
 &  & =\ln\left(w\right)-\ln\left(\alpha\right)+\ln\left(1+u^{2}-2u\cos\left(t\right)\right),\end{eqnarray*}
and \begin{eqnarray*}
\bar{\mathcal{I}}_{\infty} & = & \left(\ln\left(w\right)-\ln\left(\alpha\right)\right)\bar{s}_{\infty}\\
 &  & +\frac{1}{2}\int_{I_{R}}\ln\left(1+u^{2}-2u\cos\left(t\right)\right)f_{T}\left(t\right)dt.\end{eqnarray*}
Define\begin{eqnarray*}
I_{0}=\frac{1}{2}\int_{I_{R}}\ln\left(1+u^{2}-2u\cos\left(t\right)\right)f_{T}\left(t\right)dt.\end{eqnarray*}
Then \[
\bar{\mathcal{I}}=\left(\ln\left(w\right)-\ln\left(\alpha\right)\right)\bar{s}+I_{0}.\]
Note that \begin{eqnarray*}
\ln\left(1+u^{2}-2u\cos\left(t\right)\right)=\ln\left(1-ue^{it}\right)+\ln\left(1-ue^{-it}\right)\end{eqnarray*}
and \begin{eqnarray*}
\int_{I_{R}}\ln\left(1-ue^{it}\right)f_{T}\left(t\right)dt & = & \int_{I_{R}}\ln\left(1-ue^{-it}\right)f_{T}\left(t\right)dt.\end{eqnarray*}
Then\begin{eqnarray*}
I_{0}=\int_{I_{R}}\ln\left(1-ue^{it}\right)f_{T}\left(t\right)dt.\end{eqnarray*}

Calculate $I_{0}$ for the case $y<1$ and the case $y=1$ respectively.

When $y<1$, \begin{eqnarray*}
I_{0} & = & \int_{I_{R}}\ln\left(1-ue^{it}\right)\cdot\frac{1}{2\pi}\cdot\frac{2-e^{2it}-e^{-2it}}{1+r^{2}-2r\cos\left(t\right)}dt\\
 & = & \frac{1}{2\pi\left(1-r^{2}\right)}\int_{I_{R}}\ln\left(1-ue^{it}\right)\left(\frac{1}{1-re^{it}}+\frac{1}{1-re^{-it}}-1\right)\left(2-e^{2it}-e^{-2it}\right)dt.\end{eqnarray*}
It is easy to verify that \begin{eqnarray}
\frac{e^{2it}-2+e^{-2it}}{1-re^{it}}=\left(\frac{1}{r}-r\right)^{2}\frac{re^{it}}{1-re^{it}}+\left(-\frac{1}{r}e^{it}-2+r^{2}+re^{-it}+e^{-2it}\right)\label{eq:fraction-expansion-1}\end{eqnarray}
and\begin{eqnarray}
\frac{e^{2it}-2+e^{-2it}}{1-re^{-it}}=\left(\frac{1}{r}-r\right)^{2}\frac{1}{1-re^{it}}+\left(e^{2it}+re^{it}-\frac{1}{r^{2}}-\frac{1}{r}e^{-it}\right).\label{eq:fraction-expansion-2}\end{eqnarray}
Then\begin{eqnarray*}
 &  & \left(\frac{1}{1-re^{it}}+\frac{1}{1-re^{-it}}-1\right)\left(e^{2it}-2+e^{-2it}\right)\\
 &  & =\left(r-\frac{1}{r}\right)\left[e^{it}+\frac{1}{r}+r+e^{-it}+\left(r-\frac{1}{r}\right)\cdot\left(\frac{re^{it}}{1-re^{it}}+\frac{1}{1-re^{-it}}\right)\right].\end{eqnarray*}
Define\begin{eqnarray}
I_{1} & = & \int_{I_{R}}\ln\left(1-ue^{it}\right)dt\label{eq:I1-def}\\
I_{2} & = & \int_{I_{R}}\ln\left(1-ue^{it}\right)\left(e^{it}+e^{-it}\right)dt\label{eq:I2-def}\\
I_{3} & = & \int_{I_{R}}\ln\left(1-ue^{it}\right)\left(\frac{re^{it}}{1-re^{it}}+\frac{1}{1-re^{-it}}\right)dt,\nonumber \end{eqnarray}
Then\begin{eqnarray*}
I_{0}=\frac{1}{2\pi r}\left[\left(\frac{1}{r}+r\right)I_{1}+I_{2}+\left(r-\frac{1}{r}\right)I_{3}\right].\end{eqnarray*}
Calculate $I_{1}$, $I_{2}$ and $I_{3}$ respectively. Because $\left|u\right|<1$,
\begin{eqnarray*}
\ln\left(1-ue^{it}\right)=-\sum_{k=1}^{\infty}\frac{u^{k}e^{ikt}}{k}.\end{eqnarray*}
Therefore,\begin{eqnarray*}
I_{1} & = & -\int_{I_{R}}\sum_{k=1}^{\infty}\frac{u^{k}e^{ikt}}{k}dt\\
 & = & -\sum_{k=1}^{\infty}\frac{u^{k}}{ik^{2}}\int_{I_{R}}de^{ikt}\\
 & = & i\sum_{k=1}^{\infty}\frac{u^{k}e^{-ika}}{k^{2}}-i\sum_{k=1}^{\infty}\frac{u^{k}e^{ika}}{k^{2}}.\end{eqnarray*}
Define \begin{eqnarray*}
\mathrm{Li}_{2}\left(x\right)=\sum_{n=1}^{\infty}\frac{x^{n}}{n^{2}},\;\;\mathrm{for}\:\left|x\right|\leq1,\end{eqnarray*}
which is usually called dilogarithm function \cite{Andrews_1999_Special_Functions}.
Then\begin{eqnarray}
I_{1} & = & i\left[\mathrm{Li}_{2}\left(ue^{-ia}\right)-\mathrm{Li}_{2}\left(ue^{ia}\right)\right].\label{eq:I1-result}\end{eqnarray}
To evaluate $I_{2}$, note that \begin{eqnarray*}
\int_{I_{R}}\ln\left(1-ue^{it}\right)e^{it}dt & = & \int_{I_{R}}\frac{ue^{2it}}{1-ue^{it}}dt+\frac{1}{i}\ln\left(1-ue^{it}\right)e^{it}\left|_{a}^{\pi}\right.\\
 &  & +\frac{1}{i}\ln\left(1-ue^{it}\right)e^{it}\left|_{-\pi}^{-a}\right.\end{eqnarray*}
and\begin{eqnarray*}
\int_{I_{R}}\ln\left(1-ue^{it}\right)e^{-it}dt & = & \int_{I_{R}}\frac{u}{1-ue^{it}}dt+\frac{1}{-i}\ln\left(1-ue^{it}\right)e^{-it}\left|_{a}^{\pi}\right.\\
 &  & +\frac{1}{-i}\ln\left(1-ue^{it}\right)e^{-it}\left|_{-\pi}^{-a}\right..\end{eqnarray*}
Then\begin{eqnarray*}
I_{2} & = & -2\sin\left(a\right)\ln\left(1+u^{2}-2u\cos\left(a\right)\right)+u\int_{I_{R}}\frac{e^{2it}-1}{1-ue^{it}}dt.\end{eqnarray*}
Note that \begin{eqnarray*}
u\frac{e^{2it}-1}{1-ue^{it}} & = & -e^{it}-u+\left(\frac{1}{u}-u\right)\frac{ue^{it}}{1-ue^{it}}\end{eqnarray*}
and\begin{eqnarray*}
\int_{I_{R}}\frac{ue^{it}}{1-ue^{it}}dt=-2\theta_{u},\end{eqnarray*}
where\begin{eqnarray*}
\theta_{u}=\tan^{-1}\left(\frac{u\sin\left(a\right)}{1-u\cos\left(a\right)}\right)\end{eqnarray*}
by similar analysis that we did in (\ref{eq:int-theta-r}). Then\begin{eqnarray}
I_{2} & = & 2\left[-\sin\left(a\right)\ln\left(1+u^{2}-2u\cos\left(a\right)\right)\right.\nonumber \\
 &  & \left.+\sin\left(a\right)-u\left(\pi-a\right)-\left(\frac{1}{u}-u\right)\theta_{u}\right].\label{eq:I2-result}\end{eqnarray}
To evaluate $I_{3}$, note that \begin{eqnarray*}
\ln\left(1-ue^{it}\right) & = & -\sum_{k=1}^{\infty}\frac{\left(ue^{it}\right)^{k}}{k}\end{eqnarray*}
and\begin{eqnarray*}
\frac{1}{1-re^{it}}+\frac{1}{1-re^{-it}}-1 & = & \sum_{l=-\infty}^{\infty}r^{\left|l\right|}e^{ilt}\end{eqnarray*}
because $\left|u\right|<1$ and $\left|r\right|<1$. Thus\begin{eqnarray*}
-I_{3}=\int_{I_{R}}\sum_{k=1}^{\infty}\frac{\left(ue^{it}\right)^{k}}{k}\left(\sum_{l=-\infty}^{\infty}r^{\left|l\right|}e^{ilt}\right)dt.\end{eqnarray*}
Change the order of the double summation. Then\begin{eqnarray*}
-I_{3} & = & \int_{I_{R}}\sum_{l=0}^{\infty}\left(\sum_{k=1}^{\infty}\frac{\left(ue^{it}\right)^{k}}{k}\left(re^{-it}\right)^{k+l}\right)dt\\
 &  & +\int_{I_{R}}\sum_{l=1}^{\infty}\left(\sum_{k=1}^{l-1}\frac{\left(ue^{it}\right)^{k}}{k}\left(re^{it}\right)^{l-k}+\sum_{k=l}^{\infty}\frac{\left(ue^{it}\right)^{k}}{k}\left(re^{-it}\right)^{k-l}\right)dt\\
 & = & \int_{I_{R}}\sum_{l=0}^{\infty}r^{l}e^{-ilt}\left(\sum_{k=1}^{\infty}\frac{\left(ur\right)^{k}}{k}\right)dt\\
 &  & +\int_{I_{R}}\sum_{l=1}^{\infty}e^{ilt}\left(r^{l}\sum_{k=1}^{l-1}\frac{\left(\frac{u}{r}\right)^{k}}{k}+r^{-l}\sum_{k=l}^{\infty}\frac{\left(ur\right)^{k}}{k}\right)dt.\end{eqnarray*}
Define \begin{eqnarray*}
I_{4} & = & \int_{I_{R}}\sum_{l=0}^{\infty}r^{l}e^{-ilt}\left(\sum_{k=1}^{\infty}\frac{\left(ur\right)^{k}}{k}\right)dt\end{eqnarray*}
and \begin{eqnarray*}
I_{5} & = & \int_{I_{R}}\sum_{l=1}^{\infty}e^{ilt}\left(r^{l}\sum_{k=1}^{l-1}\frac{\left(\frac{u}{r}\right)^{k}}{k}+r^{-l}\sum_{k=l}^{\infty}\frac{\left(ur\right)^{k}}{k}\right)dt.\end{eqnarray*}
Noting that $0<r<1$ and $0<ur<1$, $I_{4}$ is well defined and \begin{eqnarray}
I_{4} & = & \sum_{k=1}^{\infty}\frac{\left(ur\right)^{k}}{k}\left(i\sum_{l=1}^{\infty}\frac{r^{l}}{l}e^{-ilt}+t\right)\left|\begin{array}{c}
^{\pi}\\
_{a}\end{array}\right.\nonumber \\
 &  & +\sum_{k=1}^{\infty}\frac{\left(ur\right)^{k}}{k}\left(i\sum_{l=1}^{\infty}\frac{r^{l}}{l}e^{-ilt}+t\right)\left|\begin{array}{c}
^{-a}\\
_{-\pi}\end{array}\right.\nonumber \\
 & = & -\ln\left(1-ur\right)\left(t-i\ln\left(1-re^{-it}\right)\right)\left|_{a}^{\pi}\right.\nonumber \\
 &  & -\ln\left(1-ur\right)\left(t-i\ln\left(1-re^{-it}\right)\right)\left|_{-\pi}^{-a}\right.\nonumber \\
 & = & -2\ln\left(1-ur\right)\left(\pi-a-\theta_{r}\right),\label{eq:I4-result}\end{eqnarray}
where $\theta_{r}$ is obtained according to the similar analysis
in (\ref{eq:int-theta-r}). To evaluate $I_{5}$, we substitute the
definition of $u$ into $\frac{u}{r}$. It is easy to verify that\begin{eqnarray*}
\frac{u}{r} & = & \frac{1}{2r^{2}}\left(1+r^{2}+\alpha-\sqrt{\left(1+r^{2}+\alpha\right)^{2}-4r^{2}}\right)\\
 & < & \frac{1}{2r^{2}}\left(1+r^{2}+\alpha-\sqrt{\left(1-r^{2}+\alpha\right)^{2}}\right)\\
 & = & 1\end{eqnarray*}
and \begin{eqnarray*}
\left|I_{5}\right| & \leq & \int_{I_{R}}\sum_{l=1}^{\infty}r^{l}\left(\sum_{k=1}^{l-1}\frac{\left(ur\right)^{k}}{k}+r^{-2l}\sum_{k=l}^{\infty}\frac{\left(\frac{u}{r}\right)^{k}}{k}\right)dt\\
 & \leq & \int_{I_{R}}\sum_{l=1}^{\infty}r^{l}\left(\sum_{k=1}^{l-1}\frac{\left(\frac{u}{r}\right)^{k}}{k}+\sum_{k=l}^{\infty}\frac{\left(\frac{u}{r}\right)^{k}}{k}\right)dt\\
 & = & -\int_{I_{R}}\frac{r}{1-r}\ln\left(1-\frac{u}{r}\right)dt.\end{eqnarray*}
Therefore, $I_{5}$ is well defined. Further, define a special function
in the form of series as \begin{eqnarray*}
\mathrm{Sr}_{1}\left(u,r,t\right)=\sum_{l=1}^{\infty}\frac{r^{l}e^{ilt}}{l}\left(\sum_{k=1}^{l-1}\frac{\left(\frac{u}{r}\right)^{k}}{k}+r^{-2l}\sum_{k=l}^{\infty}\frac{r^{2k}\left(\frac{u}{r}\right)^{k}}{k}\right).\end{eqnarray*}
Then \begin{eqnarray}
I_{5} & = & \frac{1}{i}\mathrm{Sr}_{1}\left(u,r,t\right)\left|_{a}^{\pi}\right.+\frac{1}{i}\mathrm{Sr}_{1}\left(u,r,t\right)\left|_{-\pi}^{-a}\right.\nonumber \\
 & = & i\mathrm{Sr}_{1}\left(u,r,a\right)-i\mathrm{Sr}_{1}\left(u,r,-a\right).\label{eq:I5-result}\end{eqnarray}

In conclusion, when $y<1$, \begin{eqnarray*}
\bar{\mathcal{I}}_{\infty} & = & \left[\ln\left(w\right)-\ln\left(\alpha\right)\right]\bar{s}_{\infty}+\frac{1}{2\pi r}\left(\frac{1+r^{2}}{r}I_{1}+I_{2}-\frac{1-r^{2}}{r}I_{3}\right)\\
 & = & \left[\ln\left(w\right)-\ln\left(\alpha\right)\right]\bar{s}_{\infty}\\
 &  & +\frac{1+r^{2}}{2\pi r^{2}}I_{1}+\frac{1}{2\pi r}I_{2}+\frac{1-r^{2}}{2\pi r^{2}}I_{4}+\frac{1-r^{2}}{2\pi r^{2}}I_{5}\end{eqnarray*}
where $I_{1}$, $I_{2}$, $I_{4}$ and $I_{5}$ can be calculated
according to (\ref{eq:I1-result}-\ref{eq:I5-result}).

When $y=1$, the calculation can be highly simplified. Substitute
$f_{T}\left(t\right)$ into $I_{0}$, then\begin{eqnarray*}
I_{0} & = & \frac{1}{2\pi}\int_{I_{R}}\ln\left(1-ue^{it}\right)\left(e^{it}+2+e^{-it}\right)dt\\
 & = & \frac{1}{\pi}I_{1}+\frac{1}{2\pi}I_{2},\end{eqnarray*}
where $I_{1}$ and $I_{2}$ are defined in (\ref{eq:I1-def}) and
(\ref{eq:I2-def}) respectively. Thus \begin{eqnarray*}
\bar{\mathcal{I}}_{\infty} & = & \left[\ln\left(w\right)-\ln\left(\alpha\right)\right]\bar{s}_{\infty}+\frac{1}{\pi}I_{1}+\frac{1}{2\pi}I_{2},\end{eqnarray*}
where $I_{1}$and $I_{2}$ can be calculated by (\ref{eq:I1-result})
and (\ref{eq:I2-result}) respectively.

\subsection{\label{sub:Calculation-of-dCda}Calculation of $d\bar{\mathcal{I}}_{\infty}/da$}

Define $I_{R}$, $\alpha$, $w$ and $u$ as (\ref{eq:I_R}), (\ref{eq:alpha-def-appendix}),
(\ref{eq:w-def-appendix}) and (\ref{eq:u-def-appendix}). It is easy
to see that \begin{eqnarray}
\frac{d\alpha}{da} & = & -\frac{y}{\rho}f_{T}\left(a\right).\label{eq:dalpha-da}\end{eqnarray}
According to the formula for the normalized information rate $\bar{\mathcal{I}}_{\infty}$
in (\ref{eq:I_asymptotic_t}), \begin{eqnarray*}
\frac{d\bar{\mathcal{I}}_{\infty}}{da} & = & -\ln\left(1+\frac{1}{\alpha}\left(1+y-2\sqrt{y}\cos\left(a\right)\right)\right)f_{T}\left(a\right)\\
 &  & +\frac{1}{2}\int_{I_{R}}\frac{-\frac{1}{\alpha^{2}}\cdot\frac{d\alpha}{da}\cdot\left(1+y-2\sqrt{y}\cos\left(t\right)\right)}{1+\frac{1}{\alpha}\left(1+y-2\sqrt{y}\cos\left(t\right)\right)}f_{T}\left(t\right)dt.\end{eqnarray*}
By (\ref{eq:dalpha-da}), \begin{eqnarray*}
 &  & \frac{1}{2}\int_{I_{R}}\frac{-\frac{1}{\alpha^{2}}\cdot\frac{d\alpha}{da}\cdot\left(1+y-2\sqrt{y}\cos\left(t\right)\right)}{1+\frac{1}{\alpha}\left(1+y-2\sqrt{y}\cos\left(t\right)\right)}f_{T}\left(t\right)dt\\
 &  & =\frac{y}{2\rho\alpha}f_{T}\left(a\right)\int_{I_{R}}\frac{1+y-2\sqrt{y}\cos\left(t\right)}{1+y+\alpha-2\sqrt{y}\cos\left(t\right)}f_{T}\left(t\right)dt\\
 &  & =f_{T}\left(a\right)\left(1-\frac{y}{2\rho}\int_{I_{R}}\frac{1}{1+y+\alpha-2\sqrt{y}\cos\left(t\right)}f_{T}\left(t\right)dt\right).\end{eqnarray*}
Define\begin{eqnarray*}
I^{d} & =\frac{1}{2} & \int_{I_{R}}\frac{1}{1+y+\alpha-2\sqrt{y}\cos\left(t\right)}f_{T}\left(t\right)dt.\end{eqnarray*}
then\begin{eqnarray}
\frac{d\bar{\mathcal{I}}_{\infty}}{da} & = & f_{T}\left(a\right)\left[1-\ln\left(1+\frac{1}{\alpha}\left(1+y-2\sqrt{y}\cos\left(a\right)\right)\right)-\frac{y}{\rho}I^{d}\right].\label{eq:dC-da-Cd}\end{eqnarray}
Consider the calculation of $I^{d}$. Since $w\left(1+u^{2}-2u\cos\left(t\right)\right)=1+y+\alpha-2\sqrt{y}\cos\left(t\right)$,
\begin{eqnarray*}
I^{d} & = & \frac{1}{2w}\int_{I_{R}}\frac{1}{1+u^{2}-2u\cos\left(t\right)}f_{T}\left(t\right)dt\\
 & = & \frac{1}{2w\left(1-u^{2}\right)}\int_{I_{R}}\left(\frac{1}{1-ue^{it}}+\frac{1}{1-ue^{-it}}-1\right)f_{T}\left(t\right)dt.\end{eqnarray*}
According to the symmetry of $I_{R}$ and $f_{T}\left(t\right)$,\begin{eqnarray*}
\int_{I_{R}}\frac{1}{1-ue^{it}}f_{T}\left(t\right)dt & = & \int_{I_{R}}\frac{1}{1-ue^{-it}}f_{T}\left(t\right)dt.\end{eqnarray*}
Then\begin{eqnarray*}
I^{d} & = & \frac{1}{2w\left(1-u^{2}\right)}\int_{I_{R}}\left(\frac{2}{1-ue^{it}}-1\right)f_{T}\left(t\right)dt.\end{eqnarray*}

Calculate $I^{d}$ for the case $y<1$ and the case $y=1$ respectively.

When $y<1$, \begin{eqnarray*}
I^{d} & = & \frac{1}{4\pi w\left(1-u^{2}\right)}\int_{I_{R}}\left(\frac{2}{1-ue^{it}}-1\right)\frac{2-e^{2it}-e^{-2it}}{1+r^{2}-2r\cos\left(t\right)}dt\\
 & = & \frac{1}{4\pi w\left(1-u^{2}\right)\left(1-r^{2}\right)}\int_{I_{R}}\left[\left(\frac{2}{1-ue^{it}}-1\right)\right.\\
 &  & \left.\left(\frac{1}{1-re^{it}}+\frac{1}{1-re^{-it}}-1\right)\left(2-e^{2it}-e^{-2it}\right)\right]dt.\end{eqnarray*}
Expand the integrand. Since \begin{eqnarray*}
 &  & \frac{1}{1-ue^{it}}\cdot\frac{1}{1-re^{it}}\\
 &  & =-\frac{u}{r-u}\frac{1}{1-ue^{it}}+\frac{r}{r-u}\frac{1}{1-re^{it}}\end{eqnarray*}
and \begin{eqnarray*}
 &  & \frac{1}{1-ue^{it}}\cdot\frac{1}{1-re^{it}}\\
 &  & =\frac{1}{1-ur}\frac{1}{1-ue^{it}}-\frac{1}{1-ur}\frac{1}{1-\frac{1}{r}e^{it}},\end{eqnarray*}
$I^{d}$ can be split into four parts\begin{eqnarray*}
I^{d} & = & \frac{1}{4\pi w\left(1-u^{2}\right)\left(1-r^{2}\right)}\left(I_{6}+I_{7}+I_{8}+I_{9}\right),\end{eqnarray*}
 where\begin{eqnarray*}
I_{6} & = & 2\left(\frac{1}{1-ur}-\frac{u}{r-u}-1\right)\int_{I_{R}}\frac{2-e^{2it}-e^{-2it}}{1-ue^{it}}dt,\end{eqnarray*}
\begin{eqnarray*}
I_{7} & = & 2\left(\frac{r}{r-u}-1\right)\int_{I_{R}}\frac{2-e^{2it}-e^{-2it}}{1-re^{it}}dt,\end{eqnarray*}
\begin{eqnarray*}
I_{8} & = & -\frac{2}{1-ur}\int_{I_{R}}\frac{2-e^{2it}-e^{-2it}}{1-\frac{1}{r}e^{it}}dt\end{eqnarray*}
and \begin{eqnarray*}
I_{9} & = & \int_{I_{R}}\left(2-e^{2it}-e^{-2it}\right)dt.\end{eqnarray*}
$I_{9}$ can be easily calculated,\begin{eqnarray*}
I_{9}=4\left(\pi-a\right).\end{eqnarray*}
 To evaluate $I_{6}$, $I_{7}$ and $I_{8}$, expand the integrands
like what has been done in (\ref{eq:fraction-expansion-1}). Note
that \begin{eqnarray*}
\int_{I_{R}}\frac{\frac{1}{r}e^{it}}{1-\frac{1}{r}e^{it}}dt & = & \int_{I_{R}}\left(-1-\frac{re^{-it}}{1-re^{-it}}\right)dt\\
 & = & -\int_{I_{R}}\left(1+\frac{re^{it}}{1-re^{it}}\right)dt,\end{eqnarray*}
\begin{eqnarray*}
\int_{I_{R}}\frac{ue^{it}}{1-ue^{it}}dt=-2\theta_{u}\end{eqnarray*}
and\begin{eqnarray*}
\int_{I_{R}}\frac{re^{it}}{1-re^{it}}dt=-2\theta_{r},\end{eqnarray*}
where \begin{eqnarray*}
\theta_{u}=\tan^{-1}\left(\frac{u\sin\left(a\right)}{1-u\cos\left(a\right)}\right)\end{eqnarray*}
and \begin{eqnarray*}
\theta_{r}=\tan^{-1}\left(\frac{r\sin\left(a\right)}{1-r\cos\left(a\right)}\right).\end{eqnarray*}
according to (\ref{eq:int-theta-r}). $I_{6}$, $I_{7}$ and $I_{8}$
can be calculated and finally $I^{d}$ can be written as \begin{eqnarray*}
I^{d} & = & \frac{1}{\pi w\left(1-ur\right)}\left[\pi-a-\frac{1-u^{2}}{u\left(r-u\right)}\theta_{u}+\frac{1-r^{2}}{r\left(r-u\right)}\theta_{r}\right].\end{eqnarray*}

When $y=1$, \begin{eqnarray*}
I^{d} & = & \frac{1}{4\pi w\left(1-u^{2}\right)}\int_{I_{R}}\left(\frac{2}{1-ue^{it}}-1\right)\left(2+e^{it}+e^{-it}\right)dt.\end{eqnarray*}
The integrand can be simplified as\begin{eqnarray*}
2\frac{\left(1+u\right)^{2}}{u}\frac{ue^{it}}{1-ue^{it}}+2+2u+e^{-it}-e^{it}.\end{eqnarray*}
Therefore, \begin{eqnarray*}
I^{d} & = & \frac{\pi-a}{\pi w\left(1-u\right)}-\frac{\left(1+u\right)\theta_{u}}{\pi wu\left(1-u\right)}.\end{eqnarray*}

Substitute the value of $I^{d}$ into (\ref{eq:dC-da-Cd}), $\frac{d\bar{\mathcal{I}}}{da}$
can be evaluated.

\subsection{\label{sub:Calculation-of-rho-CSITR}Calculation of the average power
for CSITR case}

For CSITR case,

\begin{eqnarray*}
\rho & = & \int_{a}^{\pi}\left(\nu-\frac{y}{1+y-2\sqrt{y}\cos\left(t\right)}\right)f_{T}\left(t\right)dt\\
 & = & \nu\bar{s}_{\infty}-\frac{1}{2}\int_{I_{R}}\frac{y}{1+y-2\sqrt{y}\cos\left(t\right)}f_{T}\left(t\right)dt\end{eqnarray*}
where $I_{R}$ is defined in (\ref{eq:I_R}), $\bar{s}_{\infty}$
can be evaluated according to Proposition \ref{pro:calculation-s}
and the second line follows from the fact that the integrand is even.
Define \begin{eqnarray*}
I_{10}=\frac{1}{2}\int_{I_{R}}\frac{y}{1+y-2\sqrt{y}\cos\left(t\right)}f_{T}\left(t\right)dt.\end{eqnarray*}
We are going to evaluate $I_{10}$ for $y<1$ and $y=1$ respectively.

When $y<1$, \begin{eqnarray*}
I_{10} & = & -\frac{r^{2}}{4\pi}\int_{I_{R}}\frac{e^{2it}-2+e^{-2it}}{\left(1+r^{2}-2r\cos\left(t\right)\right)^{2}}dt\\
 & = & -\frac{r^{2}}{4\pi}\int_{I_{R}}\left(\frac{e^{it}-e^{-it}}{\left(1-re^{it}\right)\left(1-re^{-it}\right)}\right)^{2}dt.\end{eqnarray*}
Since $e^{it}-e^{-it}=e^{it}-\frac{1}{r}+\frac{1}{r}-e^{-it}$, \begin{eqnarray*}
I_{10} & = & -\frac{1}{4\pi}\int_{I_{R}}\left(\frac{1}{1-re^{it}}+\frac{1}{1-re^{-it}}\right)^{2}dt\\
 & = & -\frac{1}{2\pi}\int_{I_{R}}\left[\frac{1}{\left(1-re^{it}\right)^{2}}+\frac{1}{\left(1-re^{-it}\right)^{2}}\right.\\
 &  & \left.-\frac{2}{\left(1-re^{it}\right)\left(1-re^{-it}\right)}\right]dt.\end{eqnarray*}
Since \begin{eqnarray*}
\frac{1}{\left(1-re^{it}\right)^{2}} & = & 1+\frac{1}{1-re^{it}}+\frac{re^{it}}{\left(1-re^{it}\right)^{2}}\end{eqnarray*}
\begin{eqnarray*}
\int_{I_{R}}\frac{1}{\left(1-re^{it}\right)^{2}}dt & = & \int_{I_{R}}\frac{1}{\left(1-re^{-it}\right)^{2}}dt\end{eqnarray*}
\begin{eqnarray*}
\frac{1}{\left(1-re^{it}\right)\left(1-re^{-it}\right)} & = & \frac{1}{1-r^{2}}\left(\frac{1}{1-re^{it}}+\frac{re^{-it}}{1-re^{-it}}\right)\end{eqnarray*}
and\begin{eqnarray*}
\int_{I_{R}}\frac{re^{-it}}{1-re^{-it}}dt & = & \int_{I_{R}}\frac{re^{it}}{1-re^{it}}dt,\end{eqnarray*}
$I_{10}$ can be simplified as \begin{eqnarray*}
I_{10} & = & \frac{1}{2\pi}\int_{I_{R}}\frac{r^{2}}{1-r^{2}}+\frac{1+r^{2}}{1-r^{2}}\frac{re^{-it}}{1-re^{-it}}-\frac{re^{it}}{\left(1-re^{it}\right)^{2}}dt\\
 & = & \frac{1}{\pi}\left[\frac{r^{2}}{1-r^{2}}\left(\pi-a\right)-\frac{1+r^{2}}{1-r^{2}}\theta_{r}+\frac{i}{2}\left(\frac{1}{1-re^{-ia}}-\frac{1}{1-re^{ia}}\right)\right]\end{eqnarray*}
where \begin{eqnarray*}
\theta_{r} & = & \tan^{-1}\left(\frac{r\sin\left(a\right)}{1-r\cos\left(a\right)}\right).\end{eqnarray*}

When $y=1$, \begin{eqnarray*}
I_{10} & = & \frac{1}{4\pi}\int_{I_{R}}\frac{1+\cos\left(t\right)}{1-\cos\left(t\right)}dt\\
 & = & \frac{1}{2\pi}\left(-\pi+a+\frac{2}{\tan\left(\frac{a}{2}\right)}\right).\end{eqnarray*}

\subsection{\label{sub:Calculation-of-C-CSITR}Calculation of the normalized
capacity for CSITR case}

For CSITR case,\begin{eqnarray*}
\bar{C}_{\infty} & = & \int_{a}^{\pi}\ln\left(\frac{\nu}{y}\left(1+r^{2}-2r\cos\left(t\right)\right)\right)f_{T}\left(t\right)dt\\
 & = & \ln\left(\frac{\nu}{y}\right)\bar{s}_{\infty}+\frac{1}{2}\int_{I_{R}}\ln\left(1+r^{2}-2r\cos\left(t\right)\right)f_{T}\left(t\right)dt,\end{eqnarray*}
where $r=\sqrt{y}$, $I_{R}$ is defined as in (\ref{eq:I_R}), $\bar{s}_{\infty}$
can be evaluated according to Proposition \ref{pro:calculation-s}
and the second line follows from the fact that the integrand is even.
Define \begin{eqnarray*}
I_{11} & = & \frac{1}{2}\int_{I_{R}}\ln\left(1+r^{2}-2r\cos\left(t\right)\right)f_{T}\left(t\right)dt,\end{eqnarray*}
we are going to evaluate $I_{11}$ for $y<1$ and $y=1$ respectively.

When $y<1$, since\begin{eqnarray*}
1+r^{2}-2r\cos\left(t\right) & = & \left(1-re^{it}\right)\left(1-re^{-it}\right)\end{eqnarray*}
and\begin{eqnarray*}
\int_{I_{R}}\ln\left(1-re^{it}\right)f_{T}\left(t\right)dt & = & \int_{I_{R}}\ln\left(1-re^{-it}\right)f_{T}\left(t\right)dt,\end{eqnarray*}
$I_{11}$ can be expressed as \begin{eqnarray*}
I_{11} & = & \int_{I_{R}}\ln\left(1-re^{it}\right)f_{T}\left(t\right)dt\\
 & = & \frac{1}{2\pi}\int_{I_{R}}\ln\left(1-re^{it}\right)\frac{2-e^{2it}-e^{-2it}}{1+r^{2}-2r\cos\left(t\right)}dt.\end{eqnarray*}
Expand the integrand like what have been done in (\ref{eq:fraction-expansion-1})
and (\ref{eq:fraction-expansion-2}), then \begin{eqnarray*}
I_{11} & = & \frac{1}{2\pi r}\left[\left(\frac{1}{r}+r\right)I_{12}+I_{13}+\left(r-\frac{1}{r}\right)I_{14}\right],\end{eqnarray*}
where\begin{eqnarray*}
I_{12} & = & \int_{I_{R}}\ln\left(1-re^{it}\right)dt,\end{eqnarray*}
\begin{eqnarray*}
I_{13} & = & \int_{I_{R}}\ln\left(1-re^{it}\right)\left(e^{it}+e^{-it}\right)dt\end{eqnarray*}
and \begin{eqnarray*}
I_{14} & = & \int_{I_{R}}\ln\left(1-re^{it}\right)\left(\frac{re^{it}}{1-re^{it}}+\frac{1}{1-re^{-it}}\right)dt.\end{eqnarray*}
By similar analysis in (\ref{eq:I1-result}) and (\ref{eq:I2-result}),
\begin{eqnarray}
I_{12} & = & i\left[\mathrm{Li}_{2}\left(re^{-ia}\right)-\mathrm{Li}_{2}\left(re^{ia}\right)\right]\label{eq:I12-result}\end{eqnarray}
and\begin{eqnarray}
I_{13} & = & 2\left[-\sin\left(a\right)\ln\left(1+r^{2}-2r\cos\left(a\right)\right)\right.\nonumber \\
 &  & \left.+\sin\left(a\right)-r\left(\pi-a\right)-\left(\frac{1}{r}-r\right)\theta_{r}\right]\label{eq:I13-result}\end{eqnarray}
where \begin{eqnarray}
\theta_{r} & = & \tan^{-1}\left(\frac{r\sin\left(a\right)}{1-r\cos\left(a\right)}\right).\label{eq:theta-r-def}\end{eqnarray}
To evaluate $I_{14}$, define\begin{eqnarray*}
I_{15} & = & \int_{I_{R}}\ln\left(1-re^{it}\right)\frac{re^{it}}{1-re^{it}}dt\end{eqnarray*}
and \begin{eqnarray*}
I_{16} & = & \int_{I_{R}}\ln\left(1-re^{it}\right)\frac{1}{1-re^{-it}}dt,\end{eqnarray*}
then $I_{14}=I_{15}+I_{16}$. 

It is easy to evaluate $I_{15}$.

\begin{eqnarray}
I_{15} & = & i\int_{I_{R}}\ln\left(1-re^{it}\right)\frac{d\left(1-re^{it}\right)}{1-re^{it}}\nonumber \\
 & = & \frac{i}{2}\int_{I_{R}}d\ln^{2}\left(1-re^{it}\right)\nonumber \\
 & = & \frac{i}{2}\left(\ln^{2}\left(1-re^{-ia}\right)-\ln^{2}\left(1-re^{ia}\right)\right).\label{eq:I15-result}\end{eqnarray}
To evaluate $I_{16}$, express the integrand in series. Because $0<r<1$,
\begin{eqnarray*}
I_{16} & = & -\left[\int_{I_{R}}\sum_{k=1}^{\infty}\frac{\left(re^{it}\right)^{k}}{k}\sum_{l=0}^{\infty}\left(re^{-it}\right)^{l}dt\right]\\
 & = & -\int_{I_{R}}\sum_{l=0}^{\infty}\sum_{k=1}^{\infty}\frac{\left(re^{it}\right)^{k}}{k}\left(re^{-it}\right)^{k+l}dt\\
 &  & -\int_{I_{R}}\sum_{l=1}^{\infty}\sum_{k=l}^{\infty}\frac{\left(re^{it}\right)^{k}}{k}\left(re^{-it}\right)^{k-l}dt.\end{eqnarray*}
Define \begin{eqnarray*}
I_{17} & = & \int_{I_{R}}\sum_{l=0}^{\infty}\sum_{k=1}^{\infty}\frac{\left(re^{it}\right)^{k}}{k}\left(re^{-it}\right)^{k+l}dt\end{eqnarray*}
and \begin{eqnarray*}
I_{18} & = & \int_{I_{R}}\sum_{l=1}^{\infty}\sum_{k=l}^{\infty}\frac{\left(re^{it}\right)^{k}}{k}\left(re^{-it}\right)^{k-l}dt.\end{eqnarray*}
Then \begin{eqnarray}
I_{17} & = & \int_{I_{R}}\sum_{l=0}^{\infty}r^{-l}e^{ilt}\sum_{k=1}^{\infty}\frac{r^{2k}}{k}dt\nonumber \\
 & = & -\ln\left(1-r^{2}\right)\int_{I_{R}}\frac{1}{1-\frac{1}{r}e^{it}}dt\nonumber \\
 & = & -\ln\left(1-r^{2}\right)\left(\pi-a-\theta_{r}\right),\label{eq:I17-result}\end{eqnarray}
where $\theta_{r}$ is defined as in (\ref{eq:theta-r-def}) . $I_{18}$
is well defined because\begin{eqnarray*}
\left|I_{18}\right| & \leq & \int_{I_{R}}\left|\sum_{l=1}^{\infty}\sum_{k=l}^{\infty}\frac{\left(re^{it}\right)^{k}}{k}\left(re^{-it}\right)^{k-l}\right|dt\\
 & = & \int_{I_{R}}\left|r^{-2}\sum_{l=1}^{\infty}r^{l}e^{ilt}\left(\frac{1}{r^{2l-2}}\sum_{k=l}^{\infty}\frac{r^{2k}}{k}\right)\right|dt\\
 & \leq & -\int_{I_{R}}r^{-2}\sum_{l=1}^{\infty}r^{l}\ln\left(1-r^{2}\right)dt\\
 & = & -\int_{I_{R}}r^{-2}\frac{r}{1-r}\ln\left(1-r^{2}\right)dt.\end{eqnarray*}
 Define \begin{eqnarray*}
\mathrm{Sr}_{2}\left(r,t\right) & = & \sum_{l=1}^{\infty}\frac{\left(re^{it}\right)^{l}}{l}\left(\frac{1}{r^{2l}}\sum_{k=l}^{\infty}\frac{r^{2k}}{k}\right),\end{eqnarray*}
 then \begin{eqnarray}
I_{18} & = & \int_{I_{R}}\sum_{l=1}^{\infty}r^{l}e^{ilt}\frac{1}{r^{2l}}\sum_{k=l}^{\infty}\frac{r^{2k}}{k}dt\nonumber \\
 & = & i\mathrm{Sr}_{2}\left(r,a\right)-i\mathrm{Sr}_{2}\left(r,-a\right).\label{eq:I18-result}\end{eqnarray}
Conclusively, \begin{eqnarray*}
\bar{C} & = & \ln\left(\frac{\nu}{y}\right)\bar{s}+\frac{1}{2\pi r}\left[\left(\frac{1}{r}+r\right)I_{12}\right.\\
 &  & \left.+I_{13}+\left(r-\frac{1}{r}\right)\left(I_{15}-I_{17}-I_{18}\right)\right]\end{eqnarray*}
where $\bar{s}$, $I_{12}$, $I_{13}$, $I_{15}$, $I_{17}$ and $I_{18}$
can be evaluated by (\ref{eq:I12-result}, \ref{eq:I13-result}, \ref{eq:I15-result}-\ref{eq:I18-result})
respectively.

When $y=1$,\begin{eqnarray*}
I_{11} & = & \frac{1}{2}\int_{I_{R}}\ln\left(2-2\cos\left(t\right)\right)f_{T}\left(t\right)dt\\
 & = & \frac{1}{2}\int_{I_{R}}\mathrm{Re}\left[\ln\left\{ \left[-i\left(e^{i\frac{t}{2}}-e^{-i\frac{t}{2}}\right)\right]^{2}\right\} \right]f_{T}\left(t\right)dt\\
 & = & \int_{I_{R}}\mathrm{Re}\left[-\frac{\pi}{2}i+i\frac{t}{2}+\ln\left(1-e^{-it}\right)\right]f_{T}\left(t\right)dt\\
 & = & \int_{I_{R}}\ln\left(1-e^{-it}\right)f_{T}\left(t\right)dt.\end{eqnarray*}
 Substitute $f_{T}\left(t\right)$ into it, \begin{eqnarray*}
I_{11} & = & \frac{1}{2\pi}\int_{I_{R}}\ln\left(1-e^{-it}\right)\left(e^{it}+2+e^{-it}\right)dt.\end{eqnarray*}
By similar analysis in (\ref{eq:I1-result}) and (\ref{eq:I2-result}),
\begin{eqnarray*}
 &  & \frac{1}{2\pi}\int_{I_{R}}\ln\left(1-e^{-it}\right)\left(e^{it}+e^{-it}\right)dt\\
 &  & =\frac{1}{\pi}\left[-\left(\pi-a\right)+\sin\left(a\right)-\sin\left(a\right)\ln\left(2-2\cos\left(a\right)\right)\right]\end{eqnarray*}
and \begin{eqnarray*}
\frac{1}{\pi}\int_{I_{R}}\ln\left(1-e^{-it}\right)dt & = & \frac{i}{\pi}\left(\mathrm{Li}_{2}\left(e^{-ia}\right)-\mathrm{Li}_{2}\left(e^{ia}\right)\right).\end{eqnarray*}
Then the proposition is proved.

\subsection{\label{sub:Proof-of-multirank-B}Proof of Theorem \ref{thm:optimal-feedback-multirank-B}}

If the optimal number of on-beams $\tilde{s}\left(\mathbf{H}\right)$
is known, the optimal feedback function is given by \[
\tilde{\varphi}\left(\mathbf{H}\right)=\arg\;\underset{i:\;\mathbf{Q}_{i}\in\mathcal{B}_{\tilde{s}\left(\mathbf{H}\right)}}{\max}\;\ln\left|\mathbf{I}_{L_{R}}+P_{\mathrm{on}}\mathbf{H}\mathbf{Q}_{i}\mathbf{Q}_{i}^{\dagger}\mathbf{H}^{\dagger}\right|.\]
Thus the only nontrivial part is to prove the optimality of \[
\tilde{s}\left(\mathbf{H}\right)=\max\left\{ s:\;\mathcal{I}_{s}\left(\mathbf{H}\right)-\mathcal{I}_{t}\left(\mathbf{H}\right)\geq(s-t)\kappa\;\textrm{for}\;\textbf{all}\; t\;\textrm{s.t.}\;0\leq t<s\;\mathcal{B}_{t}\neq\phi\;\mathcal{B}_{s}\neq\phi\right\} \]
where \[
\mathcal{I}_{s}\left(\mathbf{H}\right)=\underset{\mathbf{Q}_{i}\in\mathcal{B}_{s}}{\max}\;\ln\left|\mathbf{I}_{L_{R}}+P_{\mathrm{on}}\mathbf{HQ}_{i}\mathbf{Q}_{i}^{\dagger}\mathbf{\mathbf{H}}^{\dagger}\right|.\]

The following lemma is useful to prove the optimality of $\tilde{s}\left(\mathbf{H}\right)$.
For simplicity, we denote $\tilde{s}\left(\mathbf{H}\right)$ by $\tilde{s}$
from now on. It is necessary to keep in mind that $\tilde{s}$ is
not a constant but a function of the channel realization $\mathbf{H}$.

\begin{lemma}
\label{lem:s-tilde-marginal-power-efficiency}For $\forall t>\tilde{s}$
such that $\mathcal{B}_{t}\neq\phi$, $\mathcal{I}_{t}\left(\mathbf{H}\right)-\mathcal{I}_{\tilde{s}}\left(\mathbf{H}\right)<\left(t-\tilde{s}\right)\kappa$.
\end{lemma}
\begin{proof}
Suppose that this lemma is not true. $\exists t>\tilde{s}$ such that
$\mathcal{B}_{t}\neq\phi$ and $\mathcal{I}_{t}\left(\mathbf{H}\right)-\mathcal{I}_{\tilde{s}}\left(\mathbf{H}\right)\geq\left(t-\tilde{s}\right)\kappa$.
Take the minimum such $t$ and denote it as $t_{0}$,\[
t_{0}=\min\;\left\{ t>\tilde{s}:\;\mathcal{B}_{t}\neq\phi\;\mathcal{I}_{t}\left(\mathbf{H}\right)-\mathcal{I}_{\tilde{s}}\left(\mathbf{H}\right)\geq\left(t-\tilde{s}\right)\kappa\right\} .\]

Then $\forall t\;\mathrm{s.t.}\;0\leq t\leq\tilde{s}$, \begin{eqnarray*}
\mathcal{I}_{t_{0}}-\mathcal{I}_{t} & = & \mathcal{I}_{t_{0}}-\mathcal{I}_{\tilde{s}}+\mathcal{I}_{\tilde{s}}-\mathcal{I}_{t}\\
 & \geq & \left(t_{0}-\tilde{s}\right)\kappa+\left(\tilde{s}-t\right)\kappa\\
 & = & \left(t_{0}-t\right)\kappa,\end{eqnarray*}
where the inequality follows from the definitions of $\tilde{s}$
and $t_{0}$. At the same time, for a $t$ s.t. $\tilde{s}<t<t_{0}$,
$\mathcal{I}_{t}-\mathcal{I}_{\tilde{s}}<\left(t-\tilde{s}\right)\kappa$
according to the definition of $t_{0}$ and the fact that $t<t_{0}$.
Then \begin{eqnarray*}
\mathcal{I}_{t_{0}}-\mathcal{I}_{t} & = & \mathcal{I}_{t_{0}}-\mathcal{I}_{\tilde{s}}+\mathcal{I}_{\tilde{s}}-\mathcal{I}_{t}\\
 & \geq & \left(t_{0}-\tilde{s}\right)\kappa-\left(t-\tilde{s}\right)\kappa\\
 & = & \left(t_{0}-t\right)\kappa.\end{eqnarray*}

Thus, $\mathcal{I}_{t_{0}}-\mathcal{I}_{t}\geq\left(t_{0}-t\right)\kappa$
for $\forall t\leq t_{0}$ and $\mathcal{B}_{t}\neq\phi$, which contradicts
with the definition of $\tilde{s}$. This lemma is proved.
\end{proof}

To prove $\tilde{s}$ is optimal, we compare $\tilde{\varphi}\left(\cdot\right)$
with an \emph{arbitrary} deterministic feedback function $\varphi'\left(\cdot\right)$
satisfying the power constraint. Let $s'=\mathrm{rank}\left(\mathbf{Q}_{\varphi'\left(\mathbf{H}\right)}\right)$
be the number of on-beams according to the feedback function $\varphi'\left(\cdot\right)$.
Let $F_{\mathbf{H}}\left(\mathbf{H}\right)$ denote the CDF of the
channel state $\mathbf{H}$. The power constraint can be expressed
as \[
\int_{\mathbb{C}^{L_{R}\times L_{T}}}s'P_{\mathrm{on}}dF_{\mathbf{H}}\left(\mathbf{H}\right)=\rho.\]
Define $\Delta s\triangleq\tilde{s}-s'$ and \[
\Omega_{\Delta s}=\left\{ \mathbf{H}\in\mathbb{C}^{L_{R}\times L_{T}}:\;\tilde{s}-s'=\Delta s\right\} \]
where $-L_{T}\leq\Delta s\leq L_{T}$. Since both $\tilde{\varphi}\left(\cdot\right)$
and $\varphi'\left(\cdot\right)$ satisfy the power constraint, we
have \[
\sum_{\Delta s=-L_{T}}^{L_{T}}\int_{\Omega_{\Delta s}}\Delta s\cdot P_{\mathrm{on}}dF_{\mathbf{H}}\left(\mathbf{H}\right)=0.\]
On the other hand, the performance difference between $\tilde{\varphi}\left(\cdot\right)$
and $\varphi'\left(\cdot\right)$ is given by \begin{eqnarray*}
 &  & \int_{\mathbb{C}^{L_{R}\times L_{T}}}\mathcal{I}_{\tilde{s}}\left(\mathbf{H}\right)dF_{\mathbf{H}}\left(\mathbf{H}\right)\\
 &  & \quad\quad-\int_{\mathbb{C}^{L_{R}\times L_{T}}}\ln\left|\mathbf{I}_{L_{R}}+P_{\mathrm{on}}\mathbf{H}\mathbf{Q}_{\varphi'\left(\mathbf{H}\right)}\mathbf{Q}_{\varphi'\left(\mathbf{H}\right)}^{\dagger}\mathbf{H}^{\dagger}\right|dF_{\mathbf{H}}\left(\mathbf{H}\right)\\
 &  & \overset{\left(a\right)}{\geq}\int_{\mathbb{C}^{L_{R}\times L_{T}}}\mathcal{I}_{\tilde{s}}\left(\mathbf{H}\right)dF_{\mathbf{H}}\left(\mathbf{H}\right)-\int_{\mathbb{C}^{L_{R}\times L_{T}}}\mathcal{I}_{s'}\left(\mathbf{H}\right)dF_{\mathbf{H}}\left(\mathbf{H}\right)\\
 &  & =\sum_{\Delta s=-L_{T}}^{L_{T}}\int_{\Omega_{\Delta s}}\left(\mathcal{I}_{\tilde{s}}\left(\mathbf{H}\right)-\mathcal{I}_{s'}\left(\mathbf{H}\right)\right)dF_{\mathbf{H}}\left(\mathbf{H}\right)\\
 &  & =\sum_{\Delta s=-L_{T}}^{-1}\int_{\Omega_{\Delta s}}\left(\mathcal{I}_{\tilde{s}}\left(\mathbf{H}\right)-\mathcal{I}_{s'}\left(\mathbf{H}\right)\right)dF_{\mathbf{H}}\left(\mathbf{H}\right)\\
 &  & \quad\quad+\sum_{\Delta s=1}^{L_{T}}\int_{\Omega_{\Delta s}}\left(\mathcal{I}_{\tilde{s}}\left(\mathbf{H}\right)-\mathcal{I}_{s'}\left(\mathbf{H}\right)\right)dF_{\mathbf{H}}\left(\mathbf{H}\right)\\
 &  & \overset{\left(b\right)}{\geq}-\sum_{\Delta s=-L_{T}}^{-1}\int_{\Omega_{\Delta s}}\left|\Delta s\right|\cdot\kappa dF_{\mathbf{H}}\left(\mathbf{H}\right)+\sum_{\Delta s=1}^{L_{T}}\int\Delta s\cdot\kappa dF_{\mathbf{H}}\left(\mathbf{H}\right)\\
 &  & =\kappa\sum_{\Delta s=-L_{T}}^{L_{T}}\int\Delta s\; dF_{\mathbf{H}}\left(\mathbf{H}\right)\\
 &  & \overset{\left(c\right)}{=}0,\end{eqnarray*}
where 

\begin{quote}
(a) follows from the fact that \begin{eqnarray*}
 &  & \ln\left|\mathbf{I}_{L_{R}}+P_{\mathrm{on}}\mathbf{H}\mathbf{Q}_{\varphi'\left(\mathbf{H}\right)}\mathbf{Q}_{\varphi'\left(\mathbf{H}\right)}^{\dagger}\mathbf{H}^{\dagger}\right|\\
 &  & \leq\underset{\mathbf{Q}_{i}\in\mathcal{B}_{s'}}{\max}\;\ln\left|\mathbf{I}_{L_{R}}+P_{\mathrm{on}}\mathbf{H}\mathbf{Q}_{i}\mathbf{Q}_{i}^{\dagger}\mathbf{H}^{\dagger}\right|\\
 &  & =\mathcal{I}_{s'}\left(\mathbf{H}\right),\end{eqnarray*}

(b) follows from Lemma \ref{lem:s-tilde-marginal-power-efficiency}
and the definition of $\tilde{s}$, and

(c) follows from the power constraint.
\end{quote}
Therefore, $\tilde{\varphi}\left(\cdot\right)$ is the optimal feedback
function.

\newpage
\bibliographystyle{IEEEtran}
\bibliography{Bib/_Blum,Bib/_Heath,Bib/_Liu_Dai,Bib/_love,Bib/_Rao,Bib/_Tse,Bib/FeedbackMIMO_append,Bib/MIMO_basic,Bib/RandomMatrix,Bib/_Zhou}

\newpage

\newpage

\newpage

\newpage

\newpage

\end{document}